\documentclass[useAMS,usenatbib,usegraphics]{mn2e}

\usepackage{graphicx}
\usepackage{latexsym,amssymb}
\usepackage{amsmath}
\usepackage{gensymb}
\usepackage[draft]{hyperref}
\usepackage{capt-of}
\usepackage{textpos}
\usepackage{journalnames}
\usepackage{url}
\usepackage{multirow}
\usepackage{color}
\usepackage[T1]{fontenc}
\usepackage{float}
\usepackage{rotating}
\usepackage{tikz}
\usepackage{tablefootnote}
\usepackage{longtable}
\usepackage{footnote}
\usepackage{framed}
\usepackage{pgfgantt,rotating}
\usepackage{subfloat}
\usepackage{multirow}
\usepackage{blindtext}
\usepackage{booktabs}
\usepackage{tabu}
\usepackage{threeparttablex}

\hypersetup{
    bookmarks=true,         
    unicode=false,          
    pdftoolbar=true,        
    pdfmenubar=true,        
    pdffitwindow=false,     
    pdfstartview={FitH},    
    pdftitle={My title},    
    pdfauthor={Author},     
    pdfsubject={Subject},   
    pdfcreator={Creator},   
    pdfproducer={Producer}, 
    pdfkeywords={keyword1} {key2} {key3}, 
    pdfnewwindow=true,      
    colorlinks=false,       
    linkcolor=black,          
    citecolor=blue,        
    filecolor=black,      
    urlcolor=black,           
    linkbordercolor={1 0 0},
    citebordercolor={0 1 0}
}

\makeatletter
\Hy@AtBeginDocument{%
  \def\@pdfborder{0 0 1}
  \def\@pdfborderstyle{/S/U/W 0}
}
\makeatother

\title[Hunting for brown dwarf binaries  and testing atmospheric models with X-Shooter]{Hunting for brown dwarf binaries  and testing atmospheric models with X-Shooter\thanks{Based on observations of ESO, using VLT/ESO, under the programs 084.C-1092(A), 085.C-0862(A) and 085.C-0862(B).}}
\author[E. Manjavacas et al.]
{\parbox{\textwidth}{E. Manjavacas$^{1,2}$	\thanks{E-mail:  manjavacas@iac.es; manjavacas@mpia.de},
B. Goldman$^{2, 3}$, 
J. M. Alcal\'a$^{4}$,
M. R. Zapatero-Osorio$^{5}$,
V. J. S. B\'ejar$^{1,6}$, 
D. Homeier$^{7, 8}$,
M. Bonnefoy$^{9,10}$,
R. L. Smart$^{11}$, 
T. Henning$^{2}$, 
F. Allard$^{8}$}\vspace{0.4cm}\\
\parbox{\textwidth}{
$^{1}$Instituto de Astrof\'isica de Canarias, C/ V\'ia L\'actea, s/n, E38205 - La Laguna (Tenerife), Spain.\\
$^{2}$Max Planck Institute f\"ur Astronomie
K\"onigstuhl, 17. D-69117 Heidelberg, Germany\\
$^{3}$Observatoire astronomique de Strasbourg, Universit\'e de Strasbourg, 11 rue de l'Universit\'e, F-67000 Strasbourg, France.\\
$^{4}$INAF - Osserv. Astronomico di Capodimonte, Via Moiariello 16, 80131, Napoli, Italy.\\
$^{5}$Centro de Astrobiolog\'ia  (CSIC-INTA).  Carretera de Ajalvir  km 4, E-28850 Torrej\'on de Ardoz, Madrid, Spain.\\
$^{6}$Universidad de La Laguna, Dpto. Astrof\'isica, E-38206 La Laguna (Tenerife), Spain.\\
$^{7}$Zentrum f{\"u}r Astronomie der Universit{\"a}t Heidelberg, Landessternwarte
K\"onigstuhl 12, 69117 Heidelberg.\\
$^{8}$CRAL-ENS, 46, All\'{e}e  d'Italie , 69364 Lyon Cedex 07, France.\\
$^{9}$Univ. Grenoble Alpes, IPAG, F-38000 Grenoble, France.\\
$^{10}$CNRS, IPAG F-38000 Grenoble, France.\\
$^{11}$Instituto Nazionale di Astrofisica, Osservatorio Astrofisico di Torino, Strada Osservatorio 20, 10025 Pino Torinese, Italy.}}

\begin{document}


\pagerange{\pageref{firstpage}--\pageref{lastpage}} \pubyear{2002}

\def\LaTeX{L\kern-.36em\raise.3ex\hbox{a}\kern-.15em
    T\kern-.1667em\lower.7ex\hbox{E}\kern-.125emX}

\maketitle

\label{firstpage}

\begin{abstract}
	{The determination of the brown dwarf binary fraction   may contribute  to the understanding of the  substellar formation mechanisms.   Unresolved brown dwarf binaries may be revealed through their peculiar spectra or the discrepancy between optical and near-infrared spectral type classification.
	We obtained medium-resolution spectra of 22  brown dwarfs  with these characteristics using the X-Shooter spectrograph at the VLT.
	We aimed to identify brown dwarf binary candidates, and to test if the BT-Settl 2014 atmospheric models reproduce their observed spectra.
	 To find binaries spanning the L-T boundary, we used spectral indices and  compared the spectra of the selected candidates to single spectra and synthetic binary spectra. We  used synthetic binary spectra with components of same spectral type to determine as well the sensitivity of the method to this class of binaries. 
	We identified three candidates to be combination of L plus T brown dwarfs. We are not able to identify  binaries with components of similar spectral type.  In our sample, we measured minimum binary fraction of $9.1^{+9.9}_{-3.0}$\%.	
	From the best fit of the BT-Settl models 2014 to the observed spectra, we derived the atmospheric parameters for  the single objects. 
The  BT-Settl  models were able to reproduce the majority of the SEDs from our objects, and the variation of the equivalent width of the Rb\,I (794.8 nm) and Cs\,I (852.0 nm) lines with the spectral type. Nonetheless, these models  did not reproduce  the evolution of the equivalent widths of the Na\,I (818.3 nm and 819.5 nm) and K\,I (1253 nm) lines with the spectral type.}

\end{abstract}


\section{Introduction}
\label{sec:introduction}

	Stars are believed to be born in large stellar nurseries, which they eventually leave to form the field population. A large number of stars remain in binary or hierarchical
	systems. {Multiplicity  provides  constraints on fundamental parameters, such
		as dynamical masses, essential to test atmospheric and substellar formation models.}
	It is well known that the binary fraction decreases when decreasing mass. This
	fraction decreases from 80\%-60\% for O and B  stars, to 40\% for the M dwarfs \citep{2012ApJ...754...44J}. {The  decreasing trend for binarity seems to extend to the substellar regime. For L- and T- brown dwarfs,  the binary fraction is estimated at about 20\% \citep{Gizis2003, Burgasser2003, Bouy2003, Close2003, Burgasser2007, Luhman2007, Goldman2008}.}

	Based on Monte Carlo simulations, \citet{Allen2007} determined that 98\% of the brown dwarf binaries have separations
	smaller than 20 AU. \citet{Burgasser2007} pointed
	out that the peak of the separation distribution of brown dwarfs is
	$\sim$ 3 AU, which is very close to the  limit of the high
	resolution imaging surveys. {\citet{Allen2007} estimated that a fraction of $\sim$6-7\% of brown dwarf binary systems have not been detected yet, as a consequence of observational
		biases.}   For instance, \citet{Joergens2008} searched for low-mass stars and brown dwarfs
	binaries in the Chamaeleon  star forming region using the radial velocity method, and concluded that the
	percentage of brown dwarfs binary systems with separations below 1 AU is less
	than $\sim$10\% in this star forming region. \citet{Blake2010} monitored a sample of 59 ultra-cool dwarfs with radial velocity and determined that the binary frequency of low-mass unresolved systems is $2.5^{+8.6}_{-1.6}$\%. 
	
	{Spectroscopic data provide also important constraints to atmospheric models.}
	These models  allow us to disentangle the effect of varying effective temperature, gravity, and metallicity on the spectral features. Below a effective temperature of $\sim$2600 K, models predict that clouds of iron and silicate
	grains begin to form, affecting the opacity (\citealt{Lunine}, \citealt{Tsuji},
	\citealt{Burrows_Sharp}, \citealt{Lodders}, \citealt{Marley2000},
	\citealt{Marley2001}, \citealt{Allard2001}).
	Self-consistent atmospheric models, such as the BT-Settl models \citep{Allard2012a}
	and the Drift-PHOENIX models \citep{Helling}, use cloud models where the dust
	properties do not require {the definition of any other} additional free
	parameters other than gravity, effective temperature and metallicity. Synthetic
	spectra for a specific set of atmospheric parameters can be compared to
	empirical spectra. For instance, these models have been tested  on spectra of
	young late-type objects \citep[late-type companions and free-floating
	objects;][]{Bonnefoy2010, 2011A&A...529A..44W, 2012A&A...540A..85P,
		Bonnefoy2013, Manjavacas2014}.

	In this paper, {we present X-Shooter optical and near infrared spectroscopy of 22 peculiar ultra-cool dwarfs, with spectral types between L3 and T7. We aim to find unresolved brown dwarf binary systems. Our sample consists of  objects with a different spectral
		classification in the optical and in the near-infrared or peculiar spectra in
		comparison with objects of the same spectral type.} 
	We want to contribute to the census of unresolved low mass dwarfs,  and at the same time  provide new constraints on the BT-Settl 2014 atmospheric models.
	In Section \ref{sample_selection} we describe the procedure for the selection of
	candidates in our sample, we explain how the observations were performed, and the data 
	reduction procedure. In Section \ref{empirical_analysis} we first perform a search for L+T binaries in our sample, then perform a simulation to estimate the efficiency and false positive rate of the spectral fitting method we apply. We estimate the sensitivity of our method to detect spectral binaries with {the same subspectral type}.  We also compare our targets with trigonometric
	distances in a color magnitude diagram (CMD) with the L, L-T transition and T
	brown dwarfs published by \citet{Dupuy_Liu2012}. CMD allows us to
	discover unresolved binaries and young brown dwarfs.  In Section \ref{revised_properties} we discuss
	the properties of the binary candidates selected in Section \ref{empirical_analysis}.  In Section \ref{binary_fraction} we update the binary sample of very low mass objects. 
	{In Section \ref{models} we investigate how the BT-Settl 2014
		atmospheric models reproduce our spectra over the
		optical and the near-infrared.  We compare with the results from the literature, the equivalent width provided by the models.  }  Finally, in
	Section \ref{conclusions} we summarize our results.

	\section{The sample, observations and data reduction}\label{sample_selection}
	
	\subsection{Sample selection}\label{sample_selection0}
	
	We selected a  sample of 22 brown dwarfs found in the literature, with optical
	spectral types between L3 and T7,  that have discrepant optical and
	near-infrared classification, or peculiar spectra. Optical subtypes are typically earlier than the near infrared {subtypes}. These objects are candidates to be unresolved binaries. {We selected the brightest such objects to ensure sufficient SNR in a reasonable integration time (J < 16).} Furthermore, to calibrate our results and confirm the reliability of our method, {we added some known brown dwarfs
		systems, LHS 102B \citep{Golimowski_2004}, formed by a L4.5 plus a L4.5, and  SDSS J042348.56-041403.4 \citep{Burgasser2005b}, formed by a L6$\pm$1 and a T2$\pm$1.}  Our list of targets and their physical properties
	taken from  the literature are compiled in Table
	\ref{literature}.

		\begin{table*}
			\scriptsize
			\caption{List of observed targets: Magnitudes are in the 2MASS system, except for object Gl 229B for which magnitudes are given by \citet{Leggett1999} in the UKIRT system.}
			\label{literature}
			\centering
\begin{minipage}{18cm}
			\begin{center}
				\begin{tabular}{llllllllll}
					\hline
					\hline 
					
					Number &Name & J [mag]& H [mag] & K [mag]  & $d_{trig}$ (pc)  & SpT OPT & SpT NIR & Remarks &  Reference\\		
					
					\hline              
					1 & LHS 102B                & 13.11$\pm$0.02 & 12.06$\pm$0.02 & 11.39$\pm$0.02 & 13.2$\pm$0.7 & L5 & L4.5 & Binary & 1, 2\\
					2 & 2MASS J00361617+1821104 & 12.47$\pm$0.02 & 11.59$\pm$0.03 & 11.31$\pm$0.02 & 8.8$\pm$0.1 & L3.5 & L4 & NR$^{a}$, V$^{b}$ & 3, 4, 36 \\
					3 & 2MASS J00531899-3631102 & 14.45$\pm$0.02 & 13.48$\pm$0.03 & 12.94$\pm$0.02 &              & L3.5 & L4 & & 5, 6\\
					4 & SIMP  01365662+0933473  & 13.46$\pm$0.03  & 12.77$\pm$0.03 & 12.56$\pm$0.02 &  6.0$\pm$0.1  &      & T2.5 &V & 7, 8\\
					5 & 2MASS J01443536-0716142 & 14.19$\pm$0.02 & 13.01$\pm$0.02 & 12.27$\pm$0.02 &             & L5   & & Red & 9, 10\\
					6 & 2MASS J02182913-3133230 & 14.73$\pm$0.04  & 13.81$\pm$0.04 & 13.15$\pm$0.04 &               & L3 & L5.5 & & 5,11 \\
					7 & DENIS-P J0255.0-4700    & 13.25$\pm$0.02 & 12.20$\pm$0.02 & 11.56$\pm$0.02 & 4.9$\pm$0.1 & L8 & L9 &V &12, 13, 35\\
					8 & 2MASS J02572581-3105523 & 14.67$\pm$0.03 & 13.52$\pm$0.03 & 12.88$\pm$0.03 & 10.0$\pm$0.7  & L8 & L8.5 & & 4, 5, 14 \\
					9 &2MASS J03480772-6022270 & 15.32$\pm$0.05 & 15.56$\pm$0.14 & 15.60$\pm$0.02 &   7.9$\pm$0.2  &  & T7 & & 15, 16 \\
					10 &2MASS J03552337+1133437 & 14.05$\pm$0.02 & 12.53$\pm$0.03 & 11.53$\pm$0.02 & 9.1$\pm$0.1  & L5 & L3 & Y$^{c}$& 1, 17, 18, 19, 20, 37\\
					11 &SDSS J0423485-041403  & 14.47$\pm$0.02  & 13.46$\pm$0.03 & 12.93$\pm$0.03 &15.2$\pm$0.4& L7.5 & T0 & Binary & 1, 21, 33 \\
					12 &2MASS J04390101-2353083 & 14.41$\pm$0.02 & 13.41$\pm$0.02 & 12.82$\pm$0.02 & 9.1$\pm$0.3 & L6.5&   & & 11, 19\\
					13 &2MASS J04532647-1751543 & 15.14$\pm$0.03 & 14.06$\pm$0.03 & 13.47$\pm$0.03 &             & L3pec & & Y? & 11, 14 \\
					14 &2MASS J05002100+0330501 & 13.67$\pm$0.02 & 12.68$\pm$0.02 & 12.06$\pm$0.02 &             & L4 & L4 & &1, 22\\
					15 &2MASS J05395200-0059019 & 14.03$\pm$0.03  & 13.10$\pm$0.02& 12.53$\pm$0.02& 13.1$\pm$0.4 & L5 & L5 & NR  &1, 4, 24\\
					16 &2MASS J06244595-4521548 & 14.48$\pm$0.02 & 13.34$\pm$0.02 & 12.59$\pm$0.02& 11.9$\pm$0.6 & L5pec  & L5 & & 1, 23\\
			        17 &Gl 229B                                  & 13.97$\pm$0.03  &  14.38$\pm$0.03  & 14.55$\pm$0.03  & 5.8$\pm$0.4 &   & T7pec & MP$^{d}$, Y &33, 34, 35 \\
					18 &2MASS J10043929-3335189 & 14.48$\pm$0.04 & 13.49$\pm$0.04 & 12.92$\pm$0.02 & 17.0$\pm$1.6 & L4 & L5 & & 25, 26 \\
					19 &2MASS J11263991-5003550 & 14.00$\pm$0.03  & 13.28$\pm$0.03 & 12.83$\pm$0.03 &             & L4.5 & L6.5 & Blue L & 27, 28, 29\\
					20 &2MASS J13411160-3052505 & 14.61$\pm$0.03  & 13.72$\pm$0.03  & 13.08$\pm$0.02 &              & L2pec &  L3 & & 22\\
					21 &2MASS J18283572-4849046 & 15.18$\pm$0.05 & 14.91$\pm$0.06 & 15.18$\pm$0.14 & 11.9$\pm$1.1  &  & T5.5 & & 23, 31\\
					22 &2MASS J21513839-4853542 & 15.73$\pm$0.07 & 15.17$\pm$0.09 & 15.43$\pm$0.18 &  16.7$\pm$1.1           &  & T4 & &30 \\
					
					\hline

				\end{tabular}

			\end{center}
									  \begin{tablenotes}
									  	\raggedright
									  \item 	 References: [1] - \citet{Reid2008_2}, [2] - \citet{Burgasser2007}, [3] - \citet{Dahn2002}, [4] - \citet{Schneider2014}, [5] - \citet{Marocco2013}, [6] - \citet{Martin2010}, [7] - \citet{Artigau2006}, [8] - \citet{Radigan2013}, [9] - \citet{Burgasser2011}, [10] - \citet{Liebert2003}, [11] - \citet{Liebert2003}, [12] - \citet{Cruz2003}, [13] - \citet{Castro2013}, [14] - \citet{Kirkpatrick2008},   [15] - \citet{Burgasser2003}, [16] - \citet{Parker2013}, [17] - \citet{Cruz},  [18] - \citet{Allers2013}, [19] - \citet{Faherty2013}, [20] - \citet{Gagne2014}, [21] - \citet{Antonova2013}, [22] - \citet{Antonova2013}, [23] - \citet{Faherty2012}, [24] - \citet{Leggett2000}, [25] - \citet{2011AJ....141...54A}, [26] - \citet{Gizis2002}, [27] - \citet{Folkes2007}, [28] - \citet{Faherty2009}, [29] - \citet{Burgasser2008}, [30] - \citet{Ellis2005}, [31] - \citet{Burgasser2004},  [32] - \citet{Vrba2004}, [33] - \citet{1995Natur.378..463N},  [34] - \citet{Oppenheimer2001},  [35] - \citet{Costa2006}, [36] - \citet{Gelino2002}, [37] - \citet{Zapatero_Osorio2014}. \\
									  	(a) {NR}: Not resolved binary; (b) {V}: Variability found; (c) {Y}: Young; (d) {MP}: Metal poor.
									  \end{tablenotes}				  
			\end{minipage}
		\end{table*}

			\subsection{Observations and data reduction}\label{data_reduction}
			
			Our targets were observed using X-Shooter (Wideband ultraviolet-infrared
			single target spectrograph) on the Very Large Telescope (VLT) between October
			2009 and June 2010. X-Shooter  
			is composed of three arms: UVB (300-550 nm), optical (550-1000 nm) and
			near-infrared (1000-2500 nm). It was operated in echelle slit nod mode, using
			the 1.6" slit width for the UVB arm, and the 1.5" slit width for the optical and the
			near-infrared arms. This setup provides resolutions of $\sim$3300 in the UVB
			and NIR, and $\sim$5400 in the VIS. We obtained an average signal to noise of $\sim$30. Observations were performed at the parallactic angle to mitigate the effect of differential chromatic refraction. We moved the object along the slit between
			two positions following an ABBA pattern with a size of 6 arcsec. The flux expected in the UVB arm is
			extremely low, therefore we chose  not use spectra taken in this range. Telluric
			standards were observed before or after every target at a close airmass ($\pm$0.1 with respect to the targets). 
			Bias, darks and flats were taken  every night. Arc frames
			were taken  every second day. The  observing log  including telluric standard stars and {the raw seeing during the observations} is shown in  Table \ref{log0}.

			The spectra were reduced using the ESO X-Shooter pipeline version 1.3.7 \citep{Vernet2011}. In
			the reduction cascade, the pipeline deletes the non-linear pixels and subtracts
			bias in the optical or dark frames in the near-infrared. It generates a
			guess order from a format-check frame, a reference list of arc line and a
			reference spectral format table. It refines the  guess order table into an
			order table from an order definition frame obtained by illuminating the
			X-Shooter pinhole with a continuum lamp. The master flat frame and the order
			tables tracing the flat edges are created. 
			Finally, the pipeline determines the instrumental response and science data
			are reduced in slit nodding mode. 
			
			In the case of the near
			infrared,  we used the spectrum of the telluric star of the corresponding 
			science target observed in the same night to obtain the response function.  We removed cosmetics and cosmic rays from the
			telluric stars, as well as the H and He absorption lines on their spectra, using a Legendre
			polynomial fit of the pseudo-continuum around the line. We then derived a response function by dividing the non-flux calibrated clean spectrum of the telluric standard by a black body synthetic spectrum with the same temperature as the
			telluric star \citep{Theodossiour_Danezis1991}. Finally, to calibrate in
			response, we used the package \textit{noao.onedspec.telluric} from the software
			\textit{Image Reduction and Analysis Facility}, (IRAF). More details on data reduction and flux calibration, as well as correction for telluric bands, are described in \citet{Alcala2014}.

			To make sure that the flux in the whole near-infrared spectra was correctly
			scaled, we calibrated the flux of our near-infrared spectra using fluxes given by
			2MASS (Two Micron All Sky Survey). We convolved our near-infrared spectra with
			J, H and Ks filter transmission curves of 2MASS. The resulting spectra were
			integrated. We calculated the flux for our targets corresponding to the J, H
			and Ks bands using 2MASS magnitudes \citep{Cohen2003}.  Finally, we calculated
			the scaling factor for J, H and Ks bands and  multiplied our near-infrared
			spectra in J, H and Ks filters to have the same flux as  given by 2MASS. We
			scaled flux from the optical spectra to be consistent with the flux
			in the near-infrared. In the overlapping wavelengths of the optical
			and near-infrared spectra (995-1020 nm), we calculated a scaling factor, which is the median of the flux in the these wavelengths of the
			near-infrared spectra, divided by the median of the flux in the overlapping
			wavelengths of the optical spectra. The reduced spectra are shown in Fig. \ref{all_spectra}\footnote{These spectra will be available in the ESO Phase 3 data release}.  Wavelengths  affected by telluric absorption are removed from the figure, as well as the optical part for object Gl229B, because it is contaminated by the flux of its companion, and the optical of 2M0144 because it is  noisy.


	\section{Search for spectral binaries}\label{empirical_analysis}
	
	In this Section, we used different methods  to reveal unresolved brown dwarf binaries through their spectra. These methods are tailored to the type of brown dwarf binaries that we aim to find.

	\subsection{Finding L plus T brown dwarf binaries}\label{L_T}
	
	The combined spectra of L plus T brown dwarf binary systems are predicted to show peculiar characteristics. Those spectra are expected to have blended atomic and molecular absorptions of L and T brown dwarf spectra. This combination may result in a peculiar spectrum.
	\citet{Burgasser2007,Burgasser2010} and \citet{Bardalez_Gagliuffi} have studied the spectral characteristics of L plus T brown dwarf binary spectra and they have designed an empirical method to identify them using spectroscopy.
	
	Spectra of L plus T binary systems show bluer
	spectral energy distribution in the near-infrared than single objects of the same spectral type \citep{Burgasser2010}. Some
	spectral features vary: the $\mathrm{CH_{4}}$ and $\mathrm{H_{2}O}$ features at 1.1~$\mu$m  are
	deeper for binaries. The $\mathrm{CH_{4}}$ feature at 1.6~$\mu$m is stronger in
	comparison to the 2.2~$\mu$m $\mathrm{CH_{4}}$ band.  At 2.1~$\mu$m the flux peak is shifted to the blue for the binaries. They also show larger flux from the T dwarf at  1.55~$\mu$m \citep{Bardalez_Gagliuffi}. Using such differences,
	\citet{Burgasser2006,Burgasser2010} and \citet{Bardalez_Gagliuffi}  defined spectral indices to identify L plus T brown dwarf binary candidates. Typically, spectral indices are defined as the ratio of spectral flux in two different wavelength intervals. The indices are specified in
	Table~\ref{spectral_indices}.
	\citet{Burgasser2006,Burgasser2010} and \citet{Bardalez_Gagliuffi}  compared all indices against each other for a  large sample of brown dwarfs, some of them L plus  T known binaries. They identified the best pairs of indices that segregated known binaries from the rest of the objects, and selected the regions in each combination of indices that delimit the known L plus T known binaries.
	In Table~\ref{criteria} and Table~\ref{criteria2} these regions are defined.  
	{There are several differences between \citet{Burgasser2006,Burgasser2010} and \citet{Bardalez_Gagliuffi} methods. \citet{Burgasser2006,Burgasser2010} published eight indices that are valid only to find L plus T dwarf binaries, and defined six binary index selection criteria. The objects that satisfied two criteria were considered as "weak candidates". Those that satisfied three or more criteria are considered "strong candidates". {\citet{Bardalez_Gagliuffi}  used the eight spectral indices defined in \citet{Burgasser2006,Burgasser2010} and developed five new indices that are sensitive to M7-L7 plus T binaries. \citet{Bardalez_Gagliuffi}  defined 12 new binary selection criteria. Objects satisfying  four to eight  criteria were considered as "weak candidates".} Those that satisfied more than eight indices were considered "strong candidates".}


	{By calculating} these spectral indices  we  selected  those objects in our sample that are L plus T binary candidates. {The result using \citet{Burgasser2006, Burgasser2010} criteria is shown in  Figures~\ref{indices}, and the result using \citet{Bardalez_Gagliuffi} criteria is shown in Figures \ref{indices1} and \ref{indices2}. In Table \ref{all_selected_candidates}, we summarize the weak and strong candidates given by each method.}
	
	 			\begin{table*}
	 				\caption{Candidates selected by \citet{Burgasser2006, Burgasser2010}  and  \citet{Bardalez_Gagliuffi} indices.}
	 				\label{all_selected_candidates}
	 				\centering
	 				\small
	 				\renewcommand{\footnoterule}{}  
	 				\begin{center}
	 					\begin{tabular}{l l l l l l }
	 						\hline
	 						\hline
	 						Number & Candidate & Number of satisfied criteria from   & Type of candidate & Number of satisfied criteria from   & Type of candidate  \\
	 						& &  \citet{Burgasser2006, Burgasser2010} &  & \citet{Bardalez_Gagliuffi}  &  \\
	 						\hline
	 						
	 						3 & 2M0053   &  2  & Weak candidate &  6 & Weak candidate  \\
	 						
	 						4 & SIMP01365 &  4     & Strong candidate &  8 &   Strong  candidate \\
	 						
	 						7 & DE0255   &  2    & Weak candidate  & 7 & Weak candidate\\
	 						
	 						8 & 2M0257   &  2 &  Weak candidate & 6  & Weak candidate \\
	 						
	 						11& SD0423   & 2    & Weak candidate & 7 & Weak candidate\\
	 						
	 						20 & 2M1341  & 2   & Weak candidate & 8   & Strong candidate \\
	 						
	 						\hline
	 					\end{tabular}
	 				\end{center}
	 				
	 			\end{table*}


	To confirm or reject the selected L plus T binary candidates, we compared our spectra with
	libraries of well characterized brown dwarf spectra, i.e. template spectra. We used as template spectra the \citet{McLean} and  \citet{Cushing} libraries, with a resolution of R$\sim$~2000, as well as the SpeX Prism Spectral Library spectra\footnote{http://pono.ucsd.edu/$\sim$adam/browndwarfs/spexprism/}, with a resolution of R~$\sim$120. In total we considered 462 spectra from SpeX Spectral Library plus 14 from \citet{Cushing} library and 47 spectra from \citet{McLean} library, {with spectral types from L0 to T7}.
	
	We degraded the resolution of our X-Shooter spectra to the
	resolution of each template. We re-interpolated the library of brown dwarf template
	spectra and X-Shooter spectra to the same grid.
	We  searched for  the best matches to  template spectra of single objects from SpeX, \citet{Cushing} and \citet{McLean} libraries and to synthetic binary spectra created using those libraries.
	To create those synthetic binaries, we calibrated the fluxes of the components to  the same distance using an absolute magnitude-color relation
	\citep{Dupuy_Liu2012} and add them together. {The final resolution of our synthetic binary templates was the same as the SpeX spectral library spectra. }
	
	{To identify the best matches to our spectra, we used the approach  explained in \citet{Cushing2008}, which is similar\footnote{{$G$ is mathematically similar to a $\chi^{2}$, but it does not follow a $\chi^{2}$ distribution as our comparison spectra have noise (see \citet{Cushing2008} for further details). We therefore do not expect to achieve $G\sim$1 for the best fits, our goal is to determine whether a binary template is fitting better than a single template for our selected binary candidates.}} to a $\chi^{2}$,}
	
	\begin{equation}
	G = \sum_{\lambda}w(\lambda)\left[\displaystyle\frac{C(\lambda)-\alpha T(\lambda)}{\sigma_{c}(\lambda)}\right]^{2},
	\label{chi}
	\end{equation}
	where $C(\lambda)$ is the spectrum of the candidate, $T(\lambda)$ is the template spectrum,   $\mathrm{w(\lambda)}$ is a vector of weights proportional to the waveband size of each pixel, $\alpha$ is a scaling factor that minimizes $G$, and $\sigma_{c}(\lambda)$ are the errors of the spectrum. To calculate the  $G$, we used the parts of the {spectra} where no {strong} telluric absorptions are {contributing, since we are confident of the telluric correction resulting from the data reduction process}: $\lambda$=~950-1350~nm, 1450-1800~nm and 2000-2350~nm. We  additionally checked the best matches by visual inspection. 
	Finally, we tested if the fit to a binary template was significantly better than  the fit to a single template using a a one-sided F-test statistic. We used as the distribution statistic ratio:

	\begin{equation}
	\eta_{SB} = \frac{min(G_{single}).df_{binary}}{min(G_{binary}).df_{single}}
	\label{F_test}
	\end{equation}
	{where min($G_{single}$) and min$(G_{binary})$ are the minimum $G$ for the best match to a single or to a composite template}, and  $df_{binary}$ and $df_{single}$ are the degrees of freedom for the binary template fit and the single template fit. The degrees of freedom are the number of data points used in the fit (n~=~296) minus one to account the scaling between our spectra and the template spectra. To rule out the null hypothesis, meaning that the candidate is not a binary with a 99\% confidence level, we require $\eta_{SB} > 1.31$ \footnote{{\citet{Burgasser2006,Burgasser2010} requires a confidence level of 99\%, and \citet{Bardalez_Gagliuffi} a confidence level of 90\%. We employed a confidence level of 99\% to be more conservative, and minimize the false positives rate. }}. 
	The F-test analysis rejected three of our candidates, namely:  SIMP0136, 2M0257, 2M1341.
	
	In Table~\ref{best_match_selected} we show the best matches of the selected candidates to single and composite brown dwarf spectra. Plots with the best matches are shown in Appendix
	\ref{L_T_best_matches}. 
	
			\begin{table*}
				\caption{Best matches to objects selected as binary candidates by spectral indices}
				\label{best_match_selected}
				\centering
				\small
				\renewcommand{\footnoterule}{}  
				\begin{center}
					\begin{tabular}{l l l l l}
						\hline
						\hline
						Candidate & Single best match spectrum  & Composite best match spectrum  & $\eta_{SB}$ & Fig. \\
						
						\hline
						
						2M0053   & 2MASS J17461199+5034036~(L5)   &  Kelu-1~(L3p) + SDSS J120602+281328~(T3) & 1.35  & \ref{2M0053}\\
						
						SIMP01365 &  SDSS J152103.24+013142~(T2)     &  DENIS-PJ225210-173013~(L7.5, bin) + SDSS J000013+255418~(T4.5) &   0.55 &\ref{SIMP0136}\\
						
						DE0255   &  SDSS J085234.90+472035.0~(L9.5)     &  SDSS J163030.53+434404.0~(L7) + SDSS J103931.35+325625.5~(T1) &  3.42 & \ref{DE0255_comp} \\
						
						2M0257   &  {SDSS J104409.43+042937.6~(L7)} &  2MASS J0028208+224905~(L7) + SDSS J204749.61-071818.3~(T0)   &  1.23  & \ref{2M0257}\\
						
						SD0423   & SDSS J105213.51+442255.7~(T0.5)    & 2MASS J15150083+4847416~(L6) + SDSS J125453.90-012247.4~(T2)  & 3.23 & \ref{SD0423}\\
						
						2M1341  &  GJ1048B~(L1)                  &  GJ1048B~(L1) + 2MASS J1217110-031113~(T7.5)    & 1.26 & \ref{2M1341} \\
						
						\hline
					\end{tabular}
				\end{center}
				
			\end{table*}
	
\begin{figure} 
			\centering
			\includegraphics[width=8.5cm]{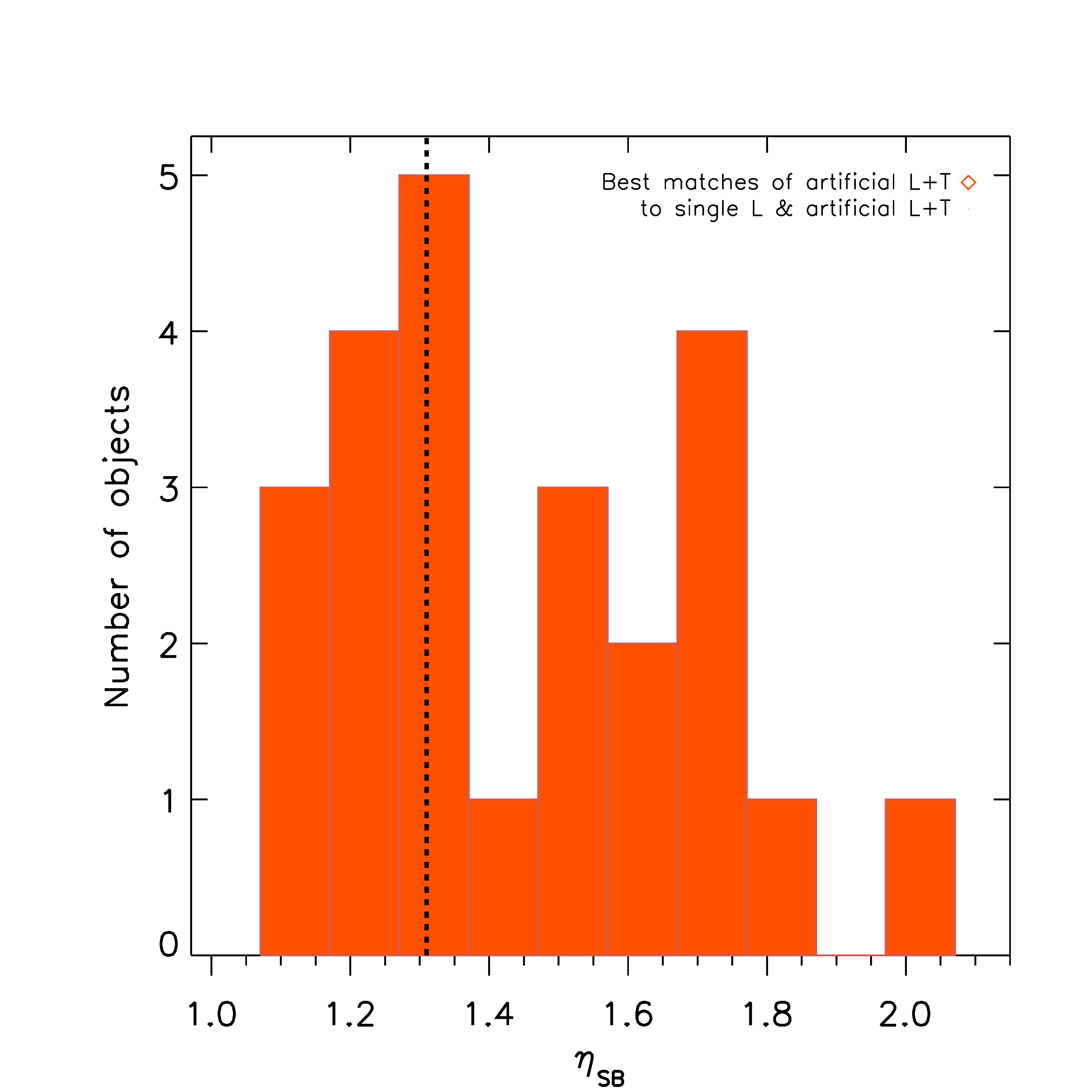}
			\caption{Histograms of $\eta_{SB}$ for comparison of L plus T synthetic brown dwarf spectra to single L and other synthetic L+T  brown dwarfs spectra.  The dashed black line indicates $\eta_{SB}$=1.31. }
			\label{histogram_binary_LT_L}
		\end{figure}

\begin{figure} 
			\centering
			\includegraphics[width=8.5cm]{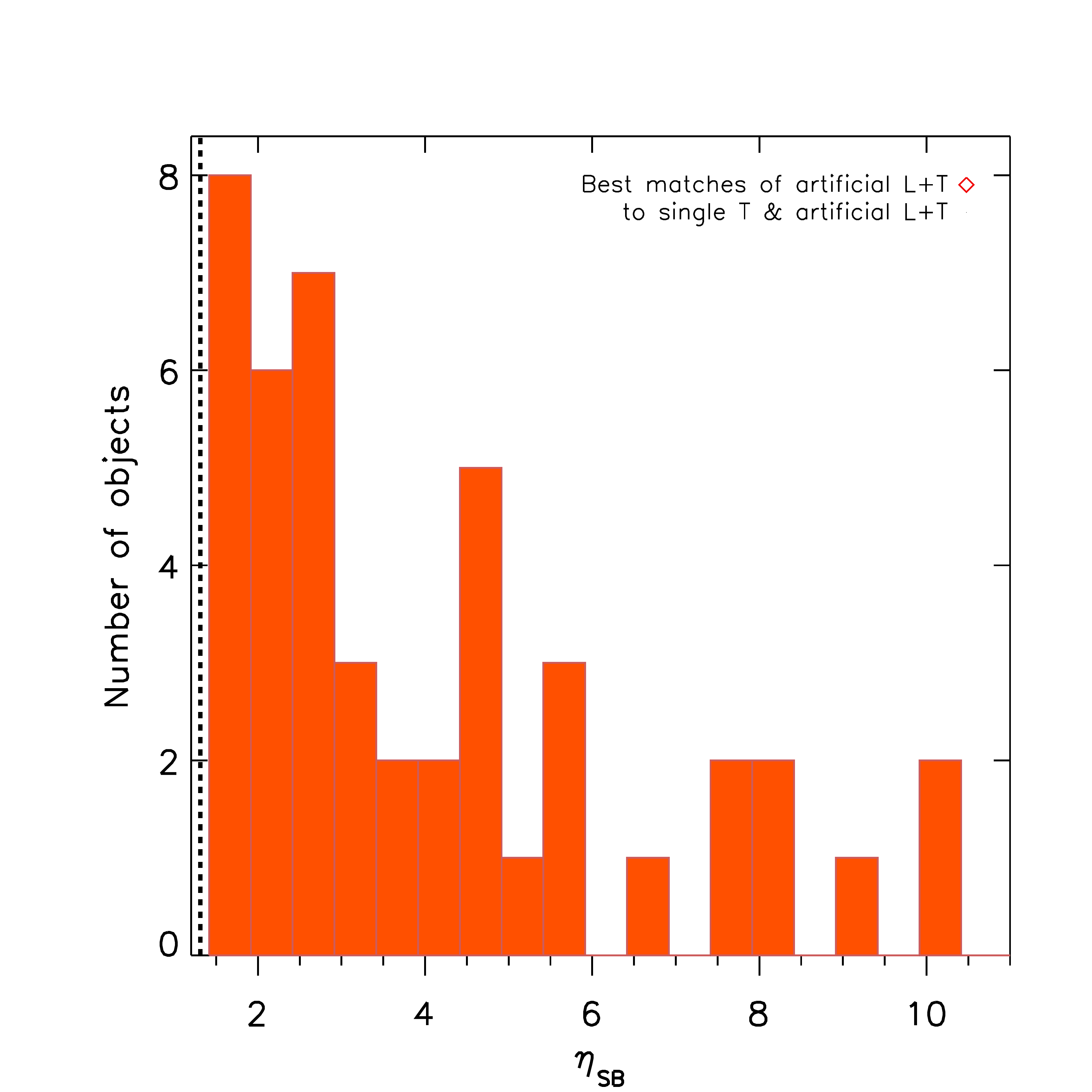}
			\caption{Histograms of $\eta_{SB}$ for comparison of L plus T synthetic brown dwarf spectra to single T spectra and other synthetic L+T  spectra.  The dashed black line indicates $\eta_{SB}$=1.31. }
			\label{histogram_binary_LT_T}
		\end{figure}

	{We intended to estimate the fraction of missed L plus T binaries applying  \citet{Burgasser2007,Burgasser2010} and \citet{Bardalez_Gagliuffi} method. To this aim, we compared 47 synthetic L plus T binaries  to single L dwarfs, single T dwarfs and to other synthetic L plus T binaries. We found that 21\% of the L plus T synthetic binaries, did not satisfy the binarity condition (i.e. they had $\eta_{SB}$ < 1.31). In particular,  for 10 L plus T artificial binaries the match to single L dwarfs was significantly better than with binaries (see Fig. \ref{histogram_binary_LT_L}). In Fig.  \ref{histogram_binary_LT_T}, for all L plus T synthetic systems the binarity criteria was satisfied. Therefore, most of the L plus T systems should be found using this method, but it must be taken into account that some binaries might be lost.}
	
\begin{figure}
				\centering
				\includegraphics[width=8.5cm]{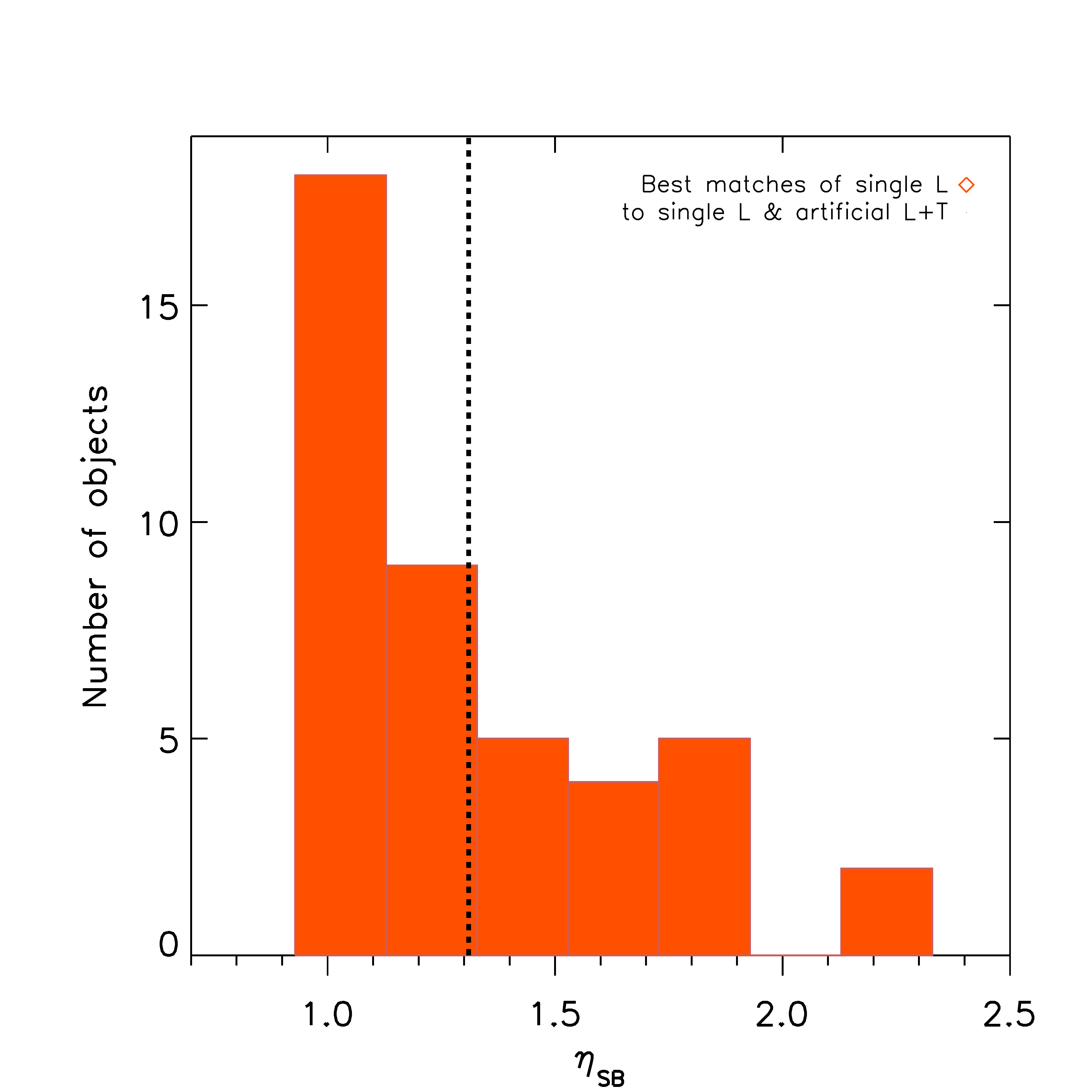}
				\caption{Histograms of  $\eta_{SB}$ for comparison of L single brown dwarf spectra to other L single dwarfs and synthetic L+T  brown dwarfs spectra.  The dashed black line indicates $\eta_{SB}$=1.31. }
				\label{histogram_Lsingle_L_T_L_T}
			\end{figure}

\begin{figure}
				\centering
				\includegraphics[width=8.5cm]{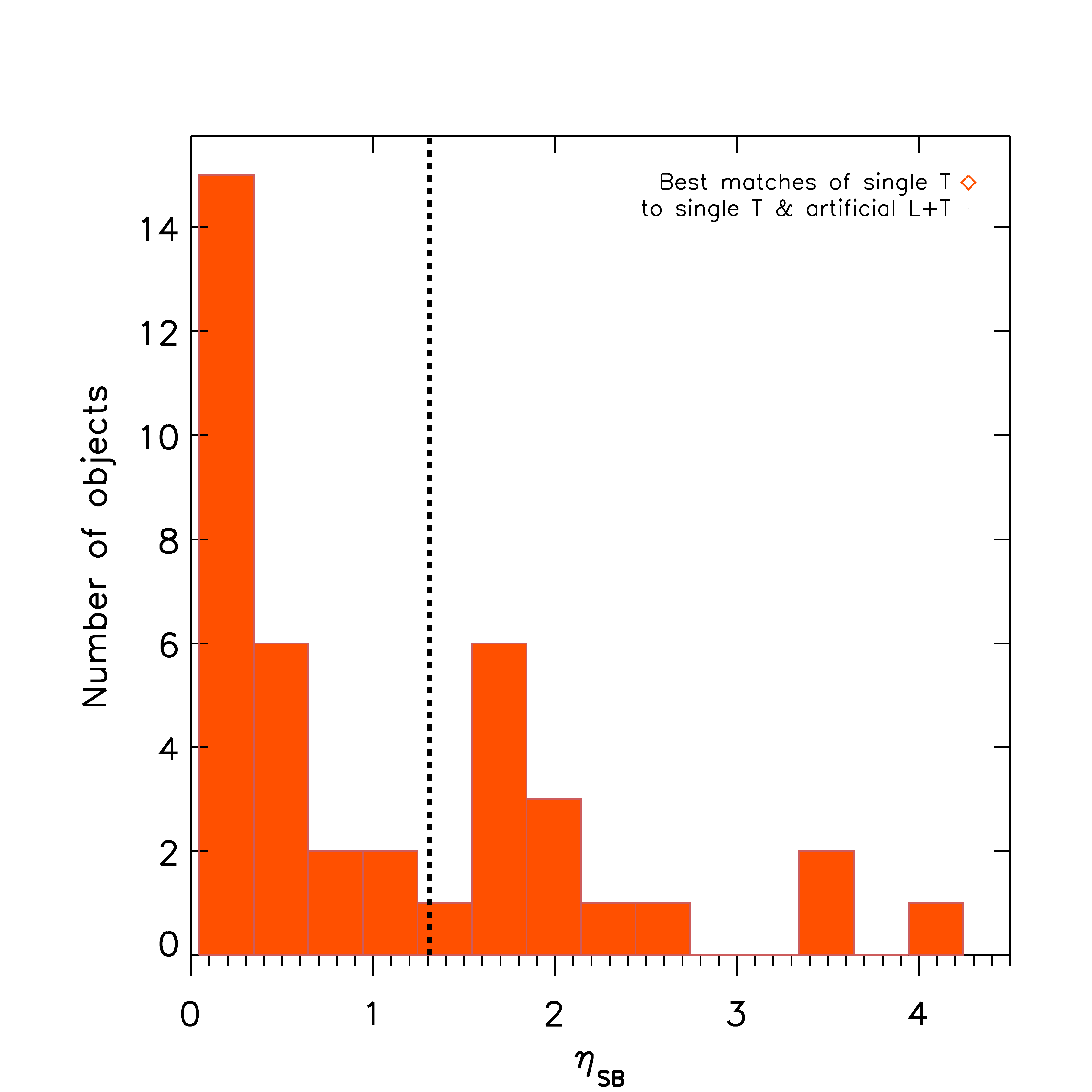}
				\caption{Histograms of $\eta_{SB}$ for comparison of T single brown dwarf spectra to other single T dwarfs and synthetic L+T  brown dwarfs spectra.  The dashed black line indicates $\eta_{SB}$=1.31. }
				\label{histogram_Tsingle_L_T_L_T}
			\end{figure}

	{Finally, we compared a sample of 43  single L dwarfs to other L single dwarfs, and to synthetic L plus T binaries (see Fig. \ref{histogram_Lsingle_L_T_L_T}). {We found that 37\% of the single L dwarfs satisfied the binarity criteria, i.e. they had significantly better matches with L plus T synthetic binaries, they are therefore false positives}. Equally, we performed a similar analysis for a sample of 40 single T dwarfs. {We obtained that  35\% of the T dwarfs are also false positives} (see Fig. \ref{histogram_Tsingle_L_T_L_T}). We examined the spectral characteristics reported in the literature for the subsamples of 16 L dwarfs and 14 T dwarfs with best matches to synthetic L plus T binaries. We found that 5 from the  16 L dwarfs, and 2 of the 14 T dwarfs had either peculiar spectra or different spectral classification in the optical and the near infrared. 
		These results suggest that this method is efficient finding different spectral type binaries, but it should be applied with caution, as some false positives might be found.}
	
		\begin{table*}
			\caption{Summary of the results obtained from Fig. \ref{histogram_binary_LT_L}, \ref{histogram_binary_LT_T}, \ref{histogram_Lsingle_L_T_L_T}, \ref{histogram_Tsingle_L_T_L_T}.}
			\label{best_match_summary}
			\centering
			\renewcommand{\footnoterule}{}  
			\begin{center}
				\begin{tabular}{l l l l l}
						\hline
						\hline
						Type of objects & First comparison objects  & Second comparison objects  & Best matches & Fig. \\
						
						\hline
						
						Synthetic L+T binaries  & Single L dwarfs   &  Synthetic L+T binaries & 21\% to single L dwarfs (false negatives) & \ref{histogram_binary_LT_L}\\
						   &  & &79\% to  L+T synthetic binaries &\\
	
						Synthetic L+T binaries &  Single T dwarfs   & Synthetic L+T binaries & 0\% to single T dwarfs  &\ref{histogram_binary_LT_T}\\
						 &     &  & 100\% to L+T  synthetic binaries &\\
	
						Single L  dwarfs &  Single L dwarfs    & Synthetic L+T binaries  & 63\% to single L dwarfs & \ref{histogram_Lsingle_L_T_L_T} \\
						   &       &   & 37\% to L+T syntetic binaries (false positives) & \\
	
						Single T dwarfs  &  Single T dwarfs &   Synthetic L+T binaries   &  65\% to single T dwarfs & \ref{histogram_Tsingle_L_T_L_T}\\
						   &   &   &  35\% L+T synthetic binaries (false positives) & \\

						\hline
					\end{tabular}
				\end{center}
	
			\end{table*}


	\subsection{Finding equal spectral type brown dwarf binaries}\label{similar_SpT}
	

	{We aimed to find equal spectral type brown dwarf binaries 
		comparing to spectral templates. To this purpose, we have chosen  
		101 L0 to L9  presumed single  brown dwarfs spectra from the SpeX
		library. {Additionally, we have created 60 synthetic brown dwarf binaries in which both components have similar  spectral type, i.e. the same spectral type, but different spectral sub-types}. To create  those, we chose several presumed single L brown dwarfs from the SpeX sample, and we created synthetic binary spectra combining single brown dwarf spectra following the same procedure as in Section \ref{L_T}.}
	
	

\begin{figure} 
		\includegraphics[width=8.5cm]{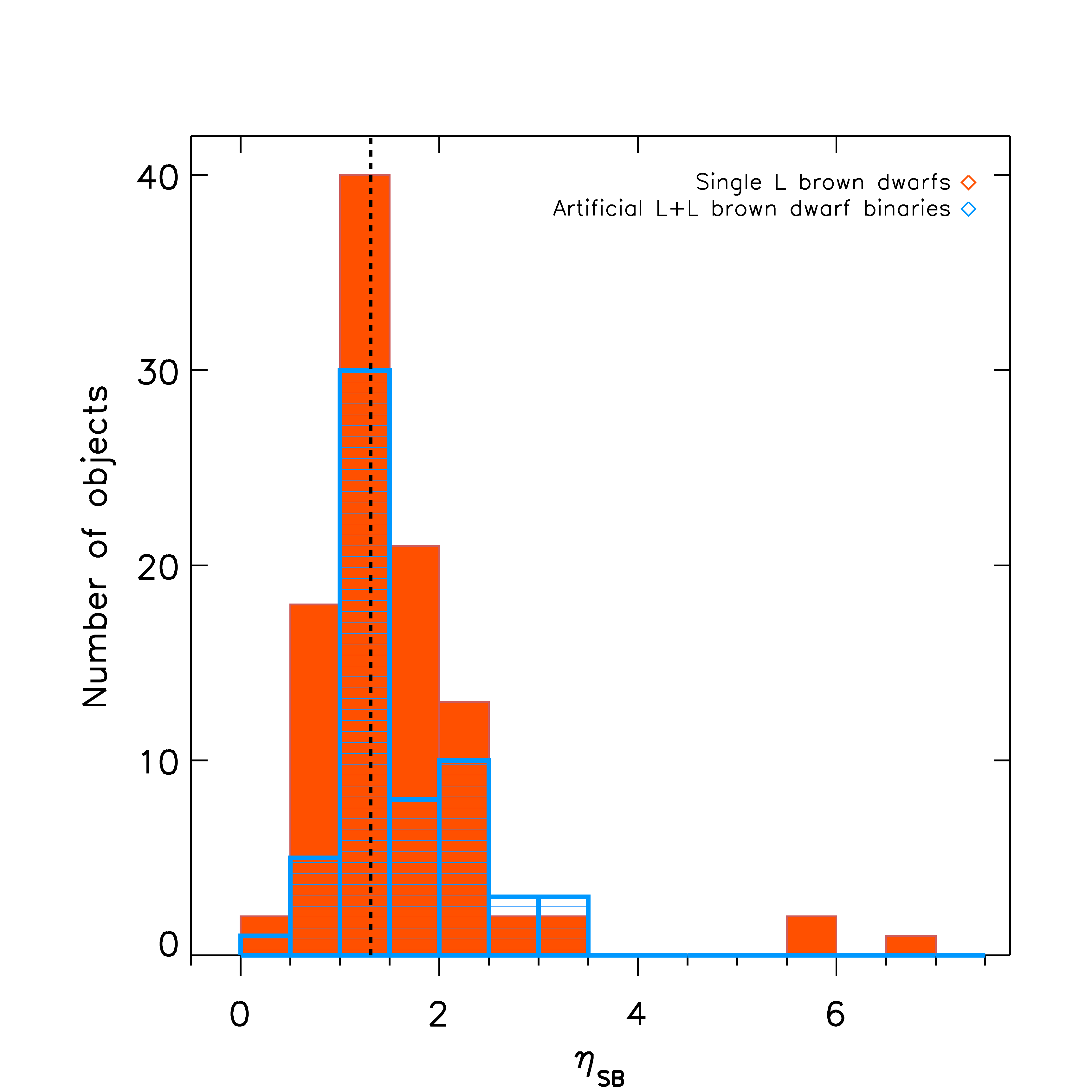}
		\caption{Histograms of  $\eta_{SB}$ for comparisons of single L brown dwarfs (in red) and  artificial L plus L binaries   (blue with lines), to other single L dwarfs and other L plus L synthetic binaries. The dashed black line indicates $\eta_{SB}$=1.31. }
		\label{histogram_Lsingle_bin}
	\end{figure}

\begin{figure} 
		\centering
		\includegraphics[width=8.5cm]{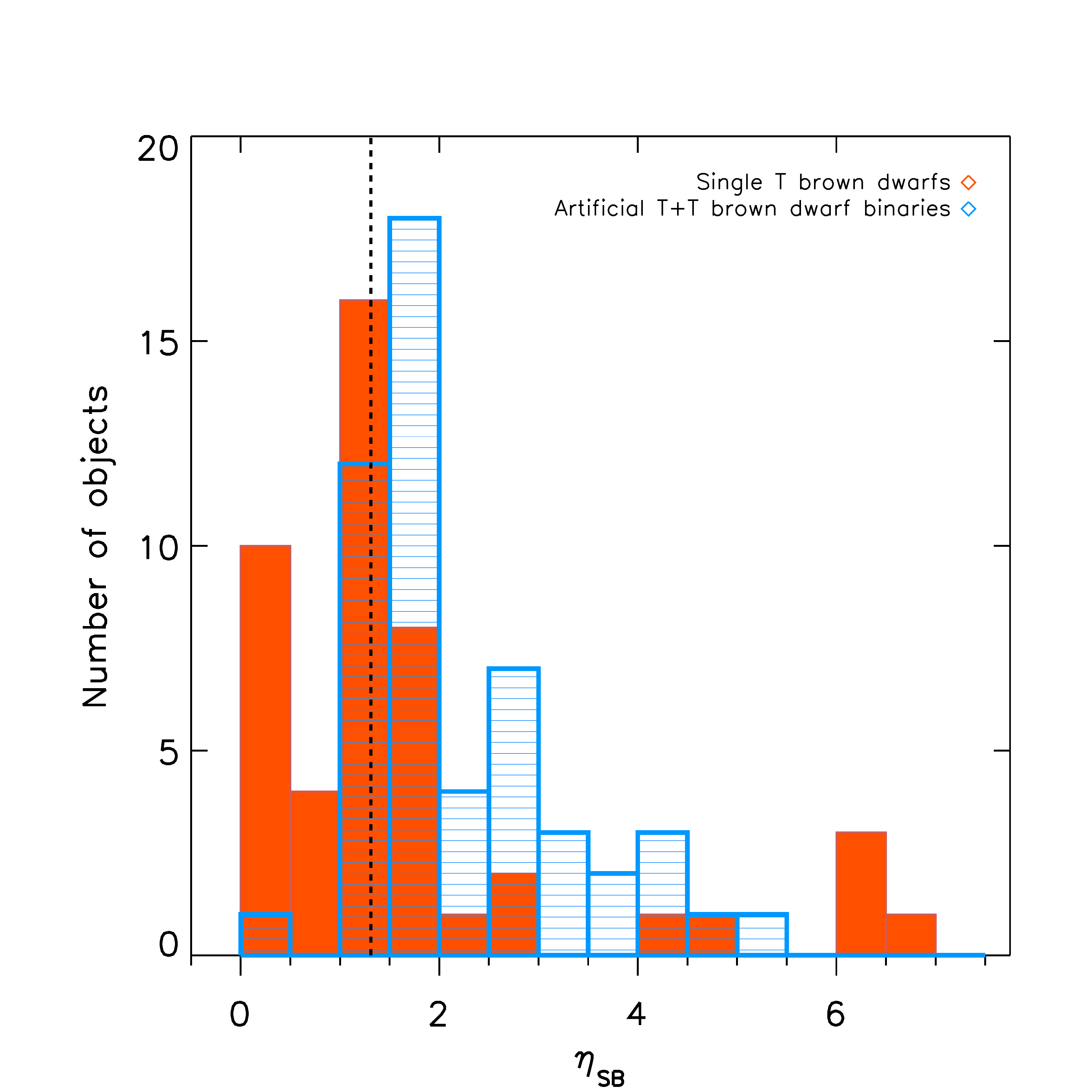}
		\caption{Histograms of  $\eta_{SB}$ for comparisons of single T brown dwarfs (in red) and  artificial T plus T binaries   (blue with lines), to other single T dwarfs and other T plus T synthetic binaries.  The dashed black line indicates $\eta_{SB}$=1.31. }
		\label{histogram_Tsingle_bin}
	\end{figure}

	{We compared  the 101  single L dwarfs and the 60 artificial L plus L dwarf binaries to other L single SpeX spectra, and to other synthetic L  binaries.  We determined the best match of the 101 single L and the 60 synthetic L plus L binaries calculating the $G$ parameter as in equation \ref{chi}, and we equally decided the significance of the best match using equation \ref{F_test}.
	We made a similar analysis for 56  T single dwarfs, and 74 T plus T synthetic binaries.
	
		{We compared the best matches to single L and T dwarf spectra, and the best matches to synthetic binaries with  similar spectral type (L plus L and T plus T, respectively).}
	 To this aim, we calculated the  $\eta_{SB}$ parameter for the 101 single L dwarfs,  56 single T and for the 60 L plus L and 63 T plus T artificial binaries created. {In} Figures \ref{histogram_Lsingle_bin} and \ref{histogram_Tsingle_bin}, we represent  in a red histogram  single brown dwarfs, and  artificial dwarf binaries in a blue histogram with horizontal lines.}

	{We found that 49\% of the 101 L single dwarfs, and  62\% of the 60 synthetic L  binaries,  satisfied the binarity criteria. In Figure \ref{histogram_Lsingle_bin}, we show that the distribution of $\eta_{SB}$ for single L and for synthetic L plus L binaries is similar.}
		 
	{Equally, we performed a similar simulation for T dwarfs. We found that 57\% of the 56 T single dwarfs, and  86\% of the 56 T plus T synthetic binaries satisfied the binarity criteria. In Figure \ref{histogram_Tsingle_bin}, we show the distributions of $\eta_{SB}$ for single T and for synthetic T plus T binaries. }
		
		{For the cases mentioned before, the distribution of the $\eta_{SB}$ value is the same for single and synthetic binaries with {the same spectral types, but different sub-spectral types}. Therefore, it is impossible to distinguish between single L or T dwarfs and synthetic L or T dwarf binaries. Furthermore, for both cases, the best fits are usually other synthetic binary spectra.  Additional data, such as parallax measurements, high-resolution imaging or high resolution spectra are necessary in order to find these systems.}

		\begin{table*}
			\caption{Summary of the results obtained from Fig. \ref{histogram_Lsingle_bin} and  \ref{histogram_Tsingle_bin}.}
			\label{best_match_summary2}
			\centering
			\renewcommand{\footnoterule}{}  
			\begin{center}
				\begin{tabular}{l l l l l}
					\hline
					\hline
					Type of objects & First comparison objects  & Second comparison objects  & Best matches & Fig. \\
					
					\hline
					
					Single L  dwarfs &  Single L dwarfs    & Synthetic L+L binaries  & 51\% single L dwarfs & \ref{histogram_Lsingle_bin} \\
					&       &   & 49\% L+L syntetic binaries (false positives) & \\
					
					Synthetic L+L binaries  & Single L dwarfs   &  Synthetic L+L binaries & 38\% single L dwarfs (false negatives)  & \ref{histogram_Lsingle_bin}\\
										&  & & 62\% L+L synthetic binaries &\\
					
					Single T dwarfs  &  Single T dwarfs &   Synthetic T+T binaries   &  43\% single T dwarfs & \ref{histogram_Tsingle_bin}\\
					&   &   &  57\% T+T synthetic binaries (false positives) & \\

					Synthetic T+T binaries &  Single T dwarfs   & Synthetic T+T binaries & 12\% single T dwarfs (false negatives) &\ref{histogram_Tsingle_bin}\\
					&     &  & 86\% T+T  synthetic binaries &\\

					\hline
				\end{tabular}
			\end{center}
			
		\end{table*}

\subsection{Photometric search for brown dwarf binaries}
	
\begin{figure}
		\hspace*{-1.3cm}
		\includegraphics[width=10.5cm]{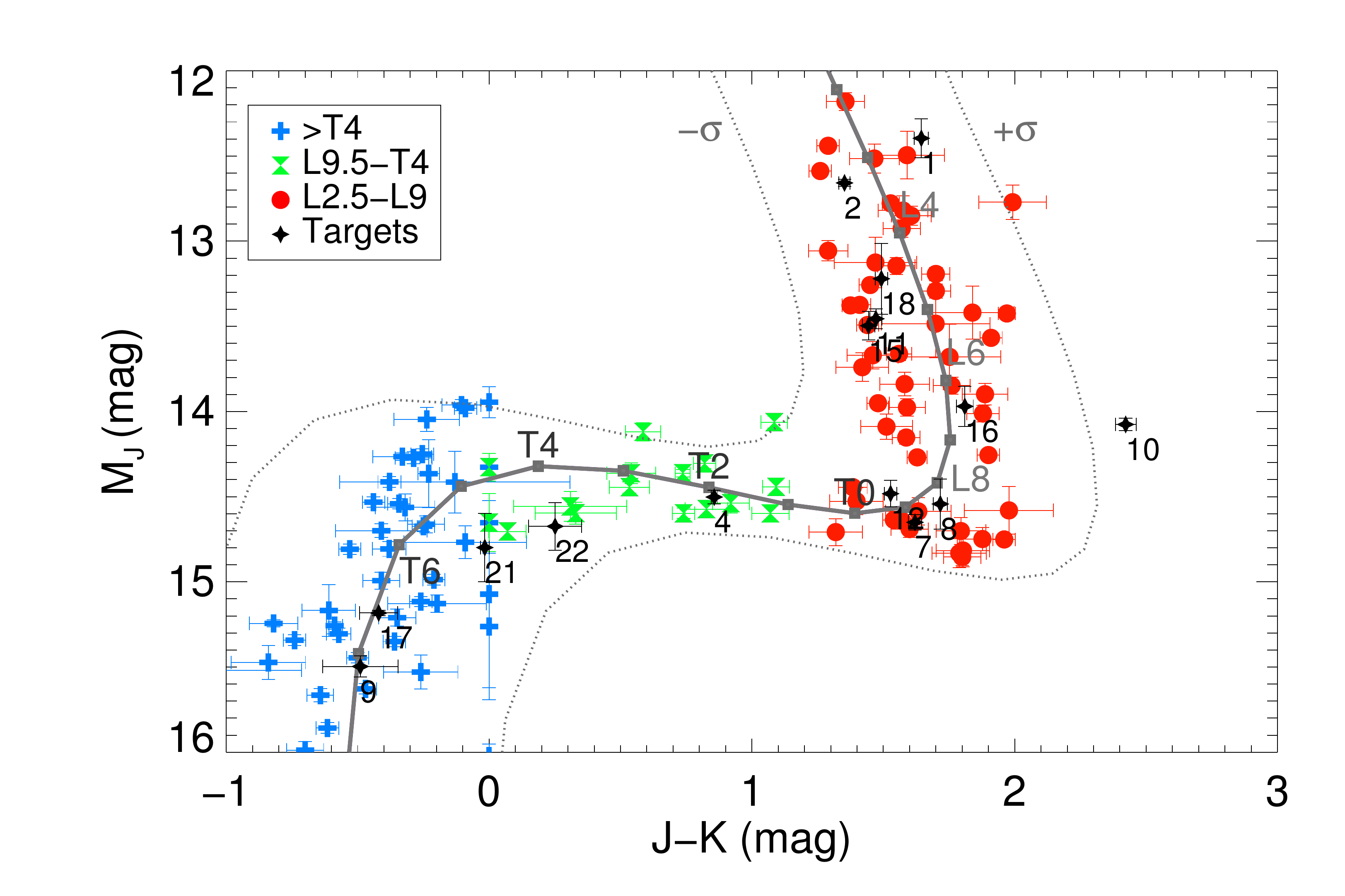}
		\caption{Color-magnitude diagram in the MKO system showing brown dwarfs with measured parallaxes from \citep{Dupuy_Liu2012}, and {its color-absolute magnitude relationship together with its plus minus one sigma curves (dotted line)}. Our targets are shown in black. Objects have the same numbers as in Table \ref{literature}.}
		\label{J_K_Mabs}
	\end{figure}

	{In our 22 object sample,  distances for 15  objects are available in the literature, with a precision of 10\% or better. In Figure \ref{J_K_Mabs} we present a CMD showing  the $J-K$ color in the MKO
		(Mauna Kea Observatory) photometric system versus absolute magnitude in the J band. In this Figure, we plot all brown dwarfs with know parallaxes \citep{Dupuy_Liu2012}, the 15 objects of our sample with {known} parallaxes, and the color-absolute magnitude relationship by \citet{Dupuy_Liu2012}. The two known
		binaries in our sample (LHS102B and SD0423, targets 1 and 11 respectively)  stand out over objects with their same spectral
		types and other one, the young object 2M0355 (target 10) is much redder as objects of its same spectral type because of its youth. For the rest of the
		objects we cannot draw clear conclusions as there are no clear outliers.}
	
	\section{Individual candidates}\label{revised_properties}
	
	Six objects in our sample were selected as binary  candidates by spectral indices. After fitting, three of our candidates were rejected due to the confidence level being lower than 99\%. This leads to a final number of three selected candidates.
	
	One of our selected candidates, the confirmed binary SD0423, was studied by \citet{Burgasser2005b}
	and was used as a test of consistency of the spectroscopic method by \citet{Burgasser2010}. We do not discuss it here.
	
	\subsection{Rejected candidates}
	
	We do not consider the following targets  as a binary candidate in the rest of the paper.
	
	\subsubsection{SIMP~01365662+0933473}\label{SIMP0136}
	
	SIMP~0136 was discovered by \citet{Artigau2006}  and 
	classified as a T2.5. \citet{Goldman2008} searched for companions using NACO/VLT, reaching a sensitivity of 0.2" (1-40~AU), but no
	companions were found. \citet{Artigau2009} detected photometric variability in the 
	J and K bands with a modulation of $\sim$2.4~h and an amplitude of
	50~mmag. \citet{Radigan2012} calculated the amplitude of the variability for an
	object similar to SIMP~0136 (2MASS J21392676+0220226, T1.5). If this variability were
	produced by a companion, it would be much
	smaller than the variability obtained. Therefore, we do not expect that it is caused by a companion. \citet{Apai2013} explained it as a mixture of thick and thin patchy iron and silicate clouds covering the
	surface of the object. 
	
	The object  SIMP~0136 was selected as a  brown dwarf binary
	candidate, but it was rejected by a F-statistic analysis in Section \ref{L_T}. Spectral indices used in Section \ref{L_T} are suitable to select peculiar spectral
	characteristics that appear usually in binary L plus T brown dwarf
	spectra. However, if variability is produced by a partial coverage of thick and
	thin clouds in the brown dwarf atmosphere, similar peculiar spectral
	characteristics would appear in brown dwarf spectra. 
	

	A preliminary parallax of 166.2 $\pm$ 2.9 mas for SIMP~0136 was obtained from
	the NPARSEC program \citep[ESO program 186.C-0756, ][]{2013MNRAS.433.2054S}.
	Using this parallax, we placed the object in a CMD together with
	other L, L-T transition and T brown dwarfs with parallaxes  \citep{Dupuy_Liu2012}, as shown in Figure \ref{J_K_Mabs}. We
	compared SIMP~0136 to objects of similar spectral type. We did not find significant overluminosity, expected in the case of late L and early T brown dwarf binaries. This result is compatible with the rejected binary hypothesis by the F-test.

		\subsubsection{2MASS J02572581-3105523}\label{2M0257}
		
		The target 2M0257 was discovered by \citet{Reid2008_2}. \citet{Kirkpatrick2008}  classified it as a L8 in the optical. \citet{Marocco2013} measured its  trigonometric parallax to be $\pi$~=~99.7$\pm$6.7~mas.
		It was selected by spectral indices as a weak brown dwarf binary candidate, and it was rejected by the F-statistic.
		When we compared this target with objects of similar spectral type in the CMD, no overluminosity  was found. This result {agrees} with the non-binarity scenario.
		
		\subsubsection{2MASS~J13411160-3052505}\label{2M1341}
		
		The target 2M1341  was
		discovered by \citet{Reid2008}. \citet{Faherty2009} published a distance of 24$\pm$2~pc. \citet{Kirkpatrick2011}
		classified it as a peculiar L2.  \citet{Bardalez_Gagliuffi} compared it to several SpeX templates and concluded that this object could be a L1.2$\pm$0.3 plus a T6.3$\pm$1.0.

		In Section \ref{L_T}, {target 2M1341 was selected as a weak candidate L plus T binary by \citet{Burgasser2006, Burgasser2010} indices, but it was selected as a strong candidate by \cite{Bardalez_Gagliuffi}} {indices}. {This candidate was rejected by our conservative F-test criterion. However, a detailed inspection of the observed spectroscopic matches (see Fig.\ref{2M1341}).  reveals no satisfactory reproduction of the near infrared features, thus leaving open the multiplicity of 2M1341. High resolution observations are required to disentangle its true nature.}   

		\subsection{Selected candidates}

			\subsubsection{2MASS J00531899-3631102}\label{2M0053}
			
			The object 2M0053 was discovered by \citet{Reid2008_2}. \citet{Kirkpatrick2008}  classified it as a L3.5 in the optical. 
			It was selected as a weak brown dwarf binary candidate. We found a best match  with a combination of a L3 dwarf and a T3 dwarf. There is not parallax measurement available for this target.
			
	\subsubsection{DENIS-P J0255.0-4700}\label{DE0255}
	
	The target DE0255 was discovered by \citet{Martin1999} and it was classified as a
	peculiar L6. \citet{Koen2005}  reported evidence of variability in different timescales (1.7 and 5~h). \citet{Morales_calderon2006} concluded that DE0255 may vary with a 7.4~h period at 4.5~$\mu$m, but it does not at 8~$\mu$m. \citet{Costa2006} reported an absolute parallax of $\pi$~=~201.4$\pm$~3.9~mas. \citet{Burgasser2008}
	classified it in the optical as a L8 and in the near infrared as a
	L9. Finally, \citet{Reid2008} searched for multiplicity for this target using
	high-resolution NICMOS NIC1 camera imaging on the Hubble Space Telescope, but
	found no evidence of multiplicity.
	
	In Section \ref{L_T}, DE0255 was selected as a L plus
	T  {weak} binary candidate. We found a best match for DE0255 to a composite spectra of a L7 plus T1 spectra (see Fig. \ref{DE0255_comp}).

	Using the \citet{Costa2006} published parallax, we plot DE0255 in a CMD as above  (see Fig. \ref{J_K_Mabs}).  In case of a late-L and early-T binary scenario, we expect to find about $\sim$0.5 mag overluminosity comparing with objects of similar spectral type on a CMD. In this case, no overluminosity was found, weakening the binarity hypothesis for object DE0255.

	\section{Very low mass binary fraction}\label{binary_fraction}

	In our peculiar sample of 22 objects, we found three L$+$T binary candidates. One of them
	{has} been  confirmed by other authors using high resolution imaging (SD0423, by \citealt{Burgasser2005b}). Other two objects were selected as  weak binary candidates (2M0053 and DE0255). The binarity hypothesis is weakened for object DE0255, due to the lack of  overluminosity in the CMD (see Figure \ref{J_K_Mabs}), expected for late L and early T brown dwarf binary scenario. 
	
	{This result  allowed us to estimate the minimum and the maximum L$+$T binary fraction for our sample, and samples selected using our same criteria (see  Section \ref{sample_selection}). The minimum   L$+$T binary fraction\footnote{The uncertainties of the binarity fraction for samples with less than 100 objects are calculated using the method explained in  \cite{Burgasser2003}.}  is estimated at $4.5^{+9.1}_{-1.4}$\% {(the only confirmed L$+$T binary is SD0423 over the whole 22 targets sample)}. The L$+$T maximum binary fraction for our sample is estimated at $13.6^{+10.4}_{-4.3}$\%~(the three L$+$T binary candidates over the whole 22 targets sample).}
	These pairs would have a mass ratio of $q \ge 0.5$ for 
	ages between 1~Gyr and 5~Gyr (expected for most of the objects in this study). Our work is not
	sensitive to smaller mass ratios. 
	The derived range for the L$+$T binarity
	coincides with the fraction of late-M stars of the solar neighborhood that host
	T-type companions. As summarized by \citet{Burgasser2015},  there are
	two late-M stars with T-type brown dwarf companions, among 14 dwarfs with spectral types between M7 and M9.5, and at 10~pc from the Sun
	($\sim$14\%). Therefore, it appears that the L dwarf primaries have T-type
	companions with a similar  frequency  to the late-M objects, despite the fact
	that the former primaries are expected to be less massive than the latter for
	typical field ages.
	
	Regarding binaries that  include L$+$L and T$+$T pairs in our sample,
	there is just one confirmed L4.5$+$L4.5 system in our target list (LHS 102B).  LHS 102B cannot
	be detected using the methods we employed in this paper, given the limitations of the spectroscopic technique. 
	
	We thus determined the minimum fraction of  L$+$L, T$+$T, and L$+$T pairs for our sample to be
	$9.1^{+9.9}_{-3.0}$\% {(the confirmed brown dwarf binaries, SD0423 and LHS102B over the whole sample of 22 objects)}. In spite of the peculiarity of our sample, this lower limit agrees with other values reported previously by
	different groups \citep{Burgasser2007aa, Goldman2008, Bardalez_Gagliuffi, Burgasser2015}.


	\section{Comparison to the BT-Settl atmospheric models}\label{models}
	
	{The X-Shooter spectra presented in this paper provide the possibility to compare with the BT-Settl atmospheric models in a wide range of wavelengths (550-2500~nm). We used 13 objects of our total sample to test the BT-Settl models 2014. We excluded brown dwarf binary candidates and spectra with low signal to noise to avoid false results}.
	
	The BT-Settl models account for the formation and gravitational settling of
	dust grains for a effective temperature ($\mathrm{T_{eff}}$) below $\approx$2700 K in the photosphere
	of the objects, following the approach described in
	\cite{Rossow}.  {The models include  180 types} of condensates via their
	interaction with the gas phase chemistry, depleting the gas from their
	vapor phase counterparts. Fifty-five of these grain species are included in
	the radiative transfer calculations to the extent to which they have not settled from the
	cloud layer. 
	Log-normal grain size distributions with a standard deviation of 1 are used,
	where the characteristic grain size for each layer is determined from the
	equilibrium size derived by the cloud model. The original timescales approach of
	the \citet{Rossow} model has further been extended to account for nucleation as
	an additional timescale, which is defined by assuming a fixed seed formation rate
	motivated by studies of cosmic ray interactions with the Earth atmosphere  \citep{2005JASTP..67.1544T}.

	The cloud model is implemented in the \texttt{PHOENIX} multipurpose atmosphere
	code version 15.5 \citep{Allard2001}, which is used to compute the model
	atmospheres and to generate synthetic spectra. Convective energy transport and
	velocities are calculated using mixing length theory with a mixing length of
	{1.6-2.0 pressure scale heights, depending on surface gravity ($\log\,g$)}, and overshoot is treated as an exponential
	velocity field with a scale height based on the RHD simulations of
	\cite{Ludwig2002,Ludwig2006} and \cite{Freytag2010,Freytag2012}; an additional
	advective mixing term due to gravity waves is included as described in
	\citet{Freytag2010}.  All relevant molecular absorbers are treated with
	line-by-line opacities in direct opacity sampling as in \citet{Allard2003};
	{regarding} this, the molecular line lists have been updated as follows:
	{water-vapor \citep[BT2,][]{2006MNRAS.368.1087B}, vanadium oxide from Plez (2004, priv. comm.), $\mathrm{TiO}$ line list from \citet{2008PhST..133a4003P} and collision-induced absorptions of $\mathrm{H_{2}}$
		\citep{abelCIA11}. Non-equilibrium chemistry for CO,
		$\mathrm{CH_{4}}$, $\mathrm{CO_{2}}$, $\mathrm{N_{2}}$, and $\mathrm{NH_{3}}$
		is treated with height-dependent diffusivity also based on the RHD simulation
		results of \cite{Freytag2010}}.
	
	\subsection{Comparison to synthetic spectra}\label{comparison-models}
	
	In this section, we compared X-Shooter  optical and near-infrared spectra  to predictions of the last version of the BT-Settl atmospheric models
	\citep{2003IauS..211..325A, 2007A&A...474L..21A, 2011ASPC..448...91A} of 2014. We exclude brown dwarf binary candidates and spectra with low signal to noise. We  derive  atmospheric parameters of the objects and to reveal non-reproducibilities
	of the models. The models are  described in \cite{2011ASPC..448...91A, Allard2012a, Allard2012b}.

	We selected subgrids of synthetic spectra with 400~K$\leq T_\mathrm{eff} \leq$
	2100~K, 3.5 $\leq$ $\log\,g$ $\leq$ 5.5 and metallicities of +0.0 and +0.3, which are the metallicities for which the latest version of the BT-Settl models are available. The solar metallicity is based on metallicities calculated by \citet{2011SoPh..268..255C}. The spacing of the model grid  is 50~K and
	0.5~dex in log\,$g$. Effective temperature, gravity, metallicity and alpha element enhancement are described in the model name strings as {\texttt{lte-LOGG+[M/H]a+[ALPHA/H]}}.

	{The} BT-Settl  2014  synthetic spectra were smoothed to the resolution of
	X-Shooter. The models were then reinterpolated on the X-Shooter wavelength
	grid. The spectra were normalized using the same method as in Section
	\ref{empirical_analysis} and explained in \citet{Cushing2008}. The results from
	the fit were always double checked visually.  The atmospheric parameters
	corresponding to the {best fit} models are reported in Table
	\ref{Tab:atmopar}. {The parameters} $T_\mathrm{eff}$, $\log\,g$, and [M/H] have 
	uncertainties of 50 K, and 0.5 dex respectively. These errors
	correspond to the sampling of the atmospheric parameters of the model
	grids. We avoid  the following objects to test models: binary candidates
	(2M1341, DE0255, SIMP0136, 2M0053, 2M0257), known binaries (LHS102B, SD0423), noisy spectra
	(2M0144) or targets with known nearby objects that may contaminate the spectra, like in the case of Gl229B.

	The $\mathrm{CH_{4}}$ and the FeH molecules opacities are still incomplete in the new BT-Settl~2014 models. {Methane line opacities are based on the semi-empirical list of \citet{HomeierCH4}, which is highly incomplete
		in the H band and only supplemented with a small set of room-temperature transitions for the Y and J bands.
		Iron hydride causes absorption features through the {$F\,^4\Delta-X\,^4\Delta$} system between 650 and
		1600 nm, but in addition to this \citet{HargreavesFeH} identified significant opacity contributions from the
		{$A\,^4\Pi-A\,^4\Pi$} system, which is not yet included in the list of FeH lines available to \texttt{PHOENIX}}. This explains that the H-band is not well reproduced for any of the L or T brown dwarf spectra, and also the J band in the case of T brown dwarfs. For three of the L brown dwarfs, the best match is found for $\log\,g$=~5.5, with solar metallicity, four of the L brown dwarfs have best matches with $\log\,g$=~5.0, but [M/H]~=~+0.3.
	
	Best matches to the BT-Setll models are shown in Fig.~\ref{model1} and \ref{model2}. There are two of the L brown dwarfs that have best matches with low gravity models, 2M0355 and 2M0624. Object 2M0355 is known previously to be young \citep{Cruz, Allers&Liu, Zapatero_Osorio2014}, so we expect gravity to be lower. The result given by the models is consistent with the {literature}. {There are no references of youth for object 2M0624. Furthermore, the target does not show extremely redder J-K color in Figure \ref{J_K_Mabs} or significantly weaker alkali lines on Figures \ref{EW_opt} and \ref{EW_nir}, as low gravity objects do. Best matches to T type brown dwarfs are always  solar metallicity models. The best match to object 2M1828 is to a model with high gravity.

	\begin{table}
		
		\caption{Atmospheric parameters corresponding to the best fit spectra or synthetic fluxes for our  targets. We give $T_\mathrm{eff}$/$\log\,g$/[M/H].}
		\label{Tab:atmopar}
		\centering
		\renewcommand{\footnoterule}{}  
		\begin{center}
			\begin{tabular}{lrll}
				\hline
				\hline 
				
				Name &  $T_\mathrm{eff}$ & $\log\:g$  &  [M/H]  \\		
				
				\hline              
				
				2MASS J00361617+1821104 & 1800 & 5.5 & +0.0 \\
				2MASS J02182913-3133230 & 1800 & 5.0 & +0.3 \\
				2MASS J03480772-6022270 &  950 & 5.0 & +0.0  \\
				2MASS J03552337+1133437 & 1700 & 4.0 & +0.3 \\
				2MASS J04390101-2353083 & 1800 & 5.0 & +0.3  \\
				{2MASS J04532647-1751543} & 1750 & 5.5 & +0.0  \\
				2MASS J05002100+0330501 & 1800 & 5.0 & +0.3    \\
				2MASS J05395200-0059019 & 1800 & 5.5 & +0.0  \\
				2MASS J06244595-4521548 & 1700 & 4.5 & +0.3  \\
				2MASS J10043929-3335189 & 1800 & 5.0 & +0.3   \\
				2MASS J11263991-5003550 & 1900 & 5.5 & +0.0  \\
				2MASS J18283572-4849046 & 1100 & 5.5 & +0.0 \\
				2MASS J21513839-4853542 & 1100 & 5.0 & +0.0   \\
				
				\hline
				
			\end{tabular}

		\end{center}
							\begin{tablenotes}
								\item The grid size in $T_\mathrm{eff}$ is 50~K and in $\log\:g$ is 0.5~dex. {Uncertainties in $T_\mathrm{eff}$  and $\log\:g$ are the same as the grid sizes for both parameters}.
							\end{tablenotes}

	\end{table}

	\subsection{Comparing predicted and observed equivalent widths}\label{equivalent_widths}

	We measured the equivalent width of a variety of alkali lines with sufficient signal-to-noise in our spectra and we compared those values to predictions of the BT-Settl 2014 models.  In the optical, we measured the equivalent width of the Rb\,I~(794.8~nm), Na\,I~(818.3~nm), Na\,I~(819.5~nm) and Cs\,I~(852.0~nm). In the near infrared, we measured the K\,I~(1253~nm) line. 
	
	In Figure \ref{EW_opt}, we plot the equivalent width of the alkali lines in the optical for the objects in our sample, for  objects from \citet{Chiu2006}, \citet{Golimowski_2004}, \citet{Knapp_2004} and \citet{Lodieu2015} versus their spectral types.  We previously degraded the spectral resolution of the spectra and the models to the X-Shooter resolution. In Figure \ref{EW_nir}, we plot equivalent widths of the K\,I line for our objects.  We overplot  field objects \citep{Cushing, McLean}, objects that belong to TW Hydrae Association~(TWA), young companions \citep{Allers&Liu, Bonnefoy2014a}, young $\beta$-dwarfs and $\gamma$-dwarfs as a comparison \citep{Allers&Liu}.  We previously degraded the resolution of the observational spectra and the models to R$\sim$700, which is the lowest resolution of all the spectra for which we calculate K\,I~(1253~nm) equivalent width.
	
	We overplot the equivalent width predicted by the BT-Settl models 2014 for those alkali lines. We transform previously the effective temperature of the models to spectral types, using the empirical relation between spectral types and effective temperature relation published in \citet{Stephens2009}. 
	
	{{In the optical, the theory of cool atmospheres predicts the disappearance of the alkali elements in neutral form at temperatures below the L-T transition. This is a consecuence of the depletion of the alkali atoms into molecular compounds, and the veiling by silicate clouds forming above the line-forming level}, where the BT-Settl models 2014 over estimate the dust scattering}. {The apparent strength of the Na\,I subordinate lines decreases with spectra types from the early Ls through the T dwarfs}. The equivalent width of Cs\,I~(852.0~nm) increases from L0 to L9 and it is maximum for the early T brown dwarfs, and it weakens progressively from the early to the late T brown dwarfs. 
	The BT-Settl models reproduce the evolution of the equivalent width with the spectral type for Rb\,I~(794.8~nm) and Cs\,I~(852.0~nm) lines, but underestimate the equivalent width of Na\,I~(818.3~nm) and Na\,I~(819.5~nm) lines, especially in early to mid-L dwarfs. These elements do not participate directly in the sedimentation and dust formation, therefore the offset between the predicted and the observed equivalent widths could be due to uncertainties in the cloud model or the dust opacity contributing to pseudo-continuum that defines the equivalent width.

	In the near infrared (see Figure~\ref{EW_nir}), {the equivalent width of the K\,I~(1253~nm) has two peaks at around L4 and T4, with a minimum at about L8. This might reflect that as for the potassium, we see different atmospheric layers for the various subtypes \citep{Faherty2014}}. Object  2M0355 (object~10)  has weaker alkali lines in the optical and in the near infrared, as it is a young object \citep{Faherty2012, Allers&Liu, Zapatero_Osorio2014}. 
	The BT-Settl models reproduce the weakening of the K\,I~(1253~nm) line for low gravity objects, but overestimates the equivalent width of this line for field objects. In the L-T transition, the K\,I~(1253~nm) depletion or molecular blanketing is overestimated.
	
	{In Table~\ref{ew_all_lines}, we report the measured equivalent widths of the alkali lines for  our sample. We do not report the equivalent widths for those alkali lines that were not detected in some of the targets.}

	\begin{figure}
		\resizebox{\hsize}{!}{\includegraphics{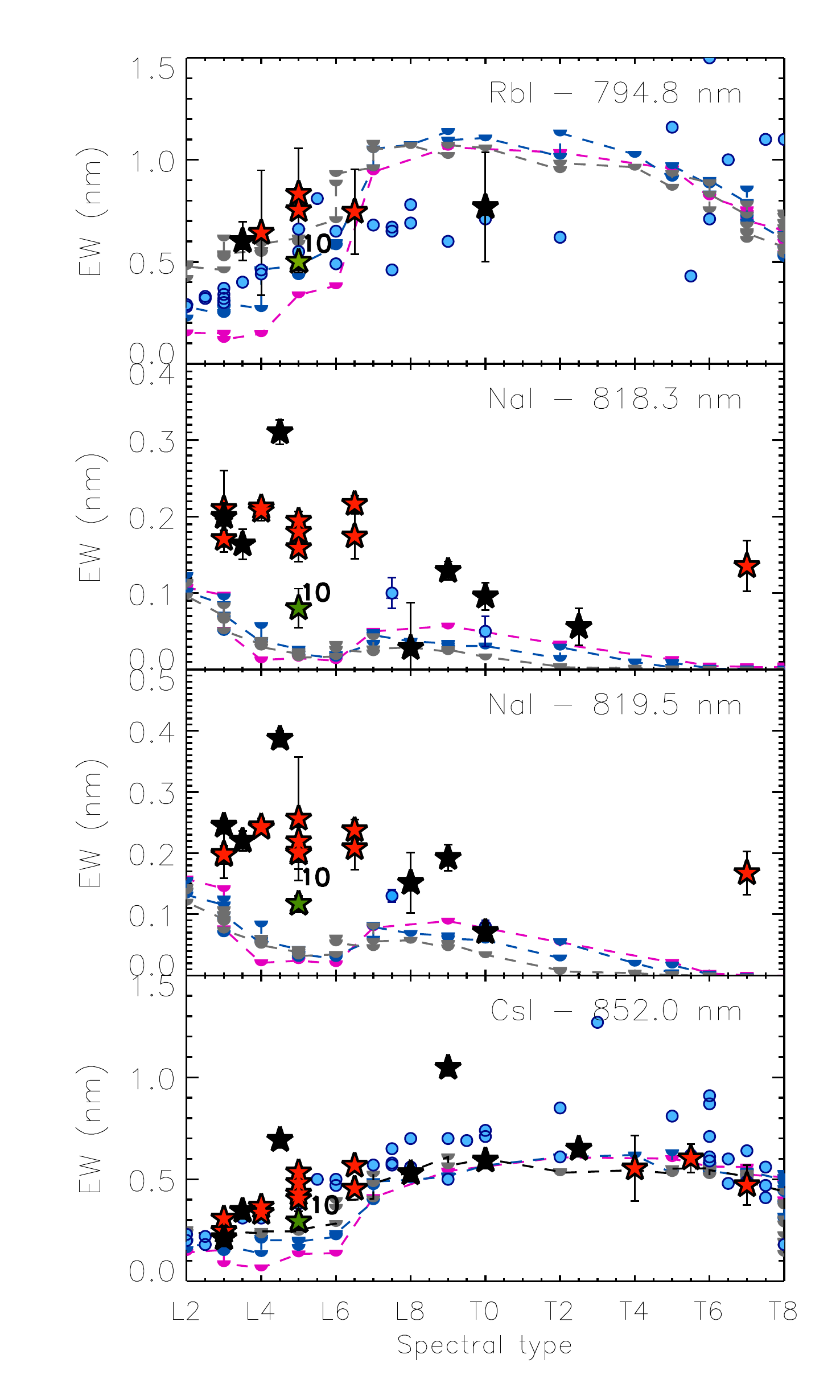}}
		\caption{Equivalent widths of the detected alkali lines in the optical for our targets (red stars), binary candidates or known binaries from our sample (black stars) and for objects with equivalent width  available in the literature (blue circles). The young object 2M0355 (number 10) is marked as green star. These equivalent widths come from \citet{Chiu2006, Golimowski_2004, Knapp_2004} and \citet{Lodieu2015}. We overplot with colored half circles joined by colored dashed lines the EW predicted by the BT-Settl models 2014  for different gravities. The  pink half filled circles and dashes lines correspond to $\log\:g$ = 4, the blue ones correspond to $\log\:g$ = 4.5 and the grey ones to $\log\:g$ = 5.0.}
		\label{EW_opt}
	\end{figure}
	
	\begin{figure}
		\resizebox{\hsize}{!}{\includegraphics{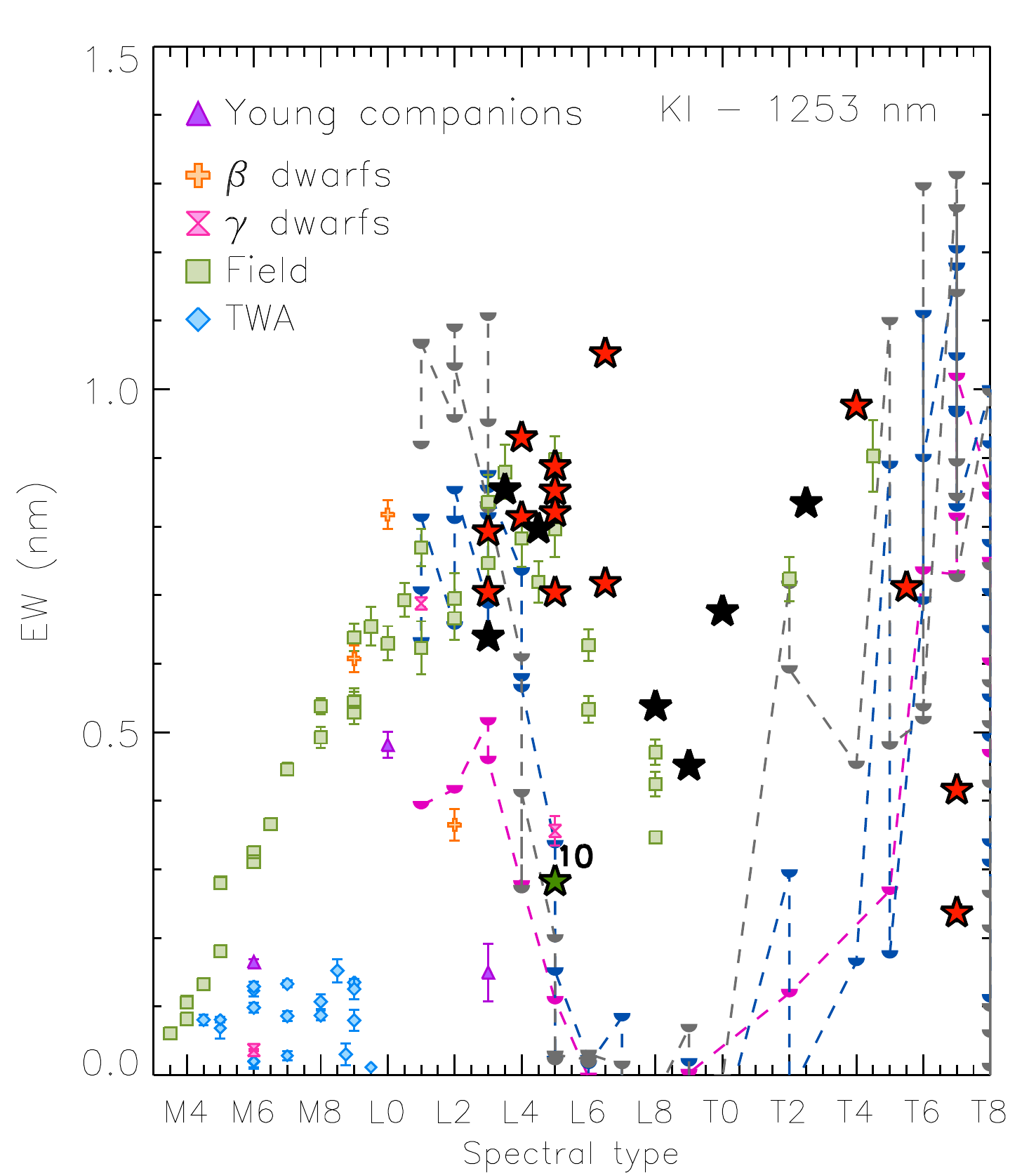}}
		\caption{Equivalent widths of the KI alkali line at 1253~nm of our objects (red stars), and  binary candidates or known binaries of our sample (black stars), compared to the equivalent widths of field brown dwarf, young companions, young brown dwarfs ($\beta$ and $\gamma$ dwarfs) and members of the TW Hydrae Association (TWA). The young object 2M0355 (number 10) is marked with a green star. We overplot with colored half circles joined by colored dashed lines the EW predicted by the BT-Settl models 2014  for differente gravities. We use the same colour code as for Figure \ref{EW_opt}.}
		\label{EW_nir}
	\end{figure}

	\begin{table*}
		\small
		\caption{Equivalent widths in nm for alkali lines measured in the optical and in the near infrared.}  
		\label{ew_all_lines}
		\centering
		\begin{center}
			\begin{tabular}{llllll}
				\hline
				\hline 
				
				Name & Rb\,I~(794.8~nm) & Na\,I~(818.3~nm) & Na\,I~(819.5~nm) & Cs\,I~(852.0~nm) & K\,I~(1253~nm)\\		
				
				\hline              
				LHS 102B                &    1.19$\pm$0.01 & 0.31$\pm$0.01 & 0.38$\pm$0.01  & 0.69$\pm$0.01& 0.79$\pm$0.01\\
				
				2MASS J00361617+1821104 &    0.53$\pm$0.01 & 0.21$\pm$0.01 & 0.24$\pm$0.01  &  0.34$\pm$0.01 &  0.93$\pm$0.01\\ 
				
				2MASS J00531899-3631102 &   0.60$\pm$0.09 & 0.16$\pm$0.02 & 0.22$\pm$0.02  &  0.35$\pm$0.01 &  0.85$\pm$0.01\\  
				
				SIMP  01365662+0933473  & <0.3   & 0.06$\pm$0.02 & <0.06   &  0.65$\pm$0.02& 0.83$\pm$0.01\\ 
				
				2MASS J01443536-0716142 &      &     &      &  0.43$\pm$0.09 &  0.82$\pm$0.02\\
				
				2MASS J02182913-3133230  & <0.09    &   0.17$\pm$0.02  &    0.19$\pm$0.02   &   0.31$\pm$0.01 &  0.79$\pm$0.01\\  
				
				DENIS-P J0255.0-4700    &  <0.18    &  0.13$\pm$0.01  &   0.19$\pm$0.02   &   1.04$\pm$0.02 &  0.45$\pm$0.01\\  
				
				2MASS J02572581-3105523  &   <0.54   &    &    0.15$\pm$0.05   &   0.53$\pm$0.03&  0.54$\pm$0.01\\ 
				
				2MASS J03480772-6022270  & <0.46     &  <0.28     &  <0.013   &   0.47$\pm$0.09 &  0.42$\pm$0.01\\   
				
				2MASS J03552337+1133437  & <0.03    & 0.08$\pm$0.03  &    0.12$\pm$0.01   &   0.29$\pm$0.01 &  0.28$\pm$0.01\\ 
				
				SDSS J0423485-041403   &   0.77$\pm$0.27   &  0.09$\pm$0.02   &  0.07$\pm$0.01  &  0.59$\pm$0.02 &  0.67$\pm$0.01\\  
				
				2MASS J04390101-2353083  &  <0.28    &   0.17$\pm$0.03   &    0.21$\pm$0.04  &   0.57$\pm$0.02 &  0.72$\pm$0.01  \\
				
				2MASS J04532647-1751543 &  <0.15   &   0.21$\pm$0.05   &    0.19$\pm$0.04   &   0.25$\pm$0.03 &  0.70$\pm$0.01 \\
				
				2MASS J05002100+0330501 &    0.64$\pm$0.31    &   0.21$\pm$0.01  &    0.24$\pm$0.01  &  0.37$\pm$0.01 &  0.81$\pm$0.01 \\
				
				2MASS J05395200-0059019  &   0.83$\pm$0.03    &   0.19$\pm$0.01  &    0.20$\pm$0.02 &   0.54$\pm$0.01 &  0.89$\pm$0.01  \\
				
				2MASS J06244595-4521548  &  <0.61    &   0.16$\pm$0.02  &  0.19$\pm$0.03  &   0.48$\pm$0.02 &  0.70$\pm$0.01  \\
				
				Gl 229B                  &      &    &    &  &  0.24$\pm$0.01  \\
				
				2MASS J10043929-3335189  &   0.75$\pm$0.30    &  0.18$\pm$0.02   &    0.22$\pm$0.02  &   0.40$\pm$0.02 &  0.85$\pm$0.01   \\
				
				2MASS J11263991-5003550  &   0.75$\pm$0.21   &  0.22$\pm$0.01    &    0.24$\pm$0.02  &   0.45$\pm$0.02 &  1.05$\pm$0.01  \\
				
				2MASS J13411160-3052505  &   0.37$\pm$0.28   &   0.19$\pm$0.02    &    0.25$\pm$0.01  &  0.21$\pm$0.01 &  0.64$\pm$0.01 \\
				
				2MASS J18283572-4849046  &       &      &     &   0.60$\pm$0.07 &  0.71$\pm$0.01  \\
				
				2MASS J21513839-4853542  &      &     &     &   0.55$\pm$0.16 &  0.98$\pm$0.01 \\

				\hline

			\end{tabular}
		\end{center}

	\end{table*}

	\section{Conclusions}\label{conclusions}
	

	We observed and analyzed medium resolution VLT/X-Shooter 
	spectra of 22 brown dwarfs with spectral types between L3 and T7. Objects in our sample have peculiar spectral characteristics or different classifications in
	the optical and in the near infrared.  Two of them   are known
	binaries, that allow us to test our analysis.

	Using \citet{Burgasser2006, Burgasser2010} and  \citet{Bardalez_Gagliuffi} empirical {methods}, we selected six objects as
	potential L plus T binary candidates: SIMP0136, SD0423, DE0255  and 2M1341, 2M0053  and 2M0257. We compared
	these six objects with single field brown dwarfs
	(\citealt{McLean, Cushing} and Spex libraries) and synthetic binaries. We found the best matches  using a {statistical} analysis similar to the $\chi^2$ analysis. Objects SIMP0136,   2M0257 and {2M1341 were discarded as candidates}. The binarity hypothesis is weakened for object DE0255, due to their lack of overluminosity  in the CMD, expected for late-L and early-T dwarf companions.
	
	We tested the efficiency of the method described in Section \ref{empirical_analysis}, and the possible proportion of false positives introduced aplying this method. To this aim, we compared single L and T dwarf spectra, and synthetic L plus T spectra, to other L and T single dwarf spectra, and to other  synthetic L plus T binaries. 
		We obtained that most of the L plus T synthetic binaries satisfy the binarity criteria.  {Nevertheless, 21\% of the L plus T synthetic binaries were not detected, i.e. 21\% of those synthetic binaries are missed when we apply the method described in Section \ref{empirical_analysis} using a 99\%  confidence level}. 37\% of the single L dwarfs, and 35\% of the single T  dwarfs satified {the binarity criteria} as well.
		The brown dwarf binary candidates found with this method   should be confirmed using additional data.

	{We  examined the possibility of finding equal spectral type  brown dwarfs
		binaries. {We compared single and synthetic binary spectra with the same subspectral type}, to other  single and synthetic binary spectra.   For both cases, we  obtained  best matches with synthetic binary spectra  for most of the cases. Therefore, we concluded that we are not able to find equal spectral type  binary systems using this method. Additional data, such as parallax measurements, high-resolution imaging or high resolution spectra are necessary in order to find these systems. }

	{We re-calculated a lower limit for the very low mass binary fraction of  $9.1^{+9.9}_{-3.0}$\% for our sample}.{ We found that at least   $4.5^{+10.4}_{-4.3}$\% of  the L and T objects in our sample may be unresolved binaries with one L and one T possible members}. This corresponds to a mass ratio of $q \ge 0.5$ for an age of a few Gyr (expected for most investigated objects). This percentage agrees with previous results.  
	
	BT-Settl models 2014 were able to reproduce the majority of the SEDs of our
	objects in the optical and in the near infrared. Nonetheless, these models
	usually failed to reproduce the shape of the H-band, due to incomplete opacities for  the
	FeH molecule in BT-Settl~2014 models. {Best matches to models gave a range of effective temperatures between {950~K (T7) and 1900~K (L6.5)}, a range of gravities between 4.0 and 5.5. Some of the best matches corresponded to supersolar metallicity.}
	
		We measured the equivalent width of alkali lines with good signal to noise (Na\,I, K\,I, Rb\,I and Cs\,I) in the optical and in the near infrared spectra. We concluded that in the transition from L to T spectral types, the Na\,I doublet at 818.3~nm and 819.5~nm in the optical  is the first to disappear, while the other alkalines are present in the optical and near infrared in the whole L to T spectral types. 
		We overploted the equivalent widths predicted by the BT-Settl models for those lines. The BT-Settl models reproduce the evolution of the equivalent width with the spectral type for the Rb\,I and Cs\,I lines, and the weakening of the K\,I line for the early L with low gravity. Nevertheless, the models underestimate the equivalent width of the Na\,I, lines in the optical, and overestimate the equivalent width for the K\,I line for field objects. These elements do not participate directly in the sedimentation and dust formation. Therefore, the differences between models and observational equivalent widths may be due to uncertainties in the cloud model or in the dust opacities.
	
	The optical and near infrared spectra reported in this paper will serve as
	templates for future studies in any of these wavelengths. In the near future, the
	Gaia satellite will release high precision parallaxes of more than one billion
	of objects in the Milky Way, including hundreds of brown dwarfs. These
	parallaxes will allow us to detect the overluminosity of brown dwarf binaries with respect to single brown dwarfs.


	\section*{Acknowledgements}
		
		{We gratefully acknowledge our referee, Daniella Bardalez-Gagliuffi, for her constructive inputs}. We thank ESO allocation time committee and Paranal
		Observatory staff for performing these observations. 
		We thank Adam Burgasser for providing and maintaining The SpeX Prism Spectral
		Libraries: http://pono.ucsd.edu/$\sim$adam/browndwarfs/spexprism/. We thank Adam
		Burgasser,  Daniella Bardalez Gagliuffi, Jackie Radigan, Bram Venemans and Esther Buenzli for
		their contribution in the development of this paper. This work was supported by Sonderforschungsbereich SFB 881 "The Milky Way System" (subprojects A4 and  B6) of the German Research Foundation (DFG). This research has made use of the SIMBAD database, operated at CDS, Strasbourg, France. 
		
		DH acknowledges support from the European Research Council under the European
		Community's Seventh Framework Programme (FP7/2007-2013 Grant Agreement No. 247060).
		The brown dwarf atmosphere models and synthetic spectra have been calculated at the P{\^o}le
		Scientifique de Mod{\'e}lisation Num{\'e}rique (PSMN) of the {\'E}cole Normale Sup{\'e}rieure de Lyon,
		and at the Gesellschaft f{\"u}r Wissenschaftliche Datenverarbeitung G{\"o}ttingen (GWDG) in
		co-operation with the Institut f{\"u}r Astrophysik G{\"o}ttingen.
		V.J.S.B. is supported by the project AYA2010-20535 from the Spanish  
		Ministry of Economy and Competitiveness (MINECO).


\bibliographystyle{mn2e_fix}
\bibliography{BD_binaries}


	
	\section*{appendix}\label{appendix}

	\section*{Observing log}\label{log0}

	\begin{table*}

		\begin{minipage}{18cm}
			\caption{Observing log{:} DIT is the integration time in each position of the slit, and {NINT} is the number of exposures. }  
			\label{log0}
			\centering
			\begin{tabular}{llllllll}
				\hline
				\hline
                Name & Date & Arm & DIT (s) & NINT & Seeing ('') & Airmass  & Notes \\
				\hline              
				LHS102B & October 16, 2009 & VIS/NIR & 290/300 & 4/4 & 1.0 & 1.05 &  \\
				Hip000349 & October 16, 2009 & VIS/NIR & 6/5 & 1/1 & 1.16& 1.31 & B9V Telluric Standard\\
				\hline	
				2M J0036+1821 & November 7, 2009 & VIS/NIR & 290/300 & 4/4 & 1.1 & 1.4 & \\
				Hip112022 & November 7, 2009 & VIS/NIR & 6/5 & 1/1 & 1.4 & 1.5 & B2IV Telluric Standard\\		
				\hline
				2M J0053-3631 & October 16, 2009 & VIS/NIR & 290/300 & 4/4 & 1.44 & 1.05& \\
				Hip000349     & October 16, 2009 & VIS/NIR & 6/5 & 1/1 & 1.15 & 1.01 & B9V Telluric Standard\\
				
				\hline
				SIMP J0136+0933 & December 14, 2009 & VIS/NIR & 290/300 & 4/4 & 0.9 & 1.2 & \\
				Hip021576       & December 14, 2009 & VIS/NIR & 6/5 & 1/1 & 1.1 & 1.05 & B3V Telluric Standard\\
				
				\hline
				2M J0144-0716  & December 14, 2009 & VIS/NIR & 290/300 & 4/4 & 1.05  & 1.15 &\\
				Hip021576      & December 14, 2009 & VIS/NIR & 6/5   & 1/1  & 1.05 & 1.15 & B6V Telluric Standard \\
				
				\hline
				2M J0218-3133 & January 4, 2010 & VIS/NIR & 290/300 & 4/4 & 1.05 & 1.15 & \\
				Hip009534     & January 4, 2010 & VIS/NIR & 6/5 & 1/1 & 1.1 & 1.1 & B6V Telluric Standard\\

				\hline
				DE J0255-4700 & October 17, 2009 & VIS/NIR & 290/300 & 4/4 & 2.2 & 1.14 & \\
				Hip009549     & October 17, 2009 & VIS/NIR & 6/5   & 1/1 & 2.2 & 1.2 & B6V Telluric Standard \\
				
				\hline
				2M J0348-6022 & October 16, 2009 & VIS/NIR & 290/300 & 12/12 & 1.7 &1.3 & \\
				Hip012389     & October 16, 2009 & VIS/NIR & 6/5 & 1/1 & 1.7 & 1.3 & B8V Telluric Standard\\
				
				\hline
				2M J0355+1133 & December 21, 2009 & VIS/NIR & 290/300 & 4/4 & 0.92 & 1.2 & \\
				Hip023060     & December 21, 2009 & VIS/NIR & 6/5 & 1/1 & 0.92 & 1.2 & B2V Telluric Standard\\
				\hline
				SD J0423-0414 & December 26, 2009 & VIS/NIR & 290/300 & 4/4 & 0.8 & 1.4 & \\
				Hip020424     & December 26, 2009 & VIS/NIR & 6/5 & 1/1 & 0.9 & 1.4 & B9V Telluric Standard\\
				\hline
				2M J0439-2353 & December 21, 2009 & VIS/NIR & 290/300 & 5/5 & 1.4 & 1.0 & \\
				Hip018926     & December 21, 2009 & VIS/NIR & 6/5   & 1/1 & 1.4 & 1.0 & B3V Telluric Standard\\
				\hline
				2M J0453-1751 & December 21, 2009 & VIS/NIR & 290/300 & 8/8 & 1.1 & 1.1 & \\
				Hip023060     & December 21, 2009 & VIS/NIR & 6/5   & 1/1 & 1.1 & 1.1 & B2V Telluric Standard\\
				\hline
				2M J0500+0330 & February 05, 2010 & VIS/NIR & 290/300 & 4/4 & 0.7 & 1.1 & \\
				Hip037623     & February 05, 2010 & VIS/NIR & 6/5   & 1/1 & 0.7 & 1.1 & B5V Telluric Standard\\
				\hline
				
				SD J0539-0059 & January 17, 2010 & VIS/NIR & 290/300 & 4/4 & 0.7 & 1.1 & \\
				Hip033007     & January 17, 2010 & VIS/NIR & 6/5   & 1/1 & 0.7 & 1.1 & B4V Telluric Standard\\
				\hline
				Gl229B        & December 14, 2009 & VIS/NIR & 290/300 & 4/4 & 1.2 & 1.4 & \\
				Hip044786     & December 14, 2009 & VIS/NIR & 6/5   & 1/1 & 1.2 & 1.4 & B6V Telluric Standard\\
				\hline
				2M J0624-4521 & December 16, 2009 & VIS/NIR & 290/300 & 5/5 & 0.8 & 1.4 & \\
				Hip030175     & December 16, 2009 & VIS/NIR & 6/5 & 5/5 & 0.8 & 1.4 & B9.5V Telluric Standard\\
				\hline
				2M J1004-3335 & February 5, 2010 & VIS/NIR & 290/300 & 4/4 & 0.9 & 1.0& \\
				Hip057861     & February 5, 2010 & VIS/NIR & 6/5   & 1/1 & 0.9 & 1.0 & B5V Telluric Standard\\
				\hline
				2M J1126-5003 & February 4, 2010 & VIS/NIR & 290/300 & 4/4 & 1.6 & 1.1 & \\
				Hip073345     & February 4, 2010 & VIS/NIR & 6/5 & 4/4 & 1.6 & 1.1 & B5V Telluric Standard\\
				\hline
				2M J2151-4853 & May 08, 2010     & VIS/NIR & 290/300 &10/10 & 1.4 & 1.2 & \\
				Hip111085     & May 08, 2010     & VIS/NIR &  6/5  &10/10 & 1.4 & 1.2& B9V Telluric Standard\\
				\hline
				2M J1341-3052 & June 02, 2010   & VIS/NIR & 290/300 &10/10 & 0.7 & 1.0& \\
				Hip068124     & June 02, 2010   & VIS/NIR & 6/5   &10/10 & 0.7 & 1.0 & B9V Telluric Standard\\
				\hline
				2M J1828-4849 & June 06, 2010   & VIS/NIR & 290/300 & 10/10 & 1.5& 1.4 & \\
				Hip092687     & June 06, 2010   & VIS/NIR & 6/5   & 1/1  & 1.5 & 1.4 & B4III Telluric Standard\\
				\hline
				
			\end{tabular}
		\end{minipage}
	\end{table*}

	\clearpage
	
	\section*{Reduced spectra}\label{reduced_spectra}
	\begin{figure*}
		\includegraphics[width=0.49\textwidth]{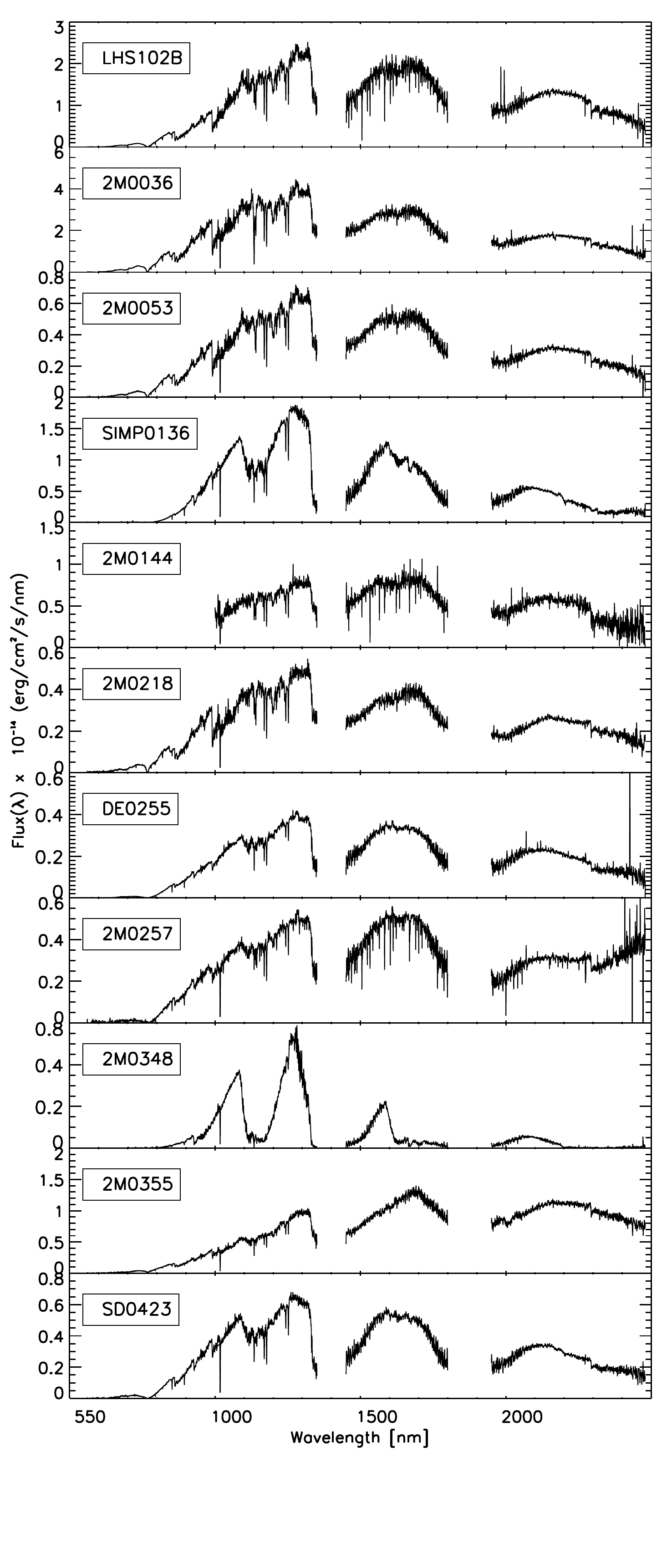}
		\includegraphics[width=0.49\textwidth]{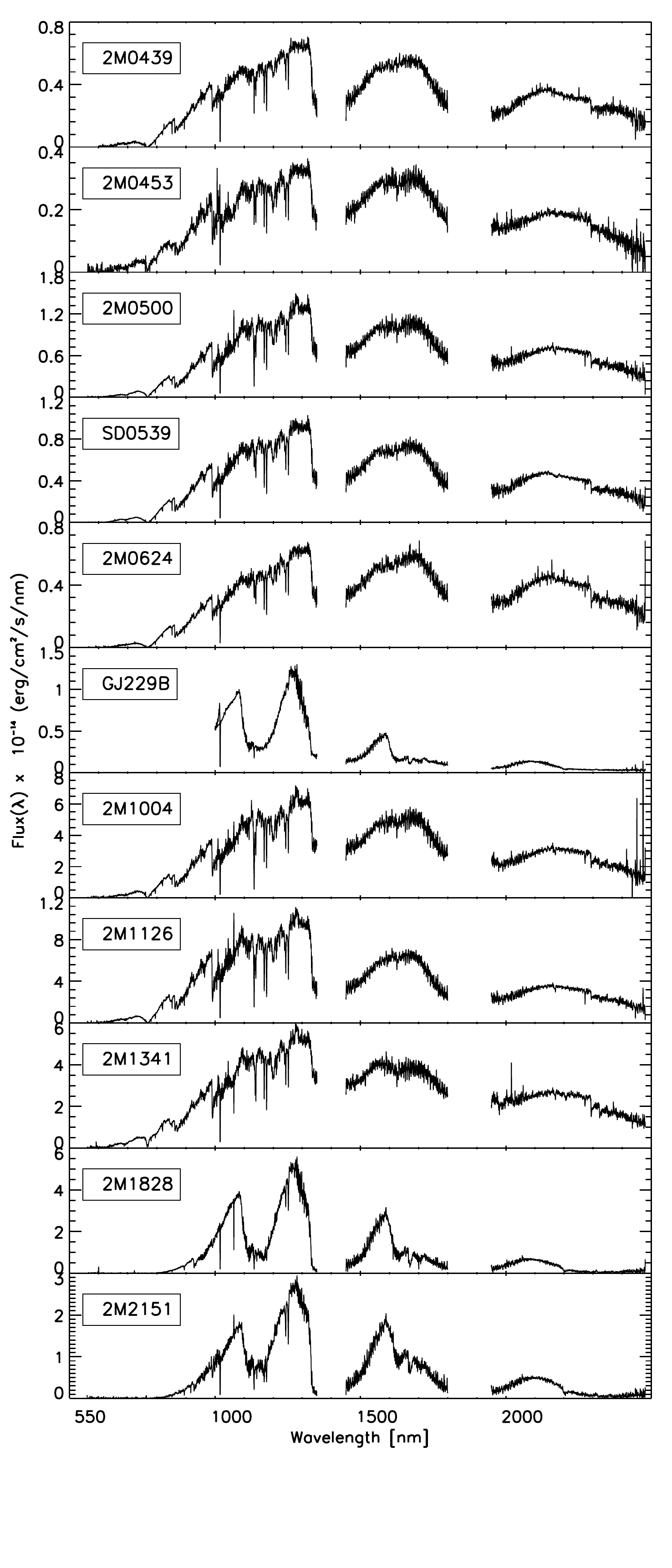}
		\caption{Spectra of our 22 targets after reduction and degrading them at R$\sim$1000.  Wavelengths largely affected by telluric absorption are removed from the figure {in the near infrared}, as well as the optical part for object Gl229B, because it is contaminated by the flux of its companion and the optical part of 2M0144 because it is noisy. {We plot spectra between 550-1350~nm, 1450-1800~nm and 1950-2500~nm to avoid telluric absortions.}}
		\label{all_spectra}
	\end{figure*}
	
			\section*{Spectral indices criteria}

			\begin{table*}
				\caption{Spectral indices to select L plus T brown dwarf binary candidates.}
				\label{spectral_indices}
				\centering
				\small
				\renewcommand{\footnoterule}{}  
				\begin{center}
					\begin{tabular}{l l l l l}
						\hline
						\hline
						Index & Numerator Range$^{a}$ & Denominator Range$^{a}$& Feature & Reference \\
						
						\hline
						
						$\mathrm{H_{2}O}$-J & 1140-1165 & 1260-1285 & 1150~nm $\mathrm{H_{2}O}$ &1 \\
						$\mathrm{CH_{4}}$-J & 1315-1340 & 1260-1285 & 1320~nm $\mathrm{CH_{4}}$ &1 \\
						
						$\mathrm{H_{2}O}$-H & 1480-1520 & 1560-1600 & 1400~nm $\mathrm{H_{2}O}$ &1 \\
						$\mathrm{CH_{4}}$-H & 1635-1675 & 1560-1600 & 1650~nm $\mathrm{CH_{4}}$ &1 \\
						
						$\mathrm{H_{2}O}$-K & 1975-1995 & 2080-2100 & 1900~nm $\mathrm{H_{2}O}$ &1 \\
						$\mathrm{CH_{4}}$-K & 2215-2255 & 2080-2120 & 2200~nm $\mathrm{CH_{4}}$ &1 \\
						
						$K/J$        & 2060-2100 & 1250-1290 & $J-K$ color & 1 \\
						H-dip      & 1610-1640 & 1560-1590 + 1660-1690$^{b}$ &1650~nm $\mathrm{CH_{4}}$& 2 \\
						
						K-slope    & 2.06-2.10  & 2.10-2.14  & K-band shape/ CIA $H_{2}$ & 3 \\
						J-slope    & 1.27-1.30  & 1.30-1.33  & 1.28~$\mu$m flux peak shape & 4 \\
						
						J-curve    & 1.04-1.07+1.26-1.29$^{c}$ & 1.14-1.17 & Curvature across J-band & 4\\
						H-bump     & 1.54-1.57  & 1.66-1.69  & Slope across H-band peak & 4\\
						
						$H_{2}O-Y$ & 1.04-1.07  & 1.14-1.17  & 1.15~$\mu$m$\mathrm{H_{2}O}$ & 4\\
						Derived NIR SpT  &             &            & Near infrared spectral type$^{d}$&1\\
						
						\hline
					\end{tabular}

				\end{center}
									\begin{tablenotes}
										\raggedright
										\item (a) Wavelength range in nm over which flux density is integrated; (b) Denominator is the sum of the flux in the two wavelength ranges; (c) Numerator is the sum of the two ranges; (d) Near infrared spectral type derived using comparison to SpeX spectra.\\ References: [1] - \citet{Burgasser2006}, [2] - \citet{Burgasser2010}, [3] - \citet{Burgasser2002}, [4] - \citet{Bardalez_Gagliuffi}.			
									\end{tablenotes}
			\end{table*}
			
			
			\begin{table*}
				\caption{Index criteria for the selection of potential brown dwarf binary systems}
				\label{criteria}
				\centering
				\begin{center}
					\begin{tabular}{l l l }
						\hline
						\hline
						Abscissa &Ordinate & Inflection Points \\
						
						\hline
						
						$\mathrm{H_{2}O}$-J & $\mathrm{H_{2}O}$-K          & (0.325,0.5),(0.65,0.7) \\
						$\mathrm{CH_{4}}$-H & $\mathrm{CH_{4}}$-K          &  (0.6,0.35),(1,0.775) \\
						$\mathrm{CH_{4}}$-H & $K/J$                 & (0.65,0.25),(1,0.375) \\
						$\mathrm{H_{2}O}$-H & H-dip               & (0.5,0.49),(0.875,0.49) \\
						Spex SpT & $\mathrm{H_{2}O}$-J/$\mathrm{H_{2}O}$-H &  (L8.5,0.925),(T1.5,0.925),(T3,0.85) \\
						Spex SpT & $\mathrm{H_{2}O}$-J/$\mathrm{CH_{4}}$-K & (L8.5,0.625),(T4.5,0.825) \\
						
						\hline
					\end{tabular}
				\end{center}
				
			\end{table*}

			\begin{table*}
				\caption{Delimiters for selection regions of potencial brown dwarf binary systems}
				\label{criteria2}
				\centering
				\renewcommand{\footnoterule}{}  
				\begin{center}
					\begin{tabular}{l l l}
						\hline
						\hline
						Abscissa & Ordinate & Limits \\
						
						\hline
						
						SpT & $\mathrm{CH_{4}}$-H &  Best fit curve: y = $-4.3x 10^{-4}x^2$+0.0253x + 0.7178       \\
						$\mathrm{H_{2}O}$-J & $\mathrm{CH_{4}}$-H & Intersection of: -0.08x+1.09 and x = 0.90         \\
						$\mathrm{H_{2}O}$-J & H-bump & Intersection of: y = 0.16x+0.806 and x = 0.90              \\
						$\mathrm{CH_{4}}$-J  & $\mathrm{CH_{4}}$-H & Intersection of: y = -0.56x + 1.41 and y = 1.04      \\
						$\mathrm{CH_{4}}$-J & H-bump & Intersection of: y = 1.00x + 0.24, x = 0.74 and y = 0.91 \\
						$\mathrm{CH_{4}}$-H & J-slope & Intersection of: y = 1.250x -0.207, x = 1.03 and y = 1.03 \\
						$\mathrm{CH_{4}}$-H & J-curve & Best fit curve: y = 1.245$x^{2}$ - 1.565x + 2.224\\
						$\mathrm{CH_{4}}$-H  & H-bump & Best fit curve: y = 1.36$x^{2}$ - 4.26x + 3.89 \\
						J-slope & H-dip & Intersection of y = 0.20x + 0.27 and x = 1.03 \\
						J-slope & H-bump & Intersection of: y = -2.75x + 3.84 and y = 0.91 \\
						K-slope & $\mathrm{H_{2}O}$-Y & Best fit curve: y = 12.036$x^{2}$ -20.000x +8.973 \\
						J-curve & H-bump & Best fit curve: y = 0.269$x^{2}$ - 1.326 + 2.479 \\

						\hline
					\end{tabular}
				\end{center}
				
			\end{table*}

			\begin{figure*}
				\resizebox{\hsize}{!}{\includegraphics{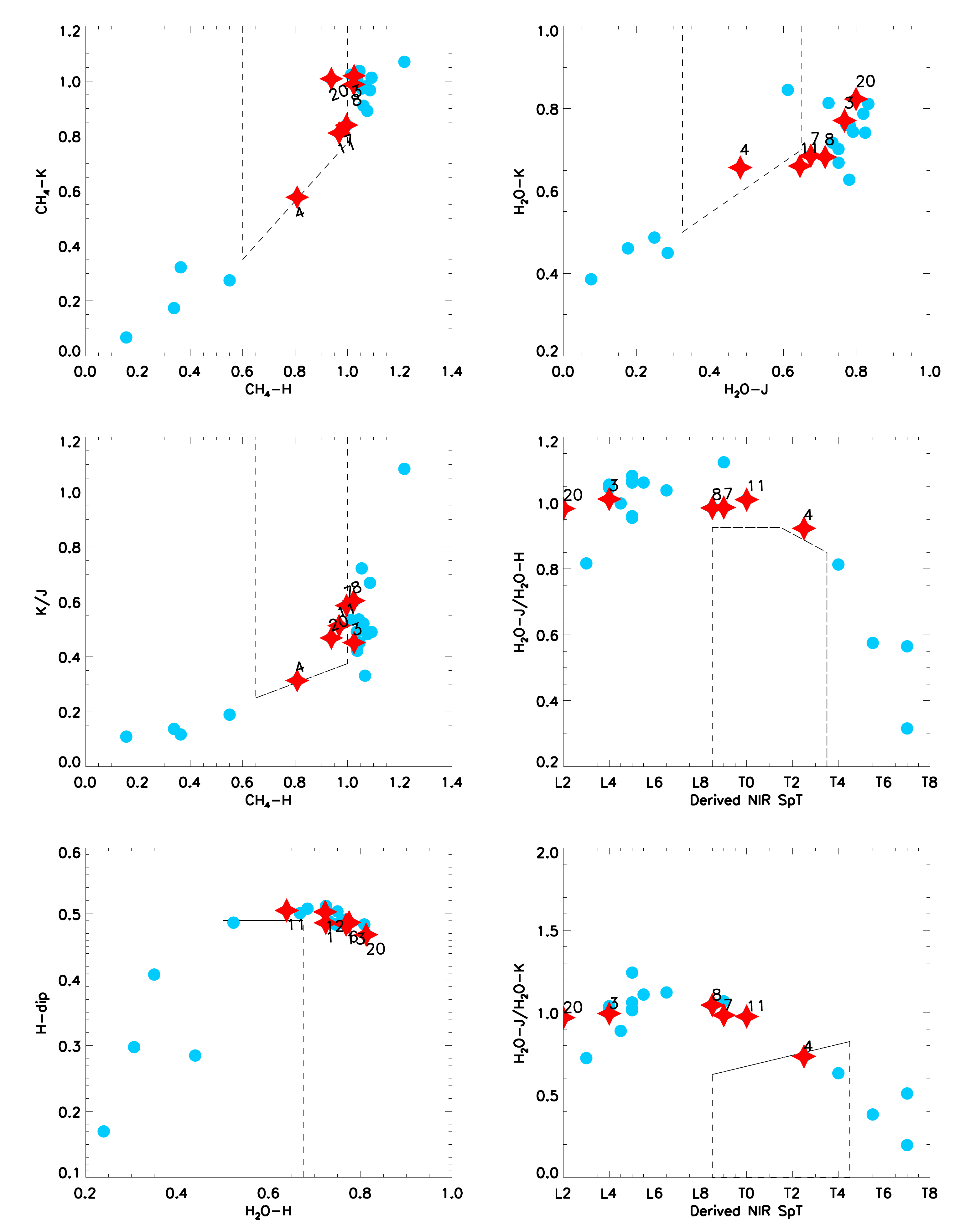}}
				
				\caption{Spectral index selection.  Numbers 1-22 correspond to our objects. The boxes shown with dashed lines mark  the areas where the selection criteria of Table~\ref{spectral_indices} are valid. The red stars represent objects satisfying more than four such criteria.}
				\label{indices}
			\end{figure*}

			\begin{figure*}
				\resizebox{\hsize}{!}{\includegraphics{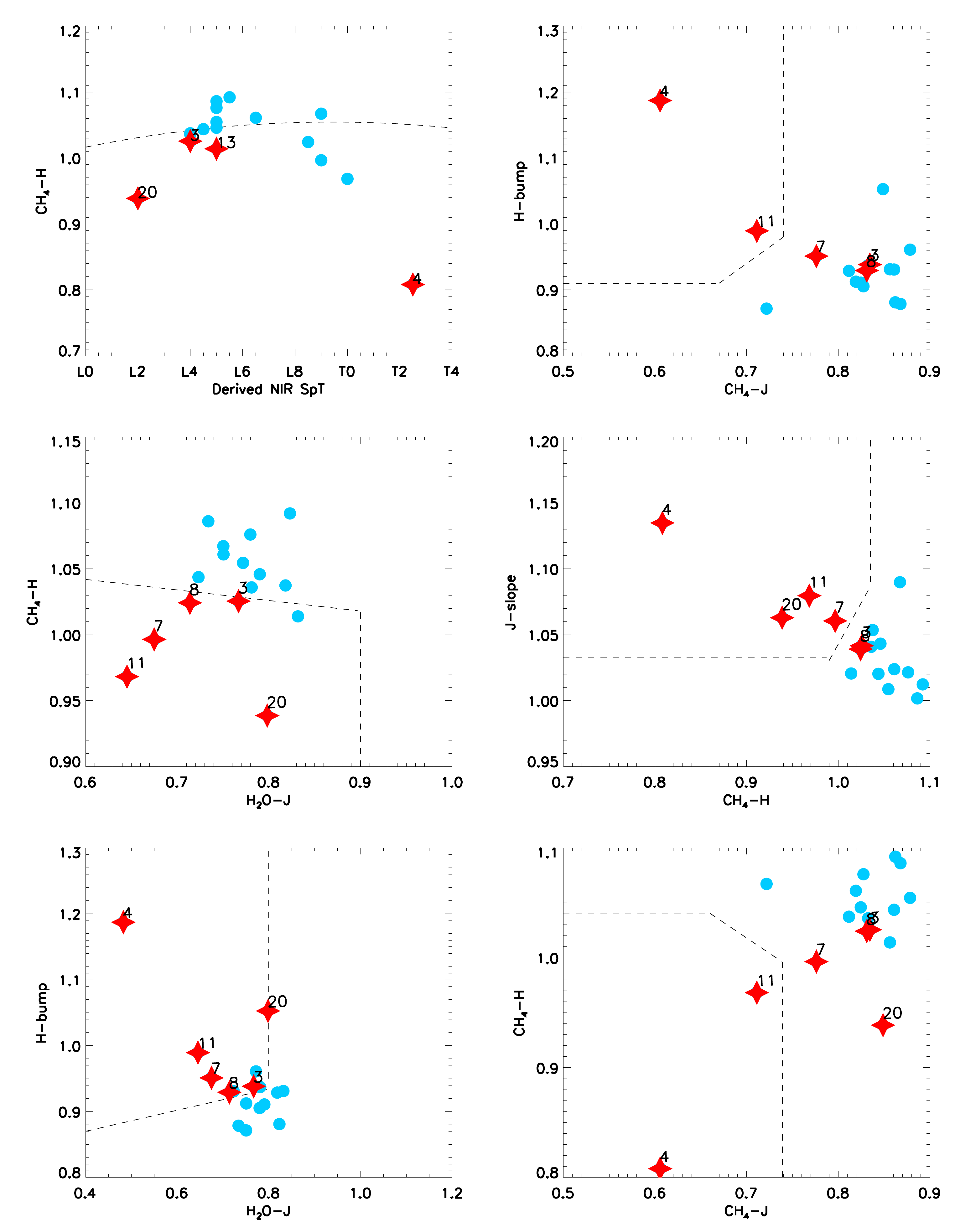}}
				\caption{Spectral index selection. }
				\label{indices1}
			\end{figure*}

			\begin{figure*}
				\resizebox{\hsize}{!}{\includegraphics{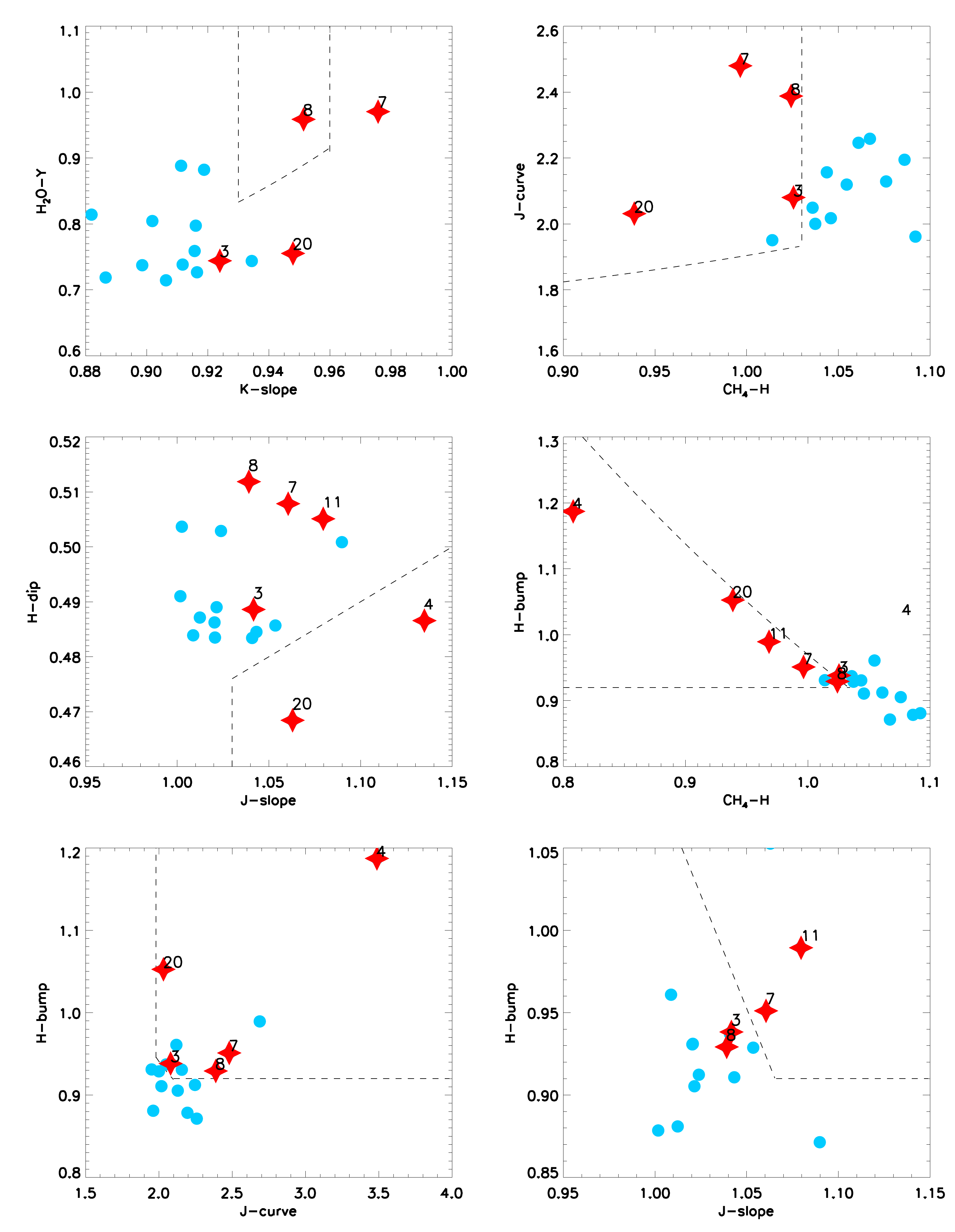}}
				\caption{Spectral index selection.  }
				\label{indices2}
			\end{figure*}


			\section*{Best matches to potential L plus T binaries}\label{L_T_best_matches}
			
			\begin{figure*}
				\centering
				\includegraphics[width=0.47\textwidth]{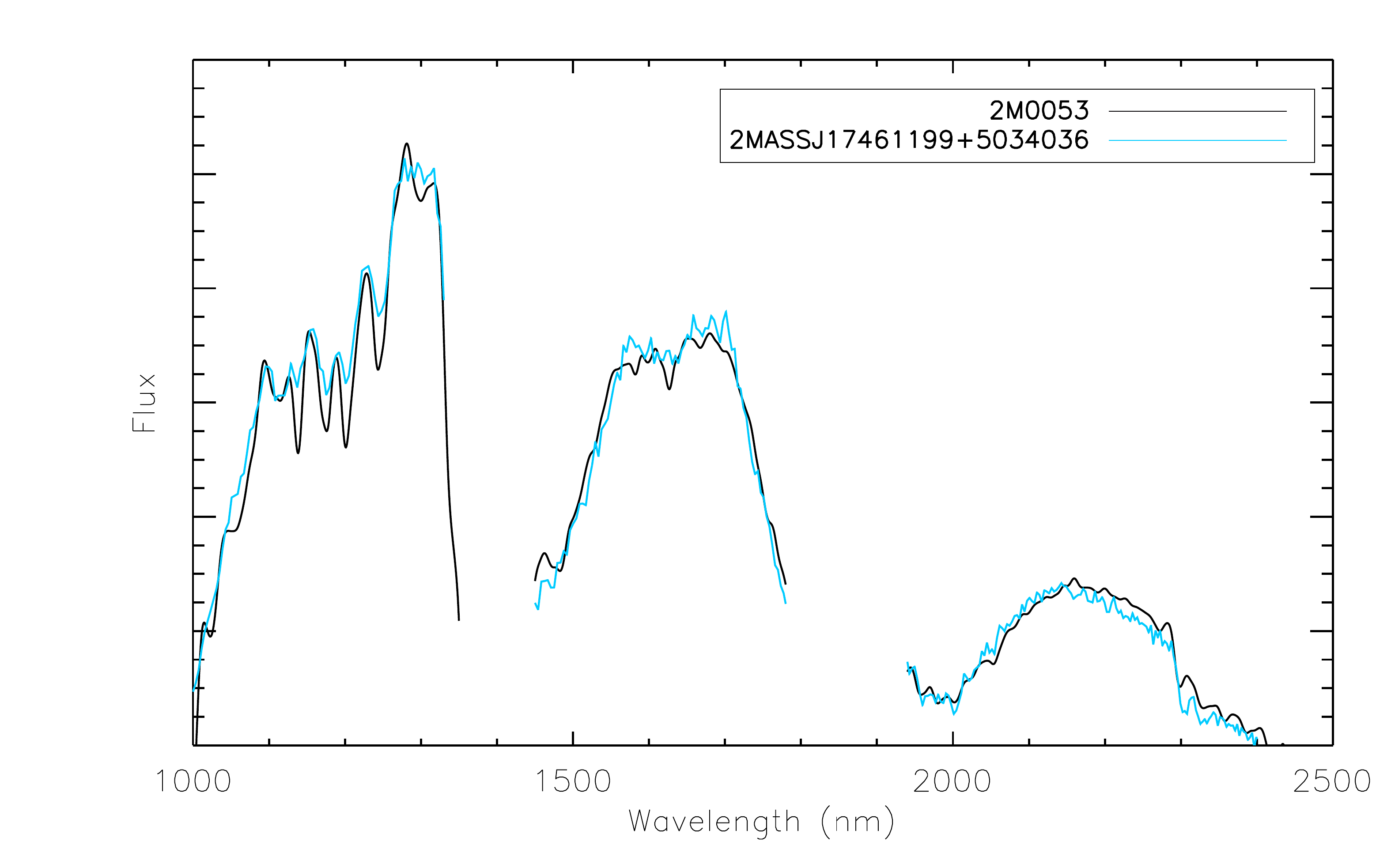}
				\includegraphics[width=0.47\textwidth]{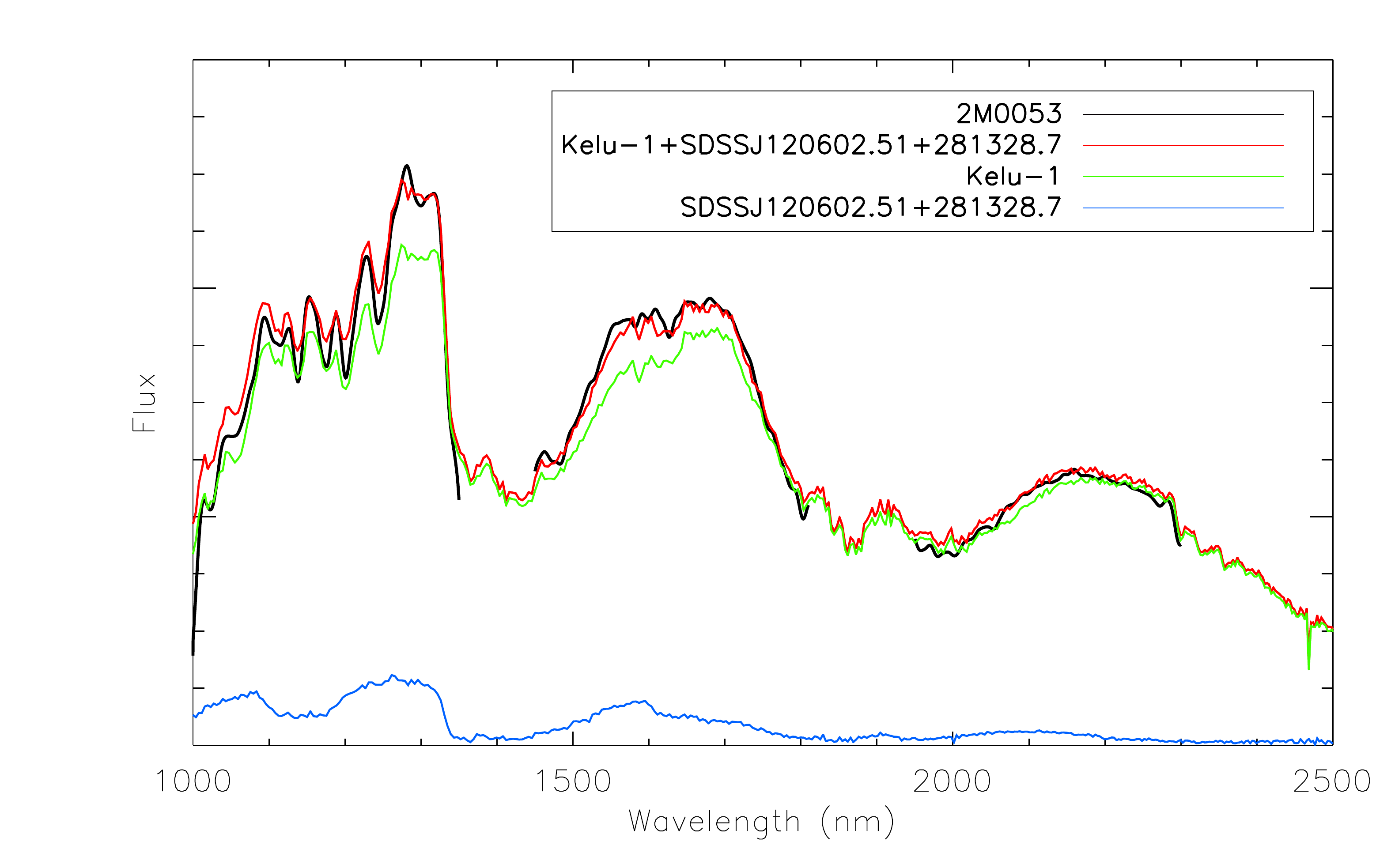}

				\caption{{Best matches for object 2M0053 to single (left plot)  and composite spectra (right plot) . We show in black our X-Shooter spectra. In the plot on the left, the blue spectrum belongs to the best single match (2MASS 17461199+5034036,  \citealt{Reid2008}). In the  plot on the right, we show in red the composite spectrum, in green the spectrum of the primary (Kelu-1,  \citealt{Ruiz1997}) and in blue the spectra of the secondary (SDSS 120602.51+281328.7, from \citealt{Chiu2006}). $\eta_{SB}$ = 1.35.  The flux is F($\lambda$).}}
				\label{2M0053}
			\end{figure*}

			\begin{figure*}
				\centering
				\includegraphics[width=0.47\textwidth]{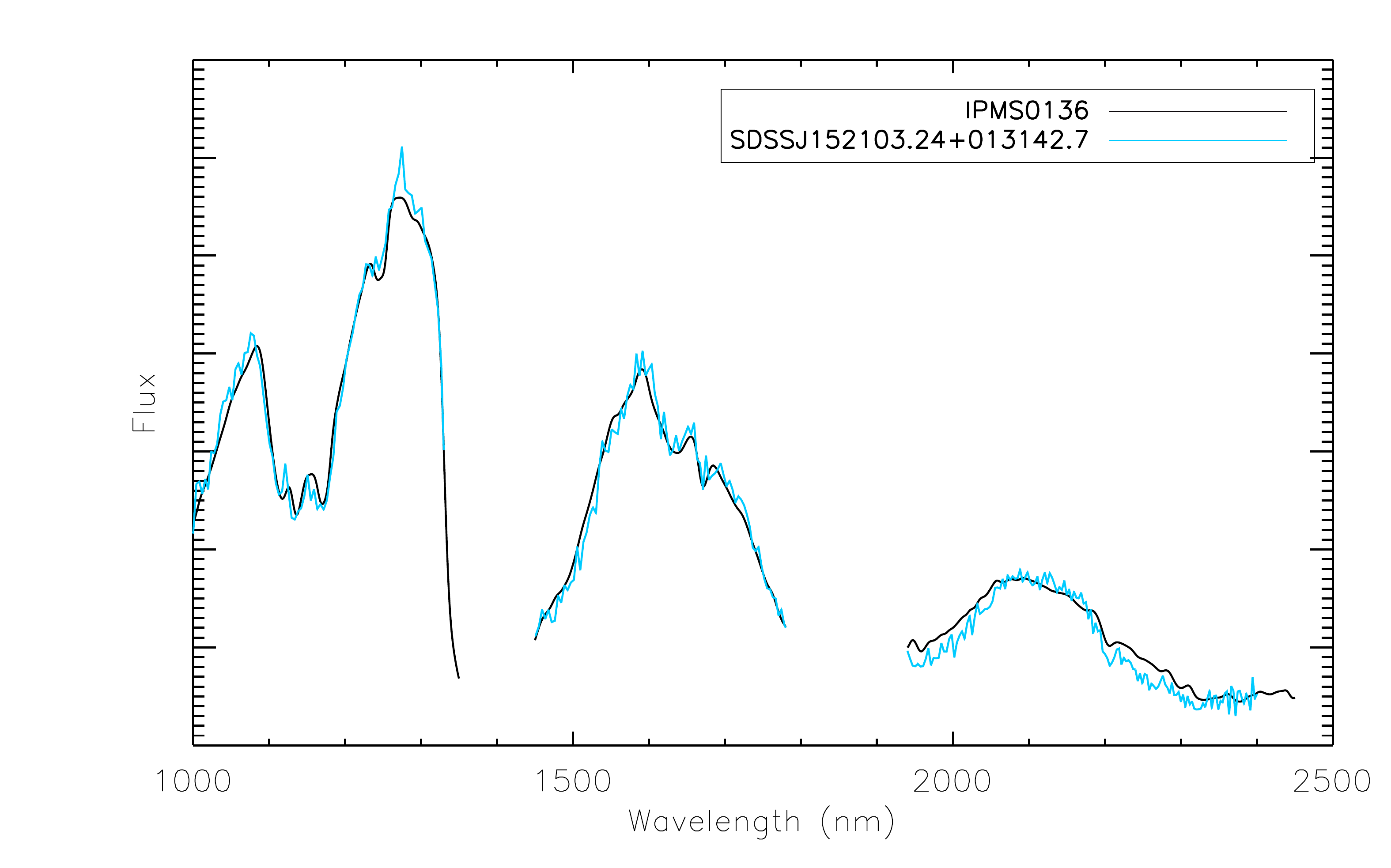}
				\includegraphics[width=0.47\textwidth]{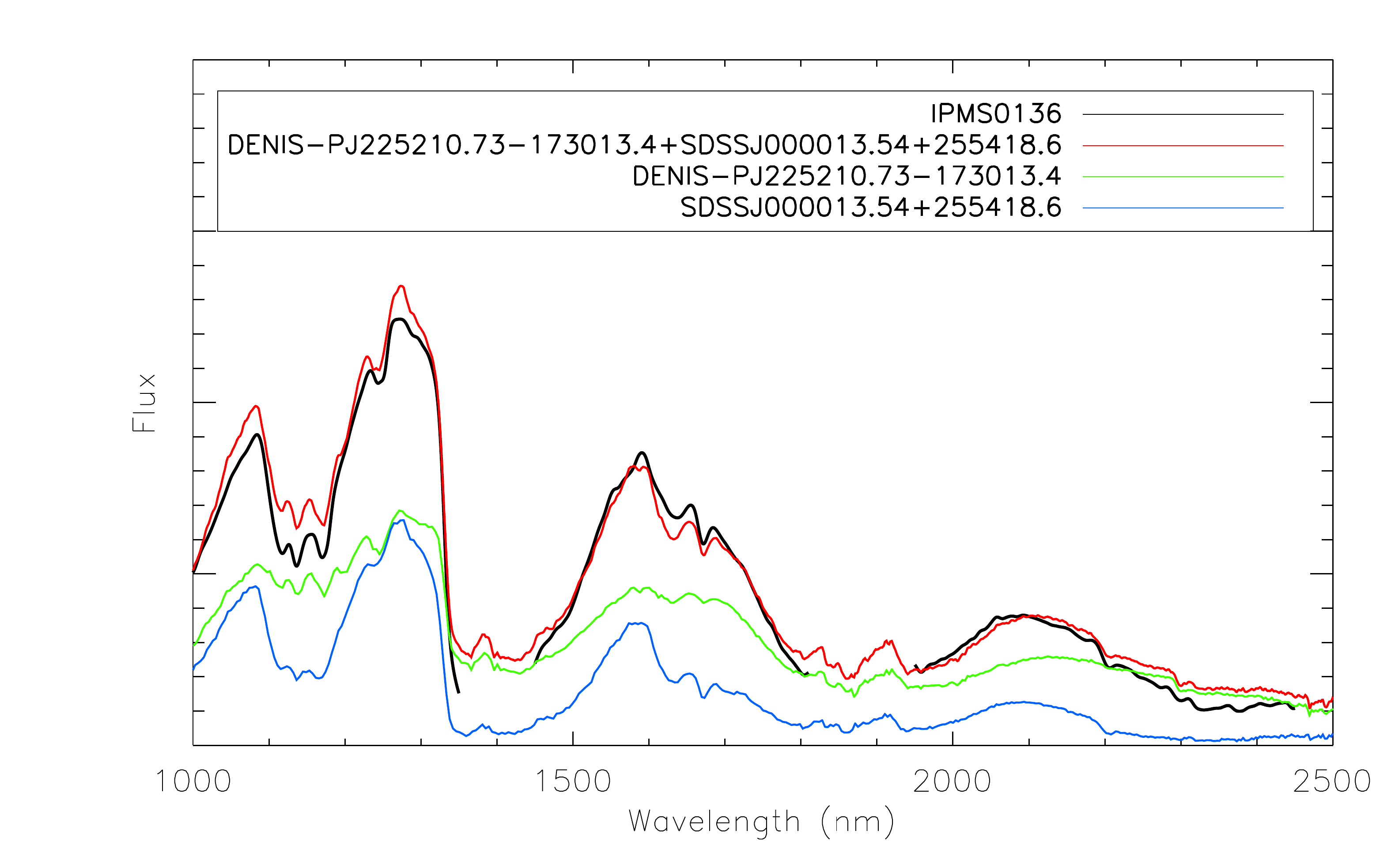}
				
				\caption{{Best matches for object SIMP0136 (T2.5) to single (SDSS J152103.24+013142.7, from \citealt{Knapp_2004}) and composite spectra (DENIS-PJ225210.73-173013, from \citealt{Kendall2004} and SDSS J000013.54+255418.6, from \citealt{Knapp_2004}). In black our smooth X-Shooter spectrum.  Colors are the same as in \ref{2M0053}. { $\eta_{SB}$ = 0.55}. The flux is F($\lambda$).}}
				\label{SIMP0136}
			\end{figure*}

			\begin{figure*}
				\centering
				\includegraphics[width=0.47\textwidth]{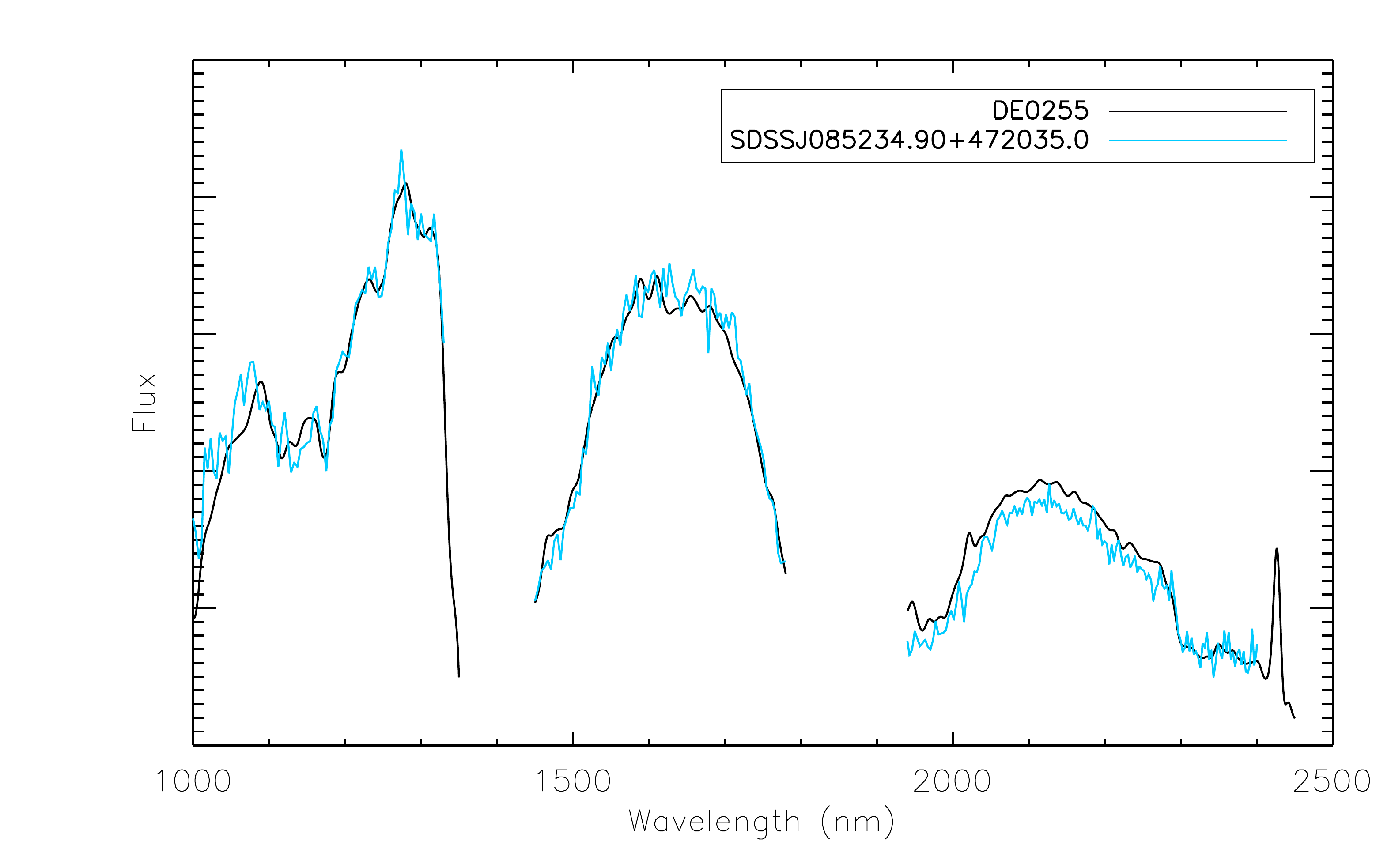}
				\includegraphics[width=0.47\textwidth]{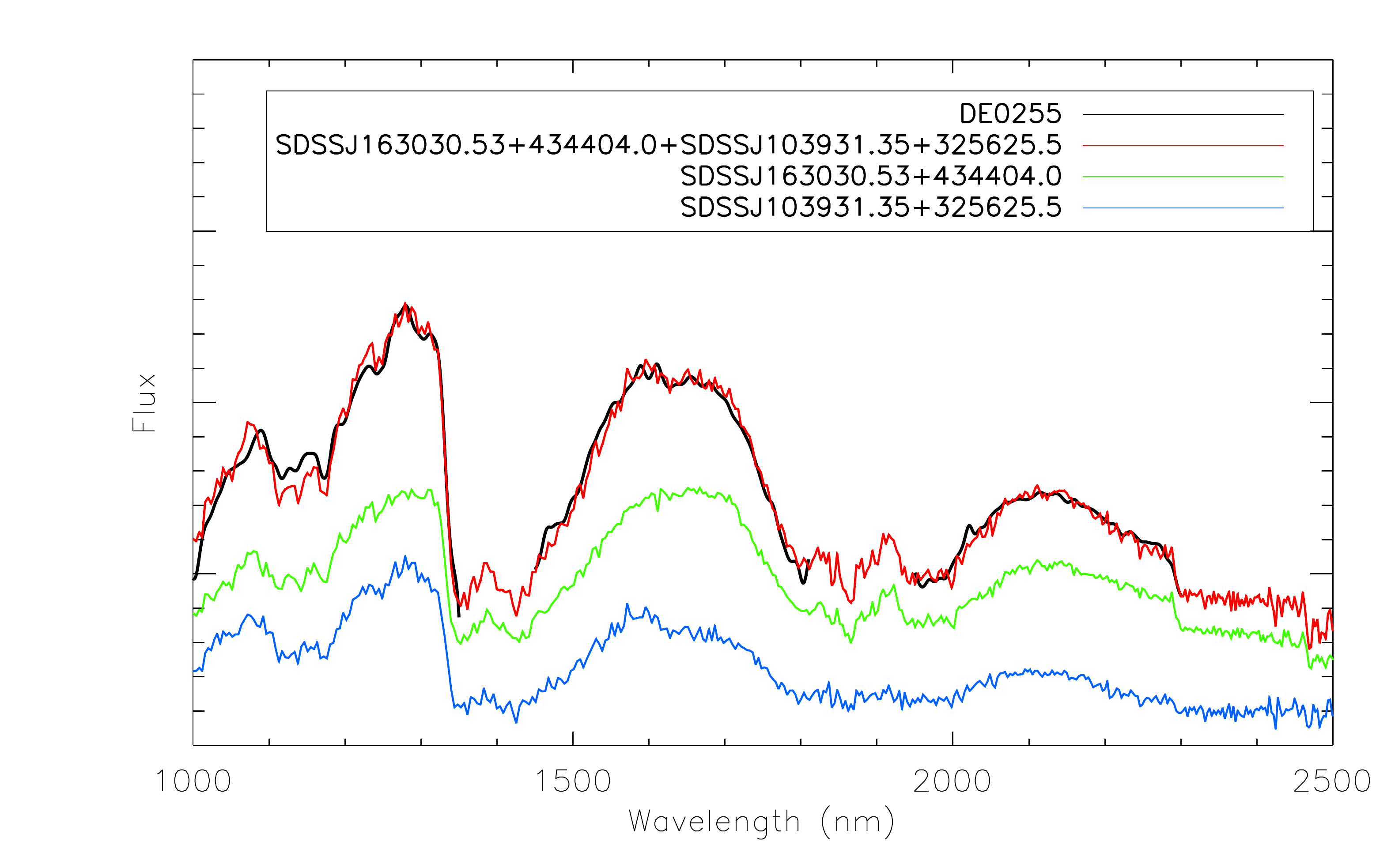}
				
				\caption{{Best matches for object DE0255 (L9) using single (SDSS J085234.90+472035.0, from \citealt{Knapp_2004})  and composite spectra (SDSS J163030.53+434404.0, from \citealt{Knapp_2004} and SDSS J103931.35+325625.5, from \citealt{Chiu2006}). Colors are the same as in \ref{2M0053}. { $\eta_{SB}$ = 3.42}. The flux is F($\lambda$).}}
				\label{DE0255_comp}
			\end{figure*}

			\begin{figure*}
				\centering
				\includegraphics[width=0.47\textwidth]{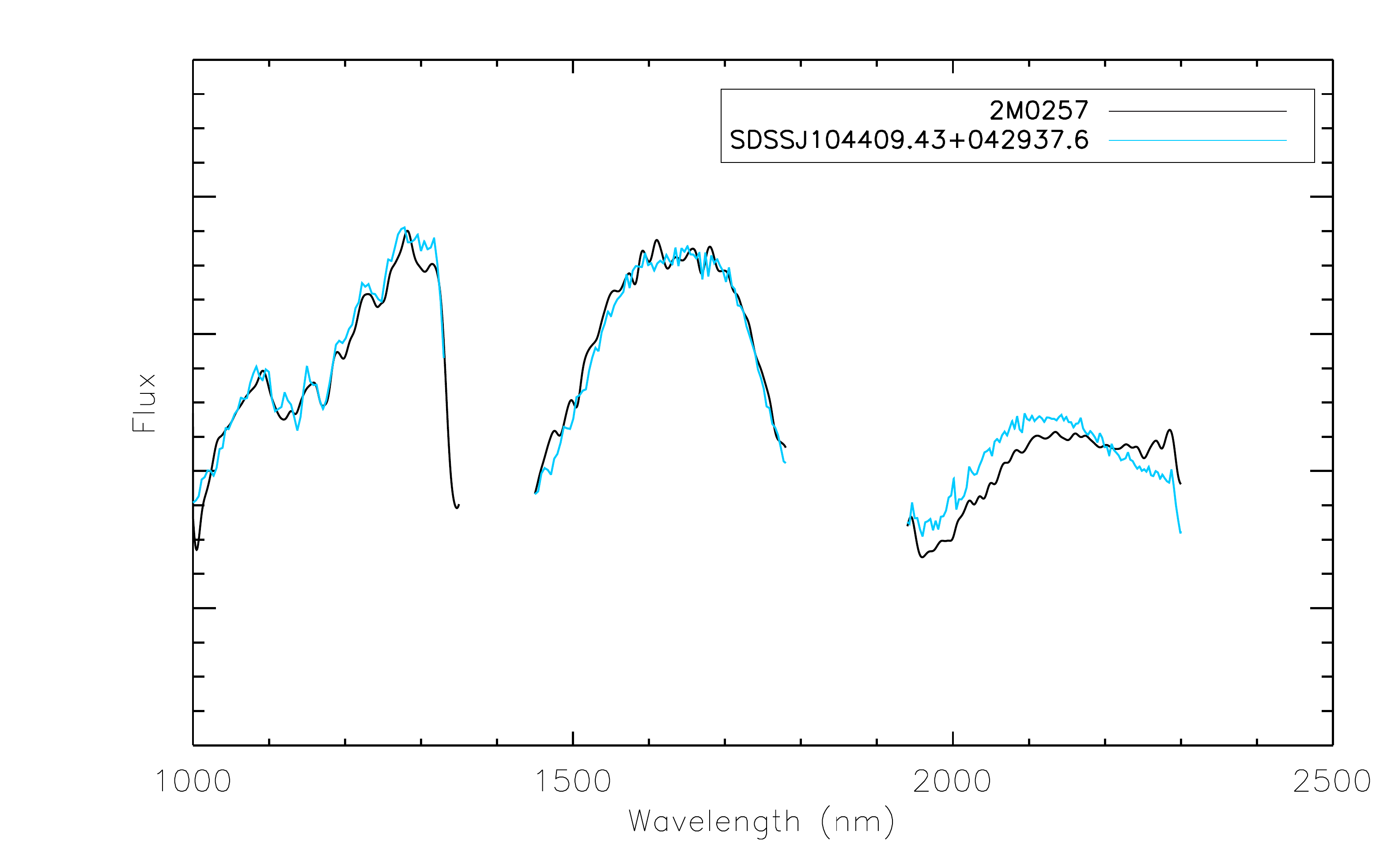}
				\includegraphics[width=0.47\textwidth]{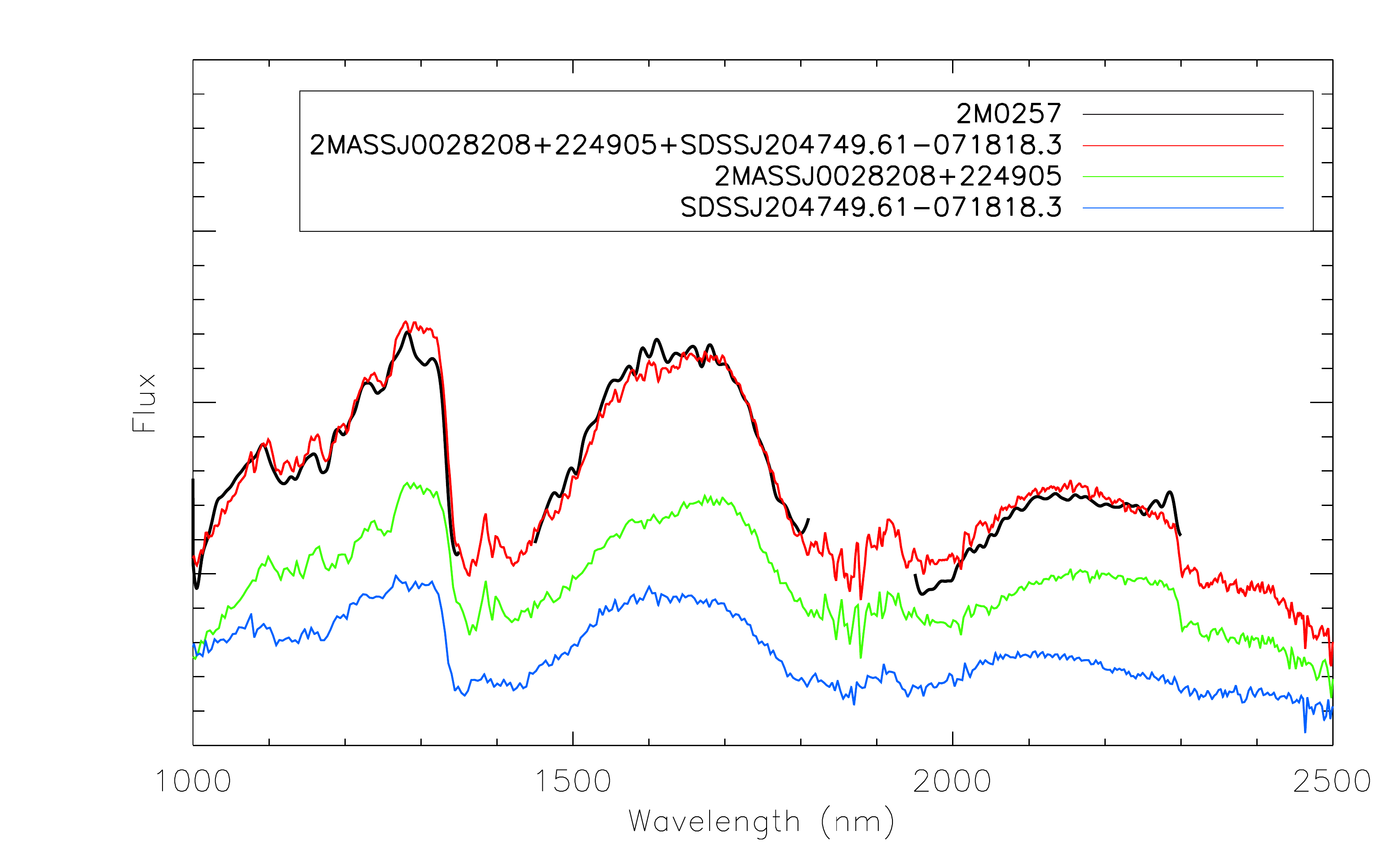}

				\caption{{Best matches for object 2M0257 to single (SDSS J104409.43+042937.6, from \citealt{Knapp_2004}) and composite spectra (2MASS J0028208+224905, from \citealt{Cutri2003},  and SDSS J204749.61-071818.3, from \citealt{Knapp_2004}). Colors are the same as in \ref{2M0053}. { $\eta_{SB}$ = 1.23}. The flux is F($\lambda$).}}
				\label{2M0257}
			\end{figure*}

			\begin{figure*}
				\centering
				\includegraphics[width=0.47\textwidth]{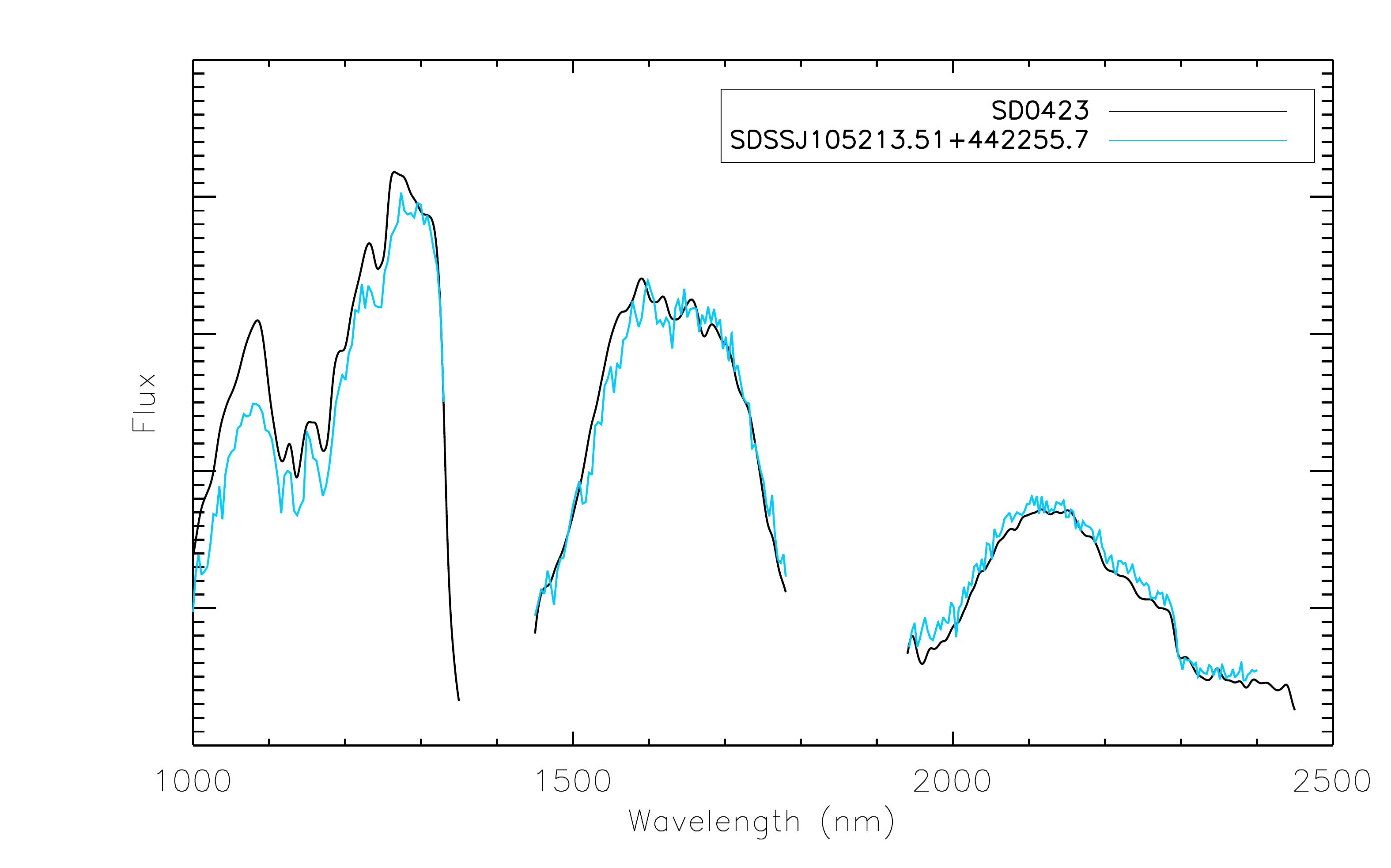}
				\includegraphics[width=0.47\textwidth]{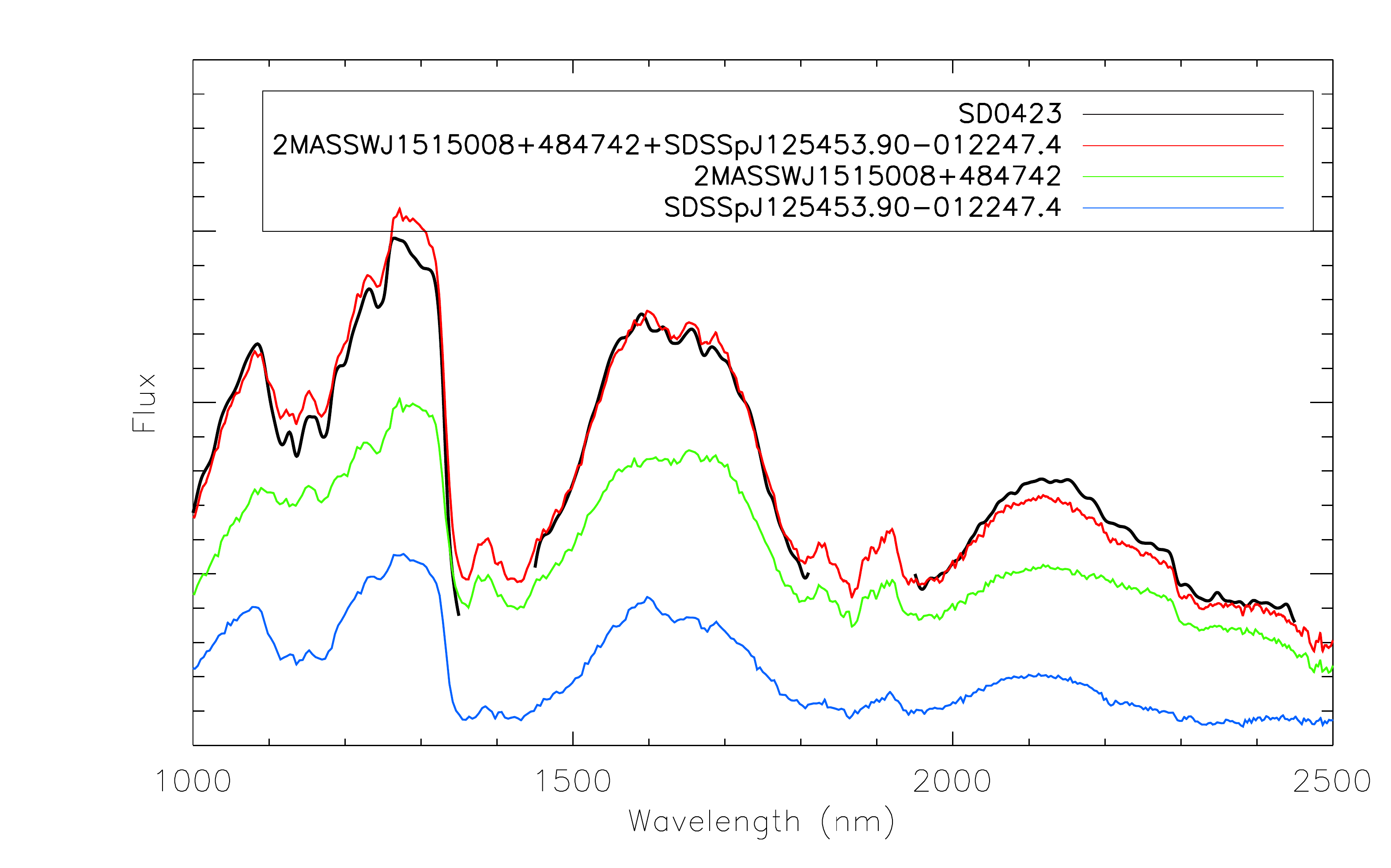}

				\caption{{Best match for object SD0423 (T0) using single (SDSS J105213.51+442255.7, from \citealt{Chiu2006}) and composite spectra (2MASS J1515008+484742, from \citealt{Wilson2003}, and SDSS J125453.90-012247.4, from \citealt{Leggett2000}). Colors are the same as in \ref{2M0053}. { $\eta_{SB}$ = 3.423.23}. The flux is F($\lambda$).}}
				\label{SD0423}
			\end{figure*}

			\begin{figure*}
				\centering
				\includegraphics[width=0.47\textwidth]{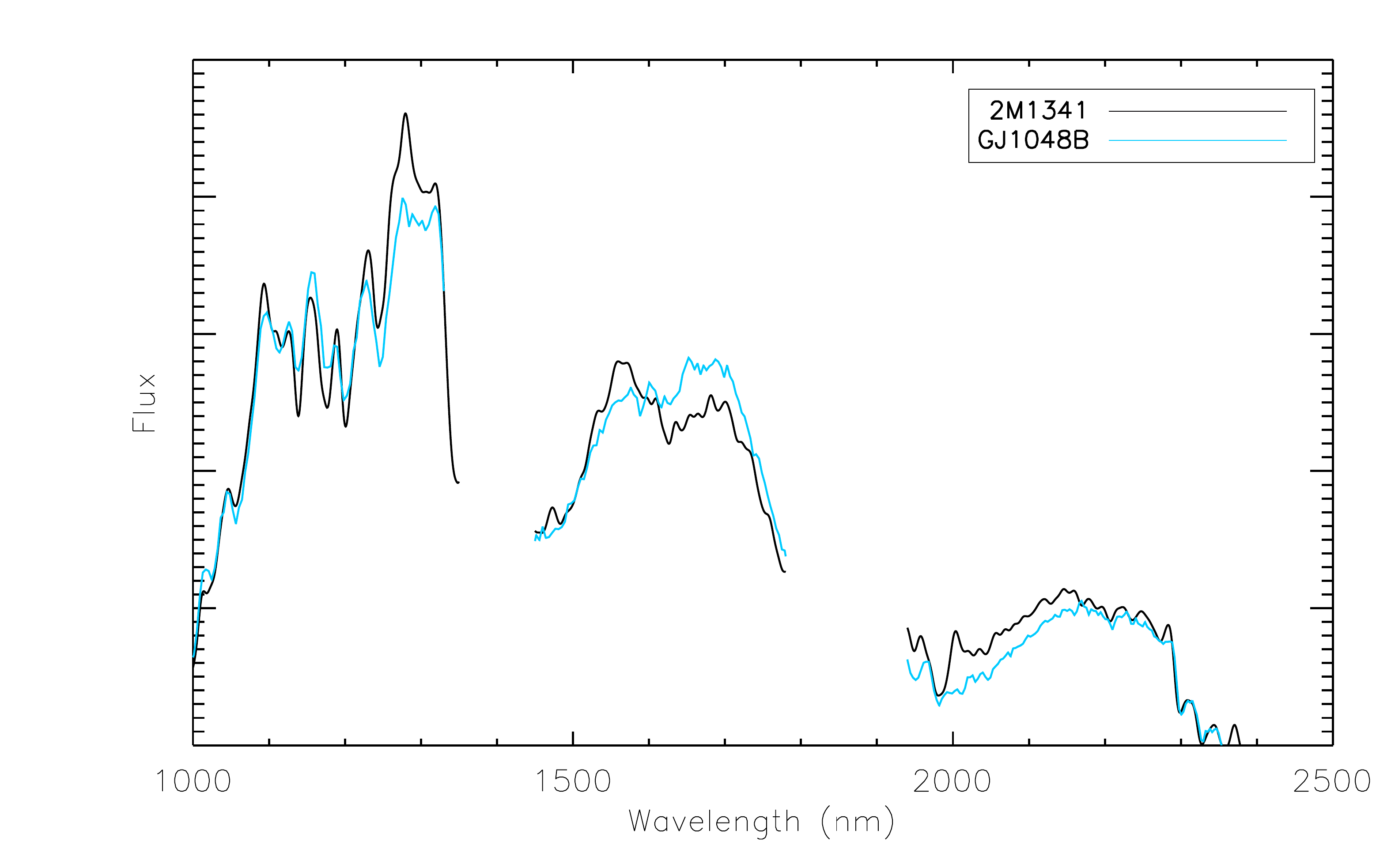}
				\includegraphics[width=0.47\textwidth]{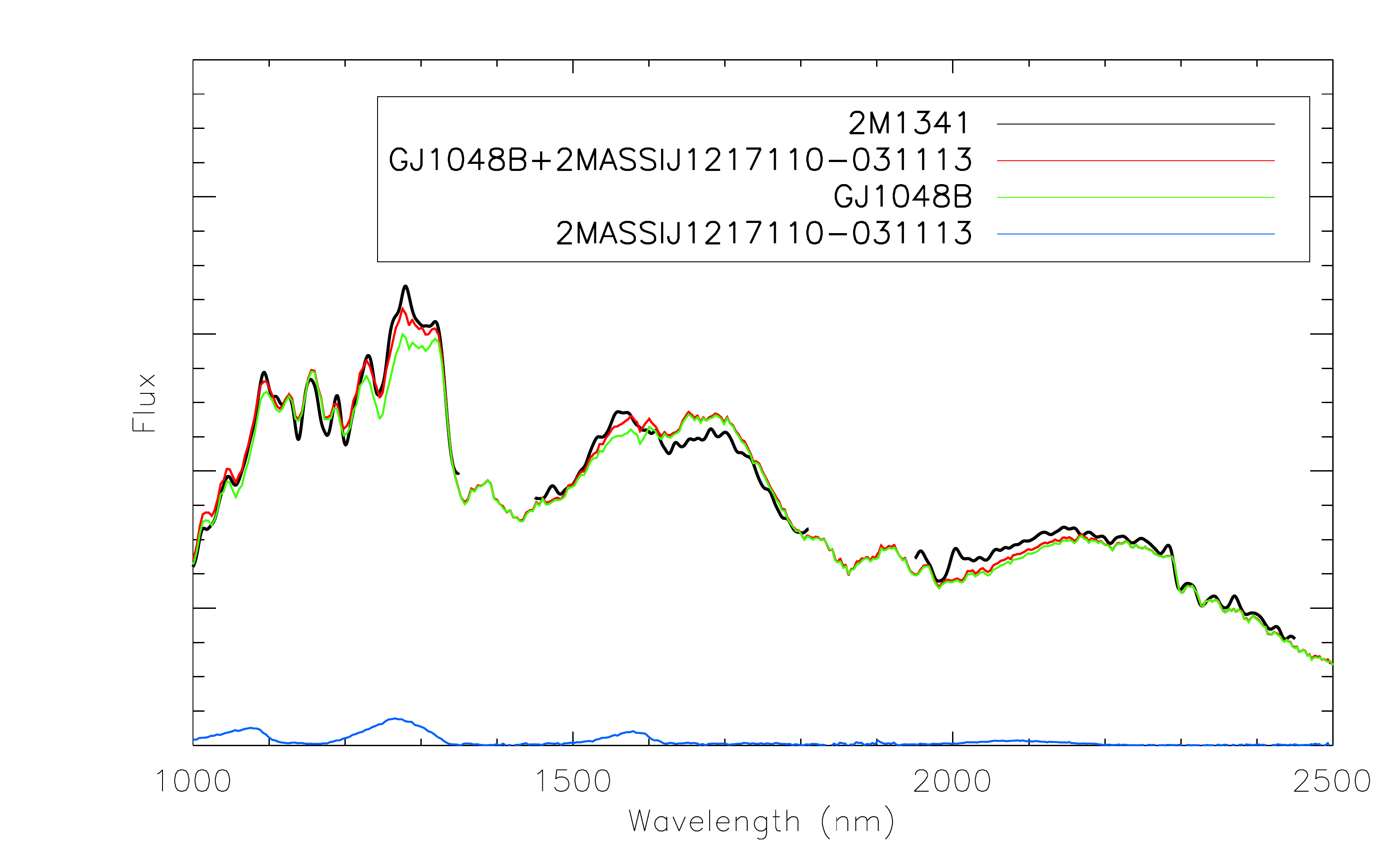}

				\caption{{Best matches for object 2M1341 (L2, peculiar) to single (GJ1048B, from \citealt{Gizis2001}) and composite spectra (GJ1048B, from \citealt{Gizis2001}, and 2MASS J1217110-031113, from \citealt{Burgasser1999}).  Colors are the same as in \ref{2M0053}. { $\eta_{SB}$ = 1.26}. The flux is F($\lambda$).}}
				\label{2M1341}
			\end{figure*}

			
			\section*{Best matches to BT-Settl models 2014}
			
			\begin{figure*}
				\raggedright
				\includegraphics[width=0.45\textwidth]{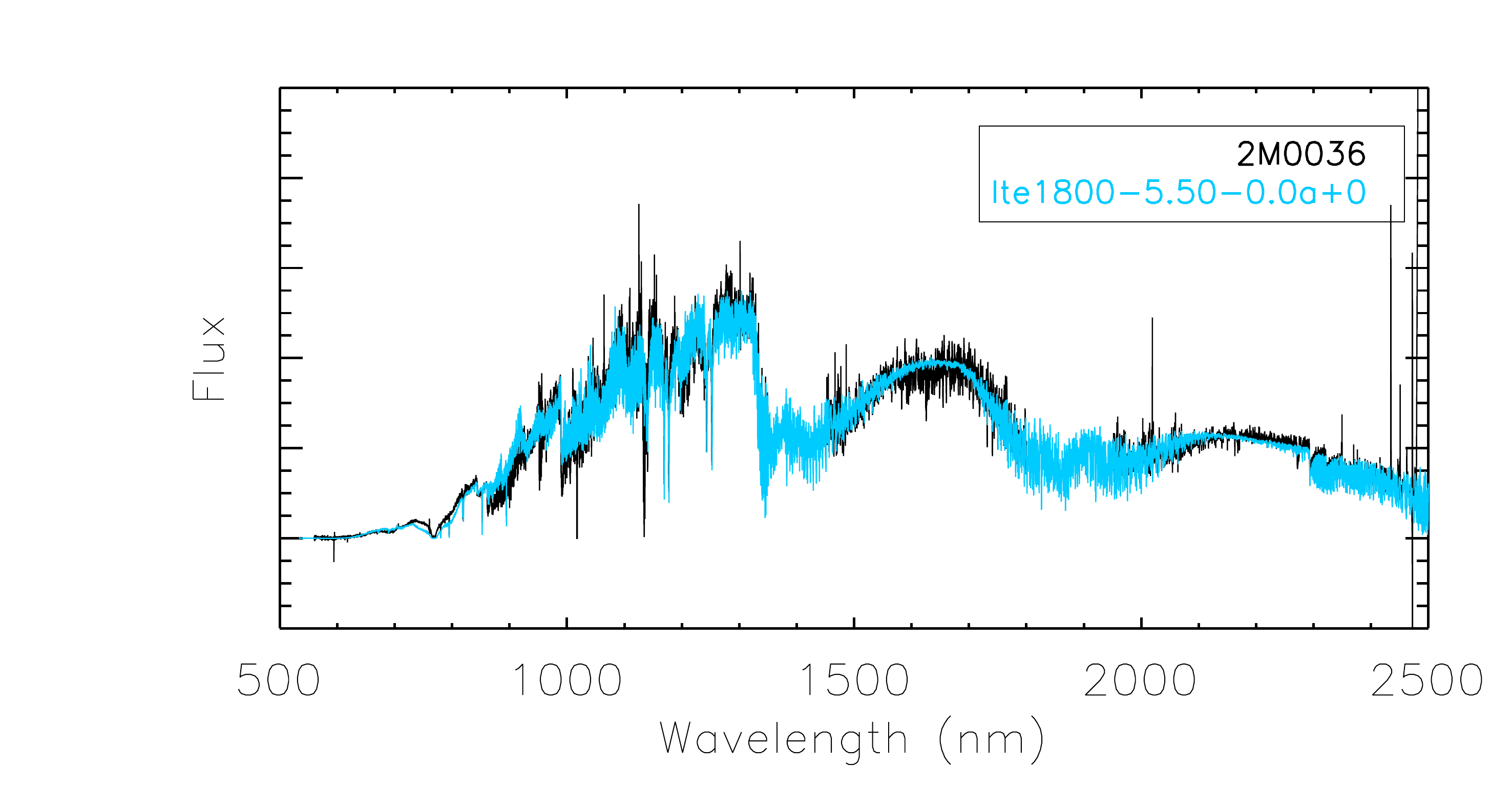}
				\includegraphics[width=0.45\textwidth]{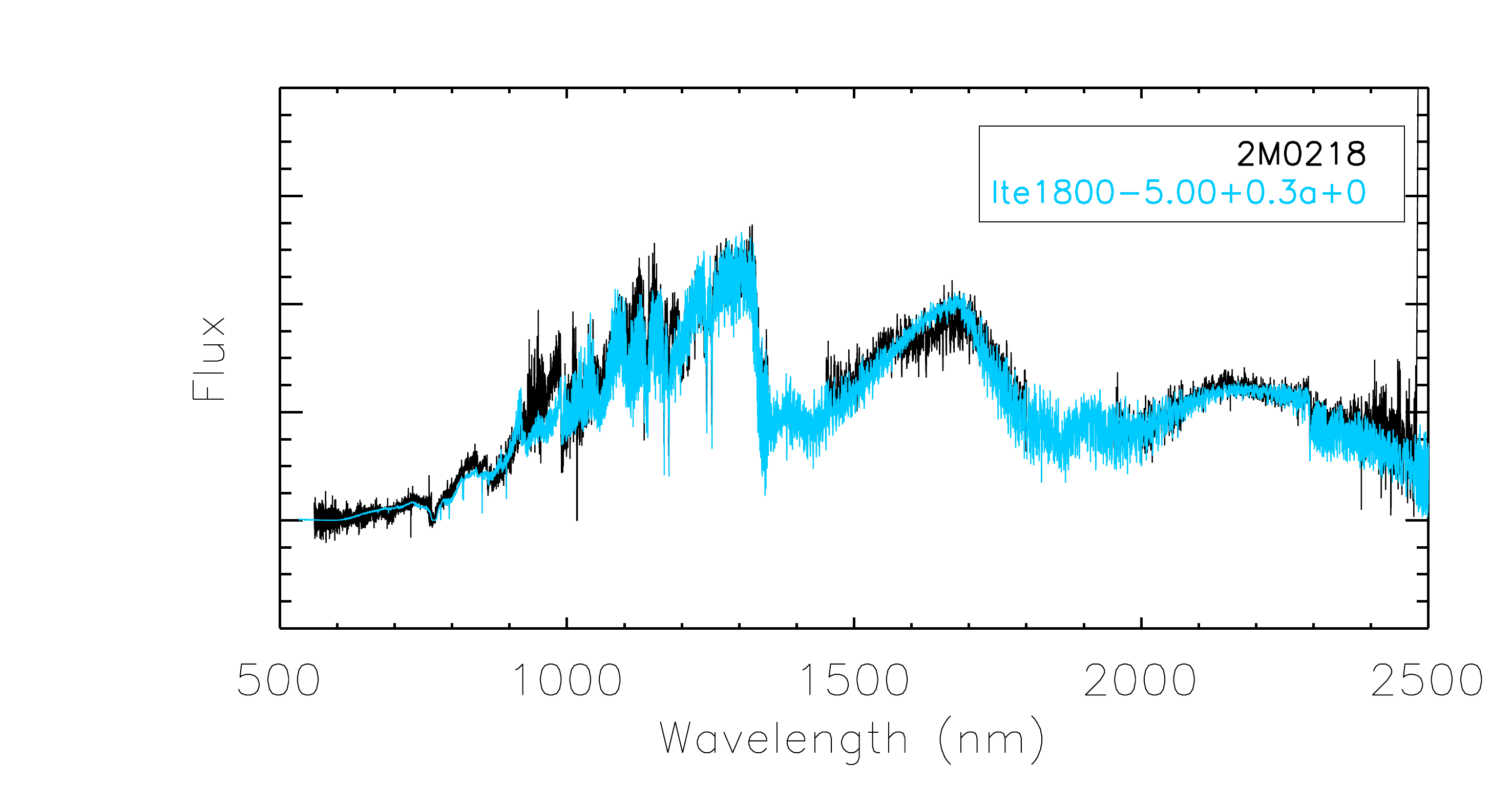}
				\includegraphics[width=0.45\textwidth]{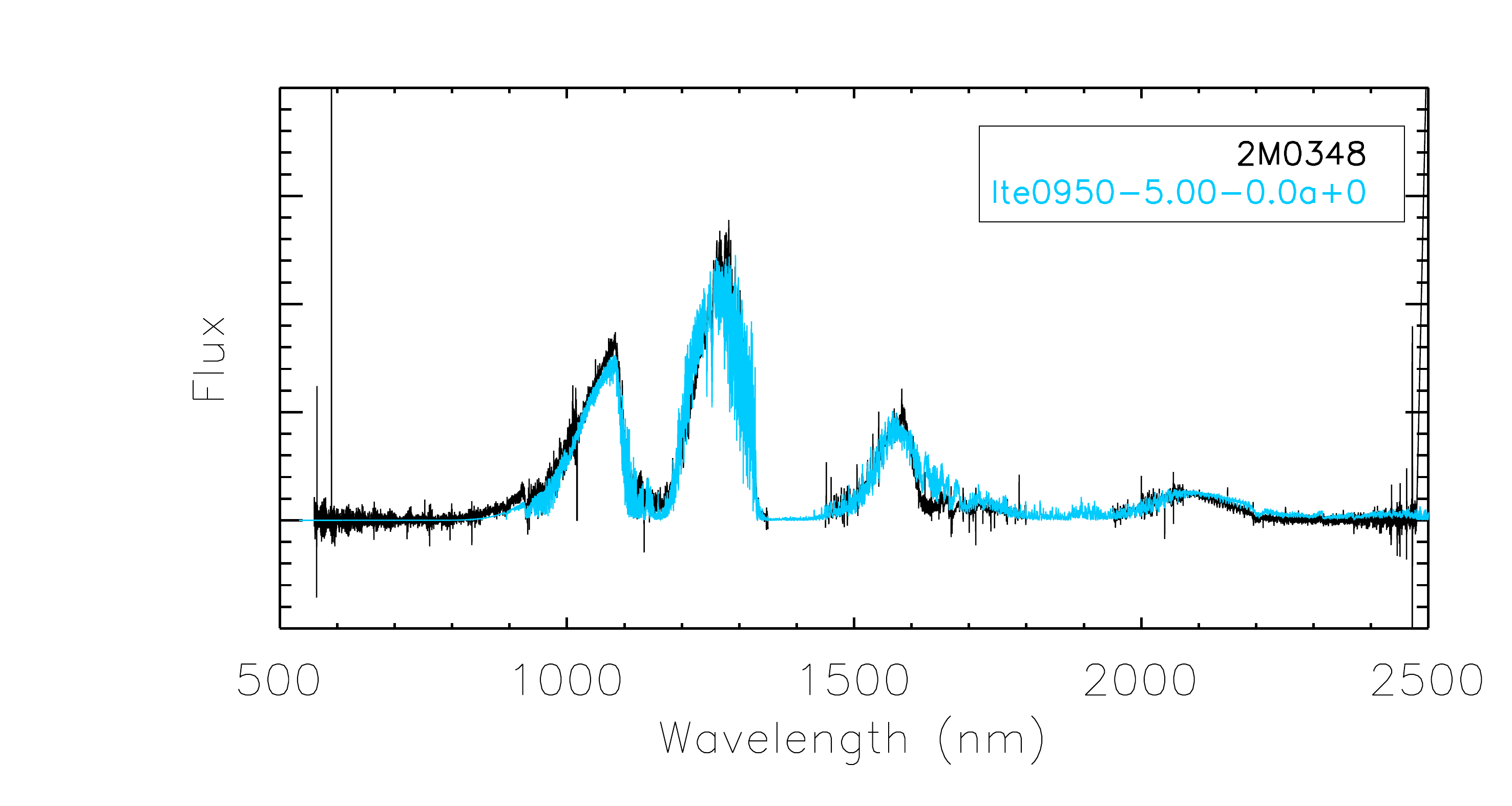}
				\includegraphics[width=0.45\textwidth]{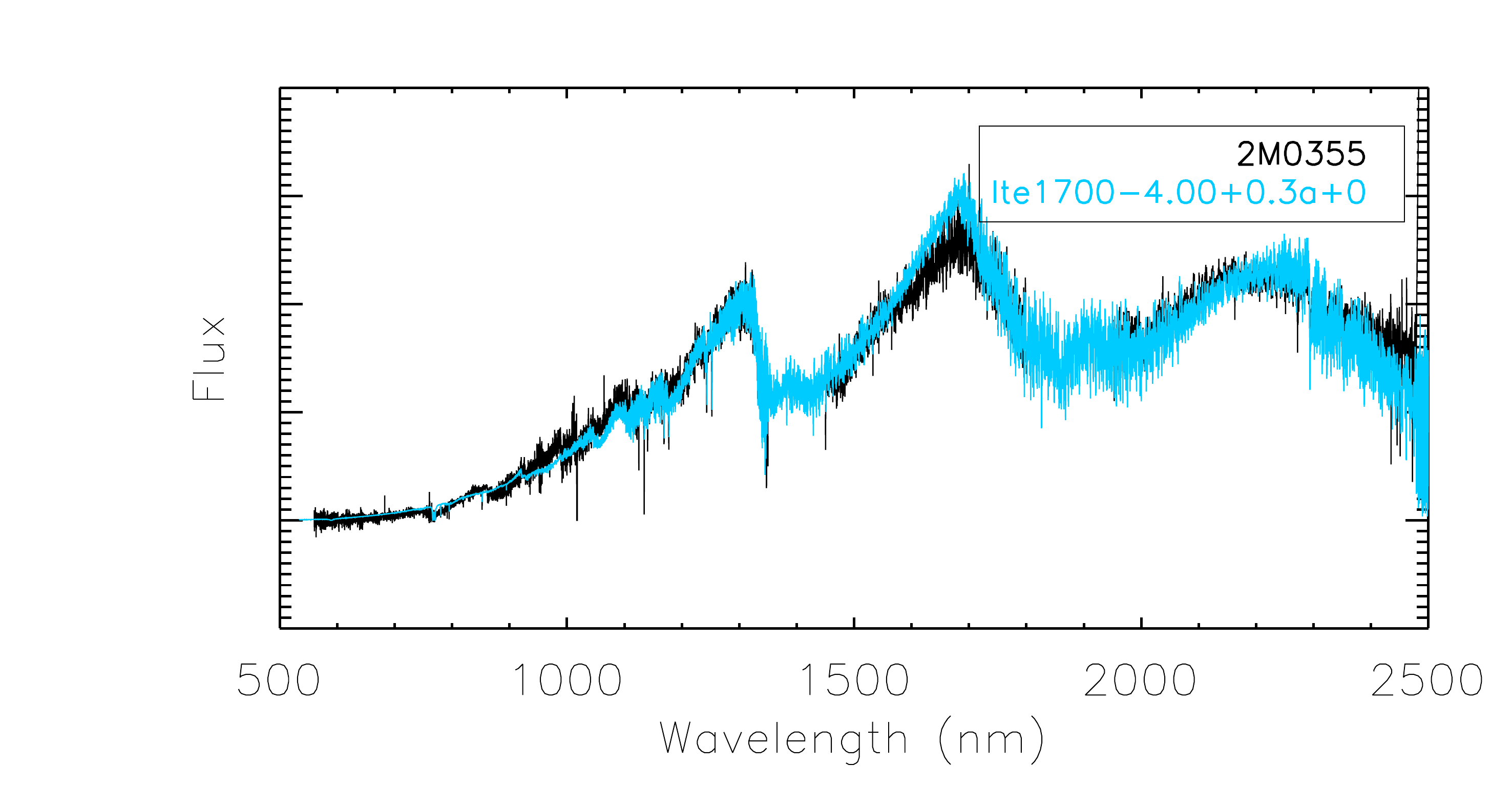}
				\includegraphics[width=0.45\textwidth]{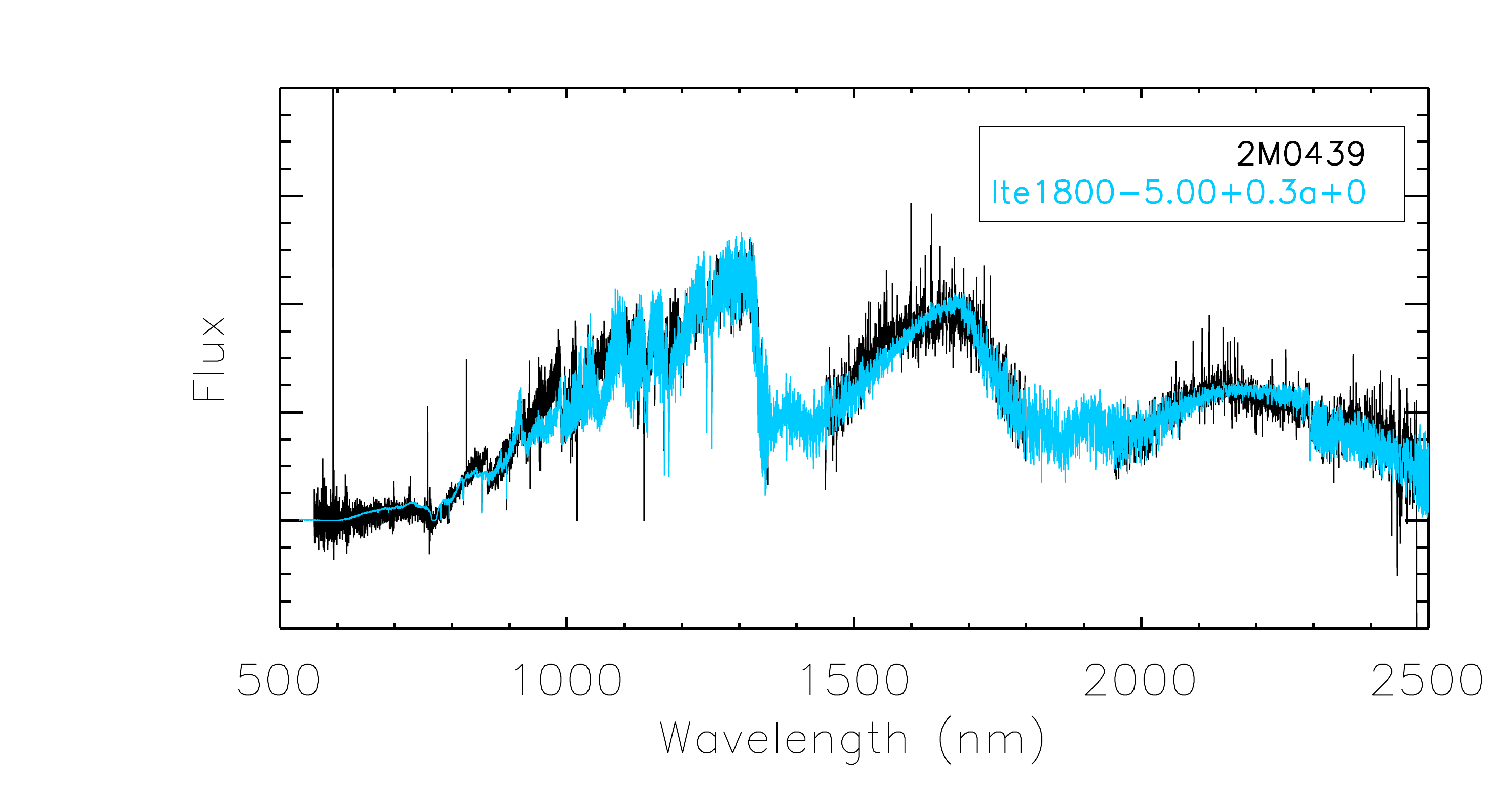}
				\includegraphics[width=0.45\textwidth]{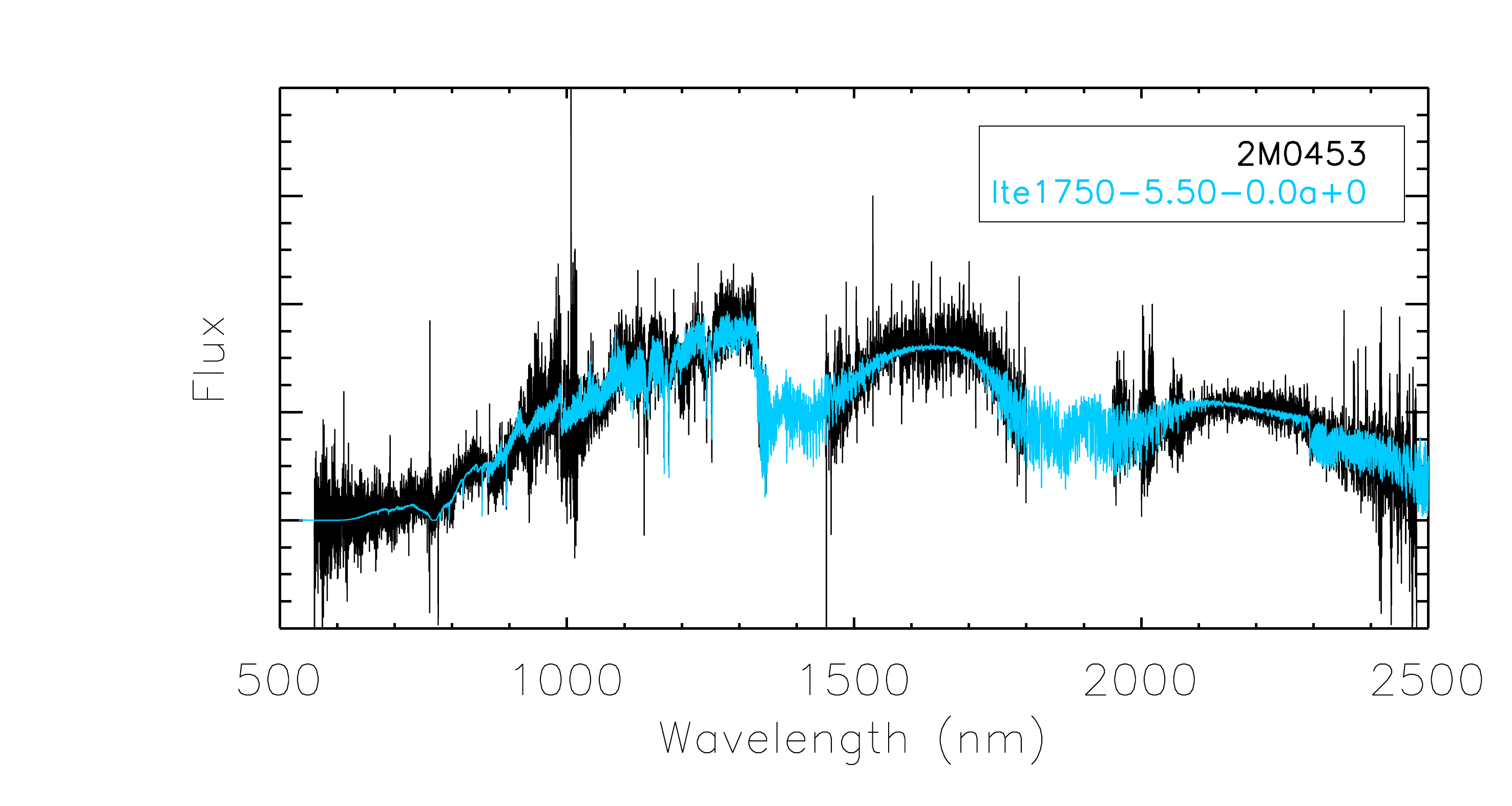}
				\includegraphics[width=0.45\textwidth]{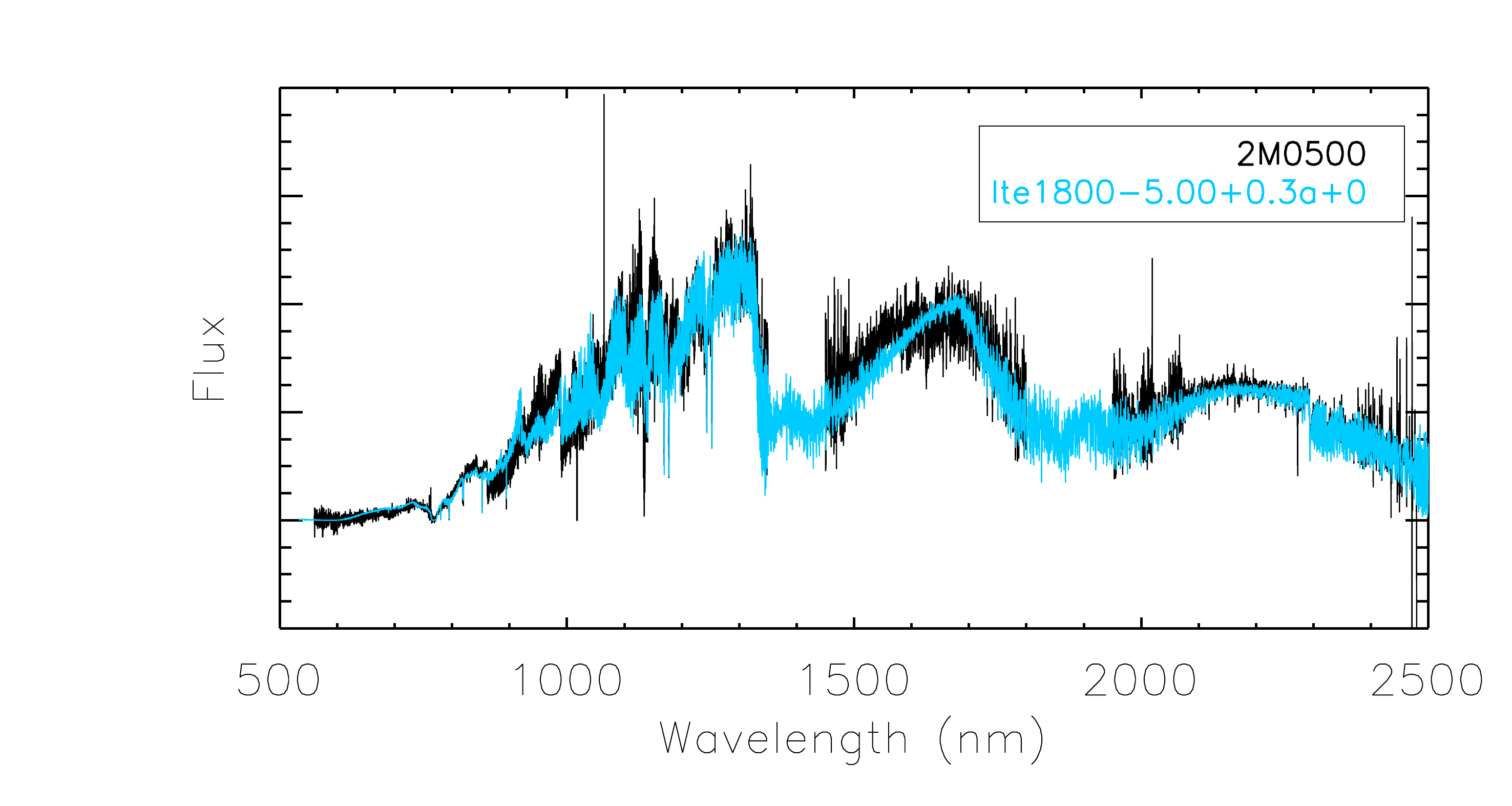}
				\includegraphics[width=0.45\textwidth]{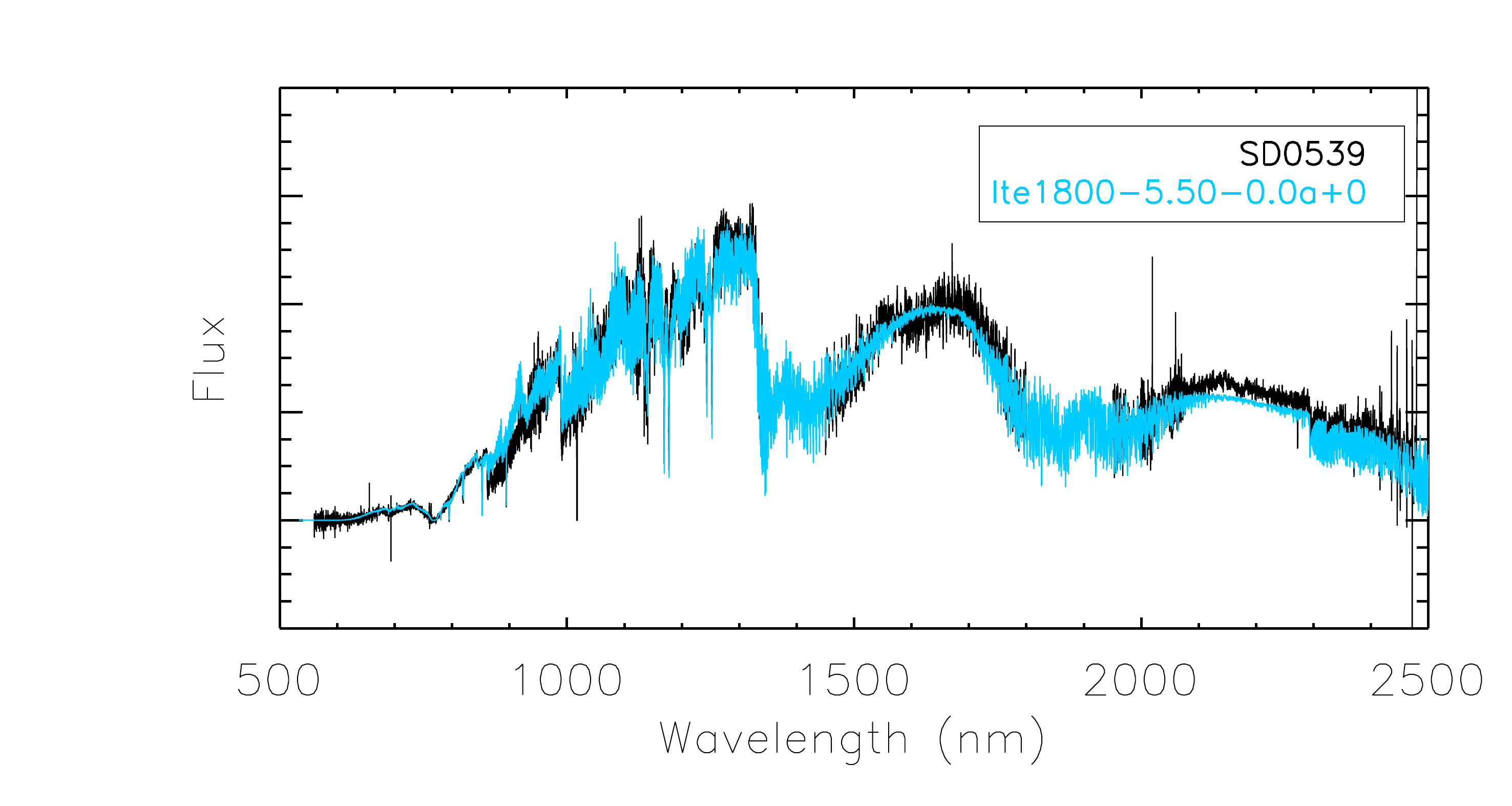}
				\includegraphics[width=0.45\textwidth]{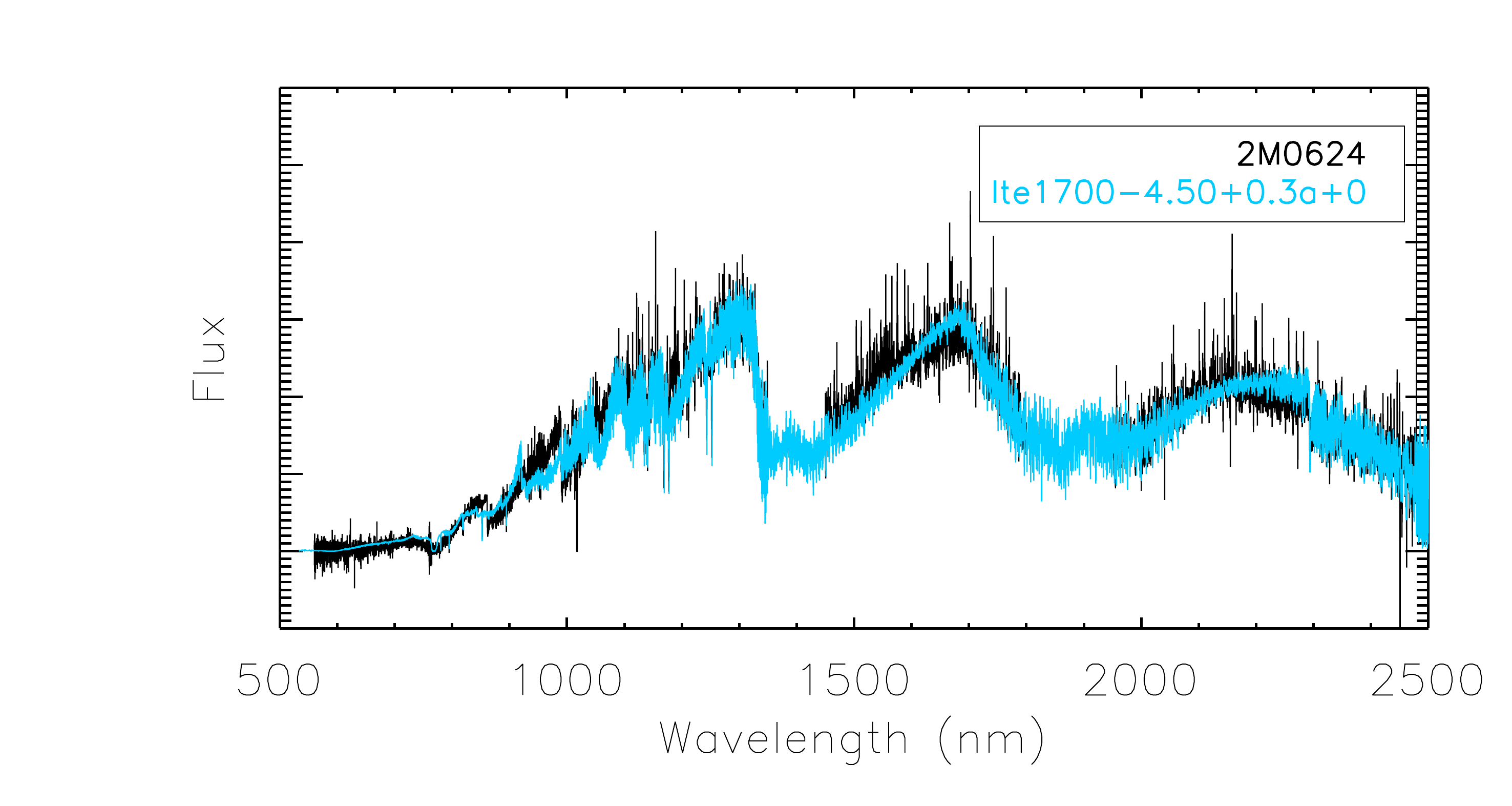}
				\includegraphics[width=0.45\textwidth]{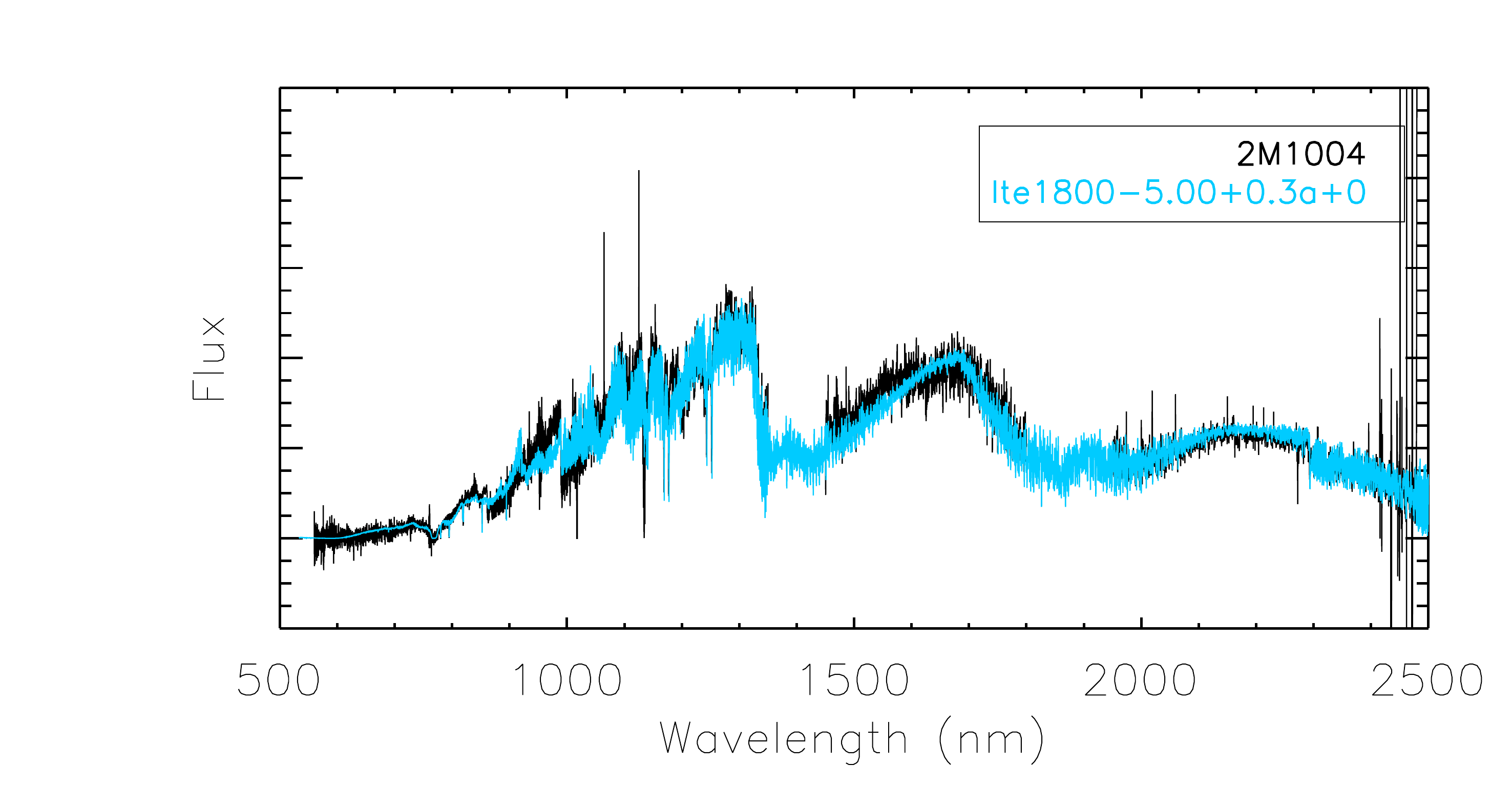}
				\caption{{Best matches to BT-Settl models 2014 found using equation \ref{chi},  as in Section \ref{L_T}}. Effective temperature, gravity, metallicity and alpha element enhancement are described in the model name strings as \texttt{lte-LOGG+[M/H]a+[ALPHA/H]}. The flux is F($\lambda$).}
				\label{model1}
			\end{figure*}

			\begin{figure*}
				
				\raggedright
				\includegraphics[width=0.45\textwidth]{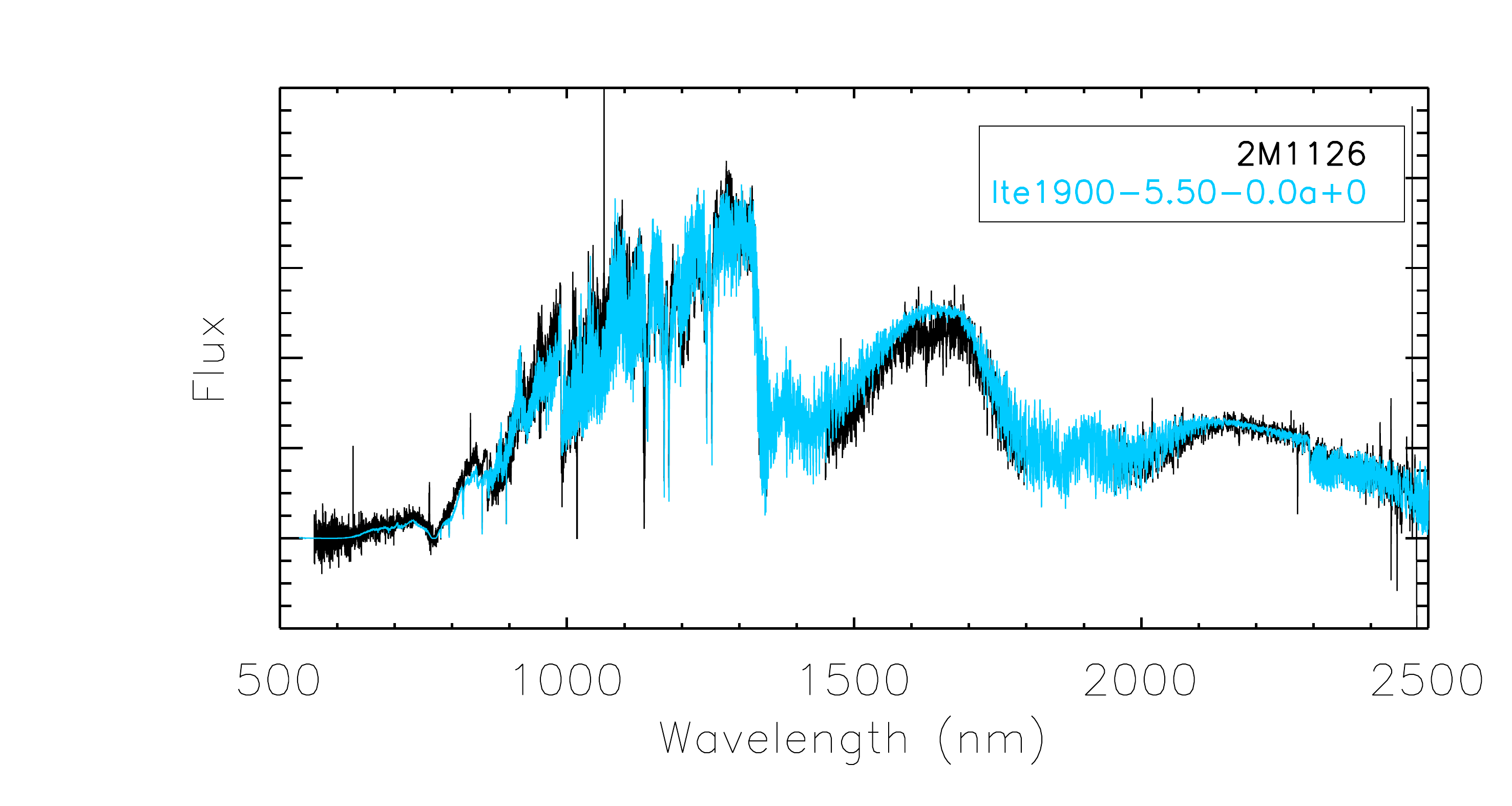}
				\includegraphics[width=0.45\textwidth]{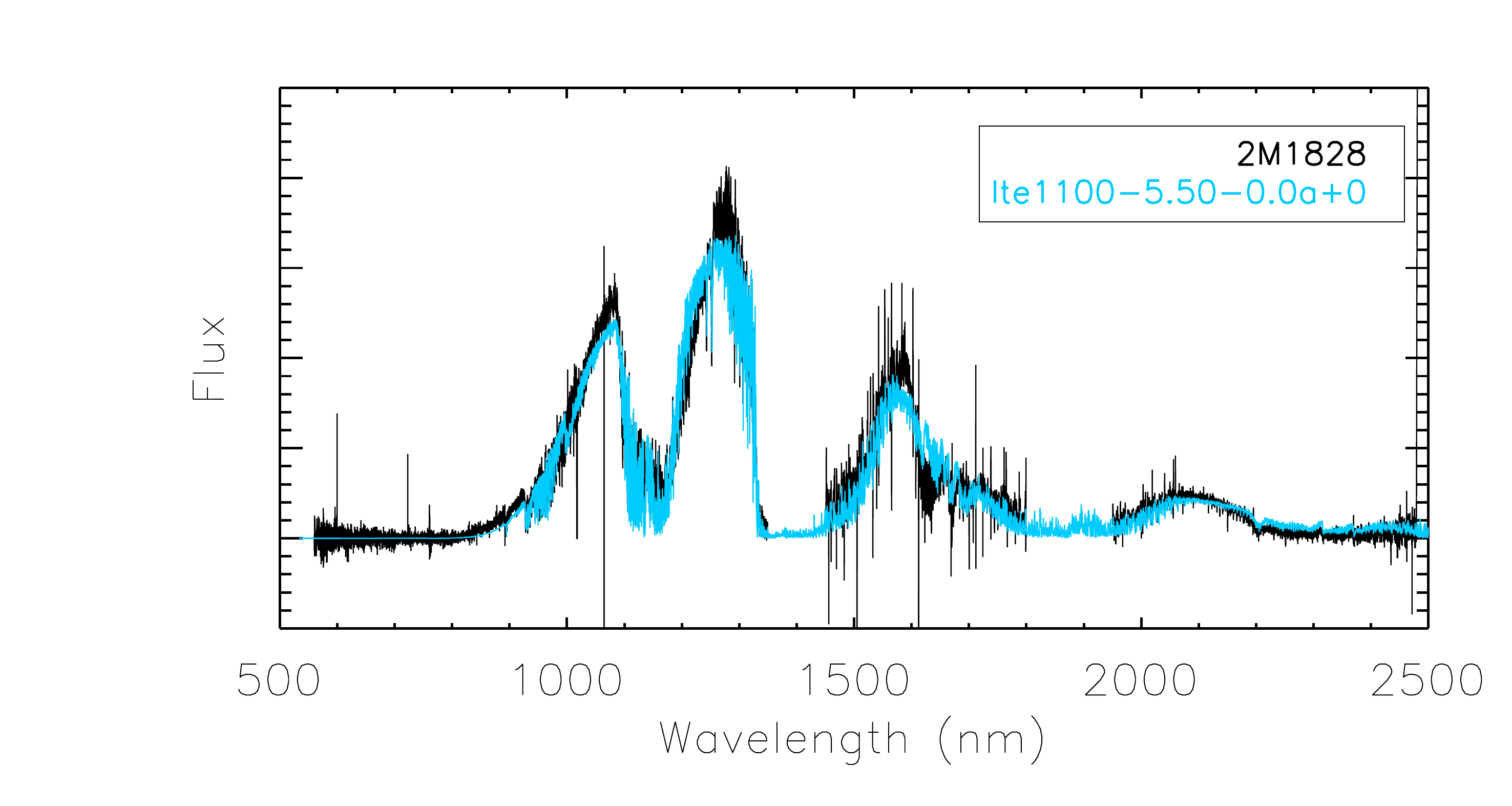}
				\includegraphics[width=0.45\textwidth]{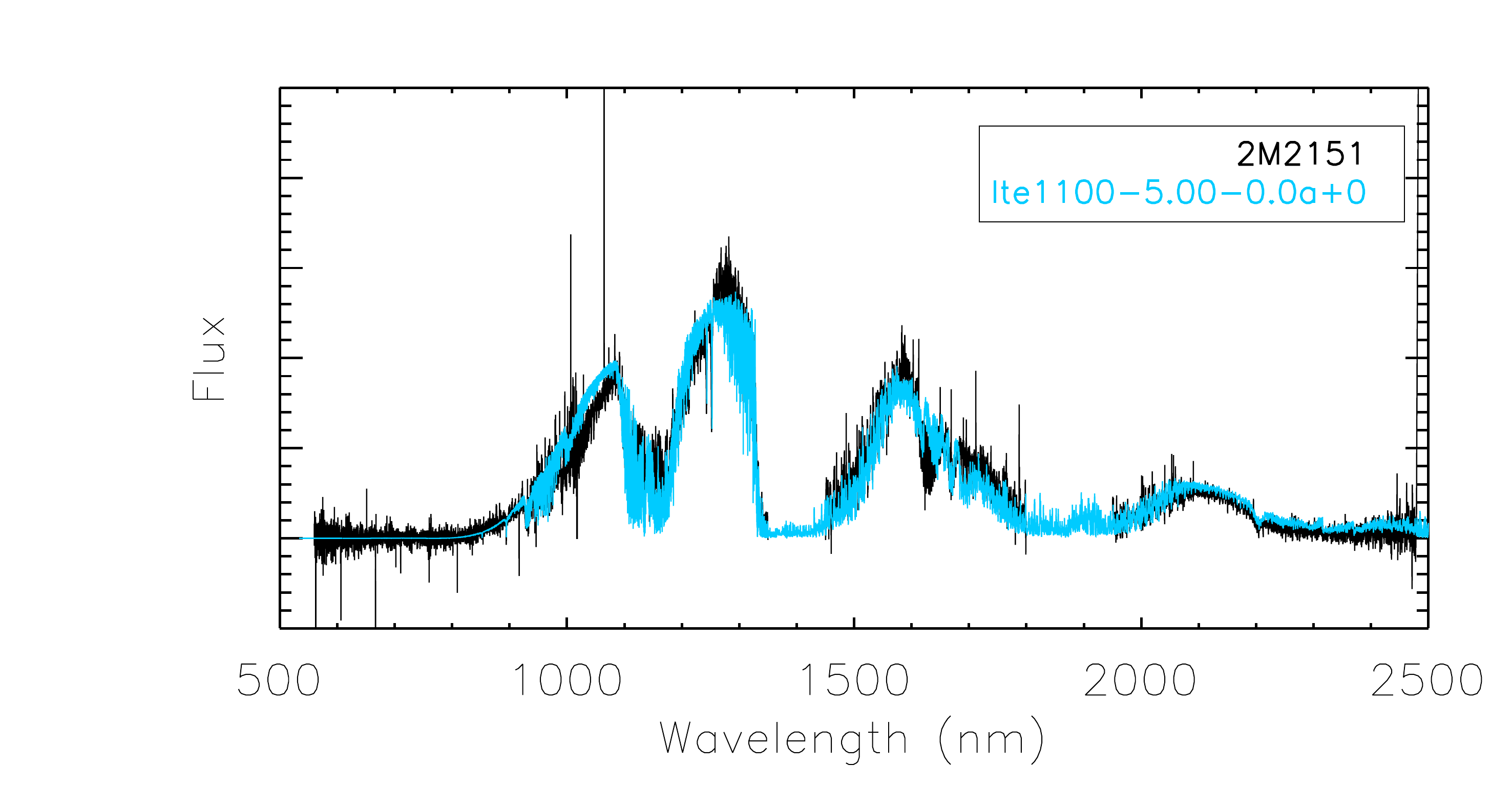}

				\caption{{Best matches to BT-Settl models 2014 found using equation \ref{chi},  as in Section \ref{L_T}}. Effective temperature, gravity, metallicity and alpha element enhancement are described in the model name strings as \texttt{lte-LOGG+[M/H]a+[ALPHA/H]}. The flux is F($\lambda$).}
				\label{model2}
			\end{figure*}

	\bsp
	
	\label{lastpage}
	
\end{document}